\documentclass[final]{iopart}
\usepackage{iopams}
\usepackage{fancyhdr}
\usepackage{graphicx}
\usepackage{bm}
\newcommand{\heb}{$^3$He-B }
\newcommand{\hea}{$^3$He-A }
\newcommand{\he}{$^3$He }

\newcommand{\heii}{$^4$He-II }

\begin{document}
\review{Dynamics of vortices and interfaces \\ in superfluid $^3$He}
\newcommand{\headstring}{Dynamics of vortices and interfaces in superfluid $^3$He}

\author{A. P.  Finne$^1$, V. B.  Eltsov$^{1,2}$, R.
H\"anninen$^{1,4}$,\\
N. B.  Kopnin$^{1,3}$, J. Kopu$^1$, M. Krusius$^1$, \\ M.
Tsubota$^4$, and G. E. Volovik$^{1,3}$}
\begin{center}

\address{$^1$Low Temperature Laboratory, Helsinki University of Technology,
P.O. Box 2200, FIN-02015 HUT, Finland}
\address{$^2$Kapitza Institute for Physical Problems, 119334, Moscow, Russia}
\address{$^3$Landau Institute for Theoretical Physics, 119334, Moscow, Russia}
\address{$^4$Department of Physics, Osaka City University, Osaka
558-8585, Japan}

\end{center}

\begin{abstract}
Rapid new developments have occurred in superfluid hydrodynamics
since the discovery of a host of unusual phenomena which arise
from the diverse structure and dynamics of quantized vortices in
$^3$He superfluids. These have been studied in rotating flow with
NMR measurements which at best provide an accurate mapping of the
different types of topological defects in the superfluid order
parameter field. Four observations are reviewed here: (1) the
interplay of different vortex structures at the first order
interface between the two major superfluid $^3$He phases, $^3$He-A
and $^3$He-B; (2) the shear flow instability of this phase
boundary, which is now known as the superfluid Kelvin-Helmholtz
instability; (3) the hydrodynamic transition from turbulent to
regular vortex dynamics as a function of increasing dissipation in
vortex motion; and (4) the peculiar propagation of vortex lines in
a long rotating column which even in the turbulent regime occurs
in the form of a helically twisted vortex state behind a
well-developed vortex front. The consequences and implications of
these observations are discussed, as inferred from measurements,
numerical calculations, and analytical work.
\end{abstract}

\submitto{\RPP}

\maketitle
\fancyhead[R]{\slshape\thepage}
\fancyhead[L]{\slshape \headstring}
\fancyfoot[L,C,R]{}
\pagestyle{fancy}

\tableofcontents
\newpage

\section{Introduction}

\subsection{Helium superfluids}\label{Sect-Helium-super}

Until 1972 the only known example of a truly inviscid fluid was
superfluid \heii at low flow velocities. Today its primacy is
challenged by the discovery of superfluid \he in 1972 and the
gaseous Bose-Einstein condensates in 1995. Nevertheless, in low
temperature physics \heii remains the epitome of a superfluid, the
benchmark to which to compare to. Its perfect inviscid flow is
known to persist only at velocities below some low critical limit
at which quantized vortex lines are formed. These are
topologically stable linear defects in the superfluid order
parameter field with extraordinary properties. One of them is
their turbulent flow, customarily known as superfluid turbulence
or quantum turbulence \cite{VinenTurbulence}, which appears, for
instance, when the applied flow velocity is suddenly increased
well above the critical limit. Recently the dynamics of quantized
vortex lines has gained renewed interest, activated by the
differences and similarities which have been discovered while
exploring the other superfluid systems, both superfluid \he and
Bose-Einstein condensates. A recent review by Vinen and Niemela
\cite{vinen_turb_review} summarizes with updated references our
understanding on vortex dynamics and turbulence in $^4$He-II.

Here we are concerned with the superfluid phases of $^3$He which
usher into superfluid hydrodynamics a broad spectrum of new
phenomena. These are associated with the structure and dynamics of
quantized vortices and other topological defects. As bulk liquid,
superfluid $^3$He can exist in three different phases, of which
the two major phases are $^3$He-A and $^3$He-B. The third phase,
$^3$He-A$_1$, exists at high magnetic fields around the zero-field
superfluid transition temperature $T_{\rm c}$ \cite{VW}. The flow
properties of \heb are isotropic in the absence of external
magnetic fields, resembling those of $^4$He-II with its quantized
vorticity. In contrast, \hea is highly anisotropic and the most
extraordinary superfluid of all that we know. Its applications as
a model system in physics have far reaching implications
\cite{volovik_droplet}. This review deals with recent observations
on vortex dynamics, primarily in $^3$He-B, which have been made in
uniformly rotating flow with noninvasive NMR measurement. Related
reviews can be found in Refs.
\cite{SalomaaVolovikRMP,neutronreview,briefreview}.

Although \heii and \heb both are isotropic helium superfluids and
in many respects rather similar, if compared to $^3$He-A,
nevertheless, important differences prevail which have profound
influence on the resulting superfluid hydrodynamics. The
implications from these differences have been appreciated only
lately. One of them concerns the vortex-core radius, whose length
scale in both cases is determined by the coherence length $\xi(T)$
of the superfluid state. In \heii the core radius is of atomic
scale $\sim 0.1$~nm, while in \heb it is $\gtrsim$ 10~nm and thus
at least two orders of magnitude larger. This difference is not
simply quantitative but has substantial impact on the interactions
of the vortex with the container wall, on critical velocities,
vortex formation, and surface pinning. The implications from this
difference became apparent in the first half of the 1990ies, when
single-vortex formation was observed in large open volume
measurements in $^3$He-B, but only in flow through micron-size
orifices in \heii.

The second major difference is the viscosity of the normal
component  in the two superfluids. In \heb it has oil-like
viscosity and is practically always in a state of laminar flow. In
contrast, the normal component of \heii is one of the least
viscous fluids known. Its flow becomes easily turbulent, which in
turn influences the flow of the superfluid component, resulting in
complicated mutual-friction coupled turbulence of the normal and
superfluid fractions. Thus the absence of turbulence in the flow
of the normal component of \heb amounts to a considerable
simplification at temperatures above the zero temperature limit
(where the normal component is present).

The third difference which influences profoundly the dynamics of
vortices is mutual friction dissipation, the damping which takes
place when a vortex moves with respect to the normal component. In
the Fermi superfluid $^3$He-B mutual friction between the vortex
and the normal component is mediated by fermionic quasiparticle
states in the vortex core
\cite{kopnin_mfhe3B,kopnin_mf,KL2tau,kopnin_rep}, the so-called
fermion zero modes. Their properties are described by a theory
similar to the BCS theory of superconductivity, according to which
the dimensionless temperature-dependent parameter $q(T)$, which is
the ratio of the dissipative and reactive components of the mutual
friction force, is a function of the normalized temperature
$T/T_{\rm c}$ and depends almost exponentially on temperature. It
crosses unity at around $T\sim 0.6\, T_{\rm c}$. At temperatures
above this division point the number of vortices is generally
found to remain constant in dynamic processes. In contrast, at
lower temperatures vortices become easily unstable in externally
applied flow which causes an increase in the vortex number owing
to superfluid turbulence. In comparison, in the Bose liquid \heii
mutual friction dissipation is small in the usual regime of
measurements and vortex dynamics is practically always turbulent.
The regular vortex number conserving flow might be expected only
within a few $\mu$K below the superfluid transition temperature
$T_\lambda$. From this temperature regime there are no experiments
available on vortex dynamics yet. Even there, the low viscosity of
the normal component might cause the coupled flow to become
turbulent.

In addition to their hydrodynamic differences, \heii and \heb
experiments often use different techniques to create and detect
vortex lines. The temperature required for superfluid \he is a
factor of 10$^3$ lower than for $^4$He-II. This sets restrictions
on the type of experiments that can be conducted on $^3$He
superfluids. Uniform rotation can be used in any temperature range
to create counterflow of the normal and superfluid components.
Owing to better control over vortex formation in $^3$He
superfluids, rotation has there proven to be a useful means to
apply flow.

As for vortex detection, in $^3$He superfluids nuclear magnetic
resonance (NMR) measurement provides a practical
\emph{noninvasive} method to count the number of vortex lines and
to study their dynamics. In \heb NMR methods can be used from
$T_{\rm c}$ down to about $0.2\,T_{\rm c}$, while in \hea
measurements at even lower temperatures should be possible. As the
temperature decreases measurements based on equilibrium state
techniques become increasingly less sensitive. This is the case
also in \heb NMR, where changes in the order parameter texture
from superfluid counterflow, vortices, and other control
parameters gradually vanish or saturate with decreasing
temperature. This is an unfortunate constraint, since today the $
T \rightarrow 0$ limit is of great interest in superfluid
hydrodynamics.

In the zero-temperature limit, where the normal component becomes
expo\-nent\-ially rarefied, the only measuring methods developed
so far for the study of vortices in \heb employ vibrating wires
\cite{turbulence_lancaster}, spheres \cite{schoepe_jltp}, or grids
\cite{TangleDecay}. These resonantly oscillating objects can be
employed as sensitive sensors of their hydrodynamic environment in
a quiescent He bath, for instance to create and detect vortices.
The oscillation is driven at amplitudes where the flow velocity at
the surface of the vibrating body exceeds the critical value for
Cooper-pair breaking \cite{VortexCreation}. In the
zero-temperature regime of ballistic quasiparticle motion, a
second resonant sensor oscillating at low drive in the linear
regime can then be used to track deviations in the exponentially
temperature-dependent equilibrium quasiparticle density
\cite{VWR-QP-Density} or the quasiparticles scattered from the
flow field around a vortex or a tangle of vortices
\cite{BeamTechniques}. These techniques have turned out to provide
efficient new tools for vortex studies \cite{LatestLancaster} and
are now in the forefront of future research. They can also be used
for constructing ultra-sensitive dark matter detectors
\cite{Dark-matter}. The lack of suitable measuring techniques has
also been an obstacle in vortex studies of $^4$He-II at the lowest
temperatures. A promising new development is here the use of
micron-size charged vortex rings for the analysis of different
vortex states. With this method both rotating arrays of
rectilinear lines and turbulent tangles can be distinguished and
monitored \cite{IonRings}.

\subsection{Novel phenomena in superfluid $^3$He hydrodynamics}\label{Novel-He}

Here we outline briefly the four main topics which are the subject
of this review. The first is concerned with the fundamental
difference in the structure of quantized vortex lines in the A and
B phases of superfluid $^3$He. This set of questions is peculiar
to superfluid $^3$He. It is the only presently known system where
vortices can be studied at a stable first order interface between
two coherent states which belong to the same order parameter
manifold. Here the phase of the order parameter is continuous
across the interface and thus vortices can, in principle, cross
the interface continuously. This is quite unlike other interfaces,
for instance between phase-separated layers of superfluid \he
above a solution of \he in superfluid $^4$He. In this latter case
the quantized vortices in the two layers belong to different
superfluid systems and can end at the interface with little
relation to each other.

$^3$He-A is an anisotropic liquid where, in a typical experimental
situation in a magnetic field, the vortex core is formed on a
length scale which is at least three orders of magnitude larger
than in $^3$He-B. This scale is not set by the pairing
interaction, but by the tiny dipolar coupling between the spin and
orbital momenta of the Cooper pairs. The structural length scale
of quantized vorticity is not the superfluid coherence length
$\xi(T) \gtrsim 10\,$nm, but the healing length $\xi_{\rm D} (T)
\gtrsim 10\,\mu$m associated with the dipolar spin-orbit coupling.
The typical A-phase vortex is doubly quantized, {\it i.e.} its
circulation is twice that of the $^3$He-B vortex. This difference
between the vortices poses a problem when they interact at the
interface between these two superfluids in a rotating sample: How
is the large core doubly-quantized A-phase vortex matched with the
small core singly-quantized B-phase vortex across the AB
interface? Measurements elucidating this question led to the
unexpected observation of dissipationless shear flow between the
two superfluids at the AB interface. The stability issue of this
superfluid shear-flow state is one the topics discussed in this
review.

The possibility of constructing the shear flow state arises from
the different conditions of vortex formation in the two
superfluids, owing to the large difference in vortex core radius.
The core of the $^3$He-B vortex is intermediate between that in
$^4$He-II and $^3$He-A, which leads to important consequences. On
one hand, being larger than the microscopic core in $^4$He-II,
pinning and surface roughness at bounding walls is not as
important as in $^4$He-II. With carefully chosen and prepared
container surfaces pinning sites can be avoided, so that pinned
remnant vortices do not exist. In such cases, substantial
vortex-free flow can be reached in a cylindrical rotating sample,
before intrinsic vortex formation starts to intervene at
relatively high critical velocities. In contrast, in $^4$He-II
vortex-free flow has generally little practical meaning because,
even at very low velocities, remnant vorticity leads to efficient
vortex formation.  An important exception is flow through a
sub-micron-size aperture in a thin membrane where vortices are
swept away from the immediate vicinity of the high-velocity flow
and do not have a chance to become pinned there \cite{varoquaux}.

On the other hand, the core radius of the $^3$He-B vortex is much
smaller than that of a continuous vortex in $^3$He-A. As a result,
the critical velocity for intrinsic vortex formation in $^3$He-B
is much larger than in $^3$He-A. This makes it possible to prepare
a flow state in which vortices are already forming on the A-phase
side of the AB interface, while on the B-phase side the
vortex-free irrotational Landau state persists. Such a situation
leads to a shear-flow state in which the superfluid components of
the two superfluids are sliding with respect to each other at the
AB interface. The relative flow of the two superfluids is
frictionless and, for the first time, provides a perfect
arrangement for the experimental investigation of the classical
Kelvin-Helmholtz instability which was theoretically predicted
hundred and fifty years ago (Sec.~\ref{ABInterfaceInRotation}).
The reason for this unique situation is that in conventional
viscous liquids the threshold for the Kelvin-Helmholtz
instability, where the formation of surface waves or ripplons on
the interface starts, is always obscured by the influence of
viscosity.

Unexpectedly, even in the perfect superfluid conditions, the
critical velocity of the AB-interface instability does not match
the classical result derived for ideal inviscid fluids
(Sec.~\ref{KHSection}). However, a modified criterion for the
onset of the instability  proved to be in excellent agreement with
the superfluid experiments, although it appears to lead to
paradoxical consequences at first glance. This instability
threshold is not determined by the velocity ${\bf v}_{\mathrm{s}2}
- \mathbf{v}_{\mathrm{s}1}$ of the relative superfluid motions in
the two liquids across the interface, but the instability would
occur even if the two liquids would have the same velocity or if
there is a single superfluid with a free surface. These new
features result from the two-fluid nature of the superfluid
liquid, from the presence of the superfluid and normal fractions.
The instability threshold is determined by the velocities
$\mathbf{v}_{\mathrm{s}1,2} - \mathbf{v}_\mathrm{n}$ of each
superfluid  with respect to the reference frame of the container
walls and thus with respect to the normal fractions of the two
liquids, which in thermodynamic equilibrium move together with the
walls. The free surface of a superfluid bath with respect to its
gas phase (or vacuum at the low temperatures) becomes unstable,
when in the reference frame of the normal component, the
superfluid velocity reaches the critical threshold value
\cite{Korshunov,Korshunov2002}. In the case of several superfluid
fractions ($i$) in the same liquid, such as neutron and proton
superfluids in a neutron star, the threshold is determined by some
combination of the superfluid velocities ${\bf v}_{\mathrm{s}i} -
\mathbf{v}_\mathrm{n}$ \cite{Abanin}.

Surprisingly, the superfluid Kelvin-Helmholtz instability has many
features in common with the instability of quantum vacuum beyond
the event horizon or even in the ergoregion of the black hole. The
ergoregion is defined as the region at the interface where the
energy of surface waves, or ripplons, is negative. The ripplon
excitations of the AB interface also provide a connection to the
presently popular idea in cosmology, according to which matter in
our Universe is confined to hypersurfaces, which are
multidimensional membranes, or {\em branes}, in a multidimensional
space. Branes can be represented by topological defects, such as
domain walls and strings, and by interfaces between different
quantum vacua. In our case, the brane is defined by the AB
interface between two quantum vacua -- the two superfluid phases
of $^3$He. The instability of the AB interface is in one-to-one
correspondence to the instability of quantum vacuum in the brane
world. It occurs in the ergoregion because of the interaction
between the matter on the brane (represented by ripplons) and the
matter in higher-dimensional space (represented by quasiparticles
in bulk superfluids).

Measurements on the AB interface instability also revealed new
properties about superfluid turbulence (Sec.~\ref{TurbulenceKH}),
which is the third main topic of this review. The later nonlinear
stage of the AB interface instability results in the injection of
a tight bundle of small vortex loops in the rapidly flowing
vortex-free $^3$He-B. It was found that the injection leads to
turbulence in $^3$He-B at temperatures below a critical onset
value $T_{\rm on} \sim 0.6\,T_{\rm c}$. The temperature of this
hydrodynamic transition turned out to depend only on the
dimensionless intrinsic parameter $q(T)$, the ratio of the
dissipative and reactive mutual friction components. The fact that
in superfluid hydrodynamics a transition to turbulence occurs as a
function of mutual friction dissipation at $q \sim 1$ was
discovered for the first time in \heb \cite{nature}. It divides
the evolution of the injected vortices to regular vortex number
conserving dynamics at $q\gtrsim 1$ and to turbulence at
$q\lesssim 1$. The main reason why the transition has not been
observed in superfluid $^4$He and only in $^3$He-B is the
favorable range of values of the parameter $q(T)$. This fortunate
coincidence, the presence of the hydrodynamic transition in the
middle of the experimentally accessible temperature range, makes
it possible to explore the dynamics in both flow regimes under
otherwise similar conditions.

Of particular significance has been the exponentially steep
temperature depend\-ence of the mutual friction dissipation. It
has allowed a whole new genre of studies on how turbulence
switches on, when one or a few vortices which are far apart, are
introduced in vortex-free flow. A further unexpected phenomenon is
the evolution and propagation of the vortices in a long rotating
cylinder or column after the turbulence has switched on. It turns
out that the propagation takes the form of a spiralling vortex
front which travels longitudinally and rotates azimuthally with
respect to the cylinder walls and thereby expands in the unstable
vortex-free state. Behind the front an ordered helically twisted
vortex bundle forms where the vortices are in a {\em force-free
configuration}. This twisted state is already close in energy to
the final state of solid-body rotation, to which it relaxes when
the vortex front has reached the end plate of the rotating
cylinder (Sec.~\ref{HelicalBundle}). Thus the front separates here
in effect the metastable vortex-free Landau state from the
equilibrium vortex state. The motion of the front and the
helically twisted state can be monitored with the NMR measurement.
These observations and their interpretation form the fourth topic
of the present review. They are not a special characteristic of
superfluid $^3$He-B, but apply equally to superfluid $^4$He-II,
for instance. However, they became possible in \heb because of
better control over vortex formation and the possibility to create
vortex-free flow at relatively high flow rate. A further
characteristic of these measurements is a longer sample cylinder
than has been used before in rotating measurements, with two
separate detectors, which made it possible to record changes in
the flow state in different parts of the rotating column as a
function of time.

The above issues have been in the forefront of recent research and
are in the focus of this review. They demonstrate new features of
superfluid hydrodynamics and often arise owing to the
multi-component order parameter of the \he superfluids. In such
cases they cannot be reproduced with the `classical' $^4$He-II
superfluid. However, some of these phenomena or their analogues
might be present in the new superfluid states of gaseous bosonic
or fermionic atom clouds.

\section{Hydrodynamics of rotating helium superfluids}

\subsection{Helium superfluids and vortex
lines}\label{sect-quantvortices}


Customarily $^4$He-II is described with a wave function
$\psi=|\psi| \, e^{i\varphi}$, where $\varphi$ is the phase
factor. The superfluid velocity is then defined as the gradient of
the phase, $\mathbf{v}_\mathrm{s}=(\kappa /2\pi)\nabla \varphi$
where $\kappa =2\pi \hbar/m_4$. Since the curl of a gradient
vanishes identically, $\nabla\times \mathbf{v}_{\rm s} =0$, the
bulk superflow is irrotational. In principle rotational flow is
thus excluded, but by forming quantized vortex lines the
condensate can accommodate to rotating flow. In its simplest form
a line vortex is a stable string-like object with a central hard
core where the order parameter vanishes in the center, thus
forming a line singularity in the coherent order-parameter field.
Since the condensate phase changes by $2\pi n$ around the core,
where $n$ is an integer number, the circulation of the superfluid
velocity around the vortex core is quantized:
\begin{equation}
\oint d{\bf r}\cdot \mathbf{v}_\mathrm{s}=n\kappa~,
\label{QuantizationCirculation}
\end{equation}
and $\kappa$ plays the role of the circulation quantum. This
persistent superfluid current around the core stores kinetic energy,
providing the vortex with an energy per unit length, or line
tension, which equals
\begin{equation}
\epsilon_{\rm v} = {1 \over 2} \, \rho_{\rm s} \int_{r_{\rm c}}^{r_{\rm
v}} dr (2\pi r) v_{\rm s}^2 = {\rho_{\rm s} \kappa^2 \over 4 \pi} \,
n^2 \, \ln{ \left(r_{\rm v} \over r_{\rm c} \right)}.
\label{VortexTension}
\end{equation}
Here the upper ($r_{\rm v}$) and lower ($r_{\rm c}$) cutoffs are
determined by the inter-vortex distance and the core size of the order
of the superfluid coherence length $\xi$, respectively.  Vortex lines
with a singly-quantized structure $n=1$ are thus energetically
favourable. The superfluid hydrodynamics which follows from the
introduction of the quantized vortex lines has been described in
textbooks \cite{tilley,donnelly}.

The order parameter in superfluid $^3$He relates to the wave function
of the Cooper pairs, and has a complicated internal
structure. Nevertheless, in \heb it still contains an explicit phase
variable $\varphi$. The above considerations remain valid with the
exception that the circulation quantum is now given by $\kappa =2\pi
\hbar/(2m_3) = 0.066$~mm$^2$/s, where $2m_3$ is the mass of the two
$^3$He atoms in the Cooper pair, rather than the single atom mass
$m_4$ in the $^4$He-II circulation quantum $\kappa =
0.099$~mm$^2$/s. In what follows we use the same symbol $\kappa$ to
denote the circulation quantum in both $^4$He and $^3$He, with
appropriate values for each particular fluid. The superfluid order
parameter does not vanish in the $^3$He-B vortex core, but the order
parameter state in the core is different from that in the bulk.  Two
different vortex-core structures are known to exist: an axisymmetric
core at high temperatures and high pressures, and a nonaxisymmetric at
low temperatures
\cite{core_trans_diag,ThunebPRL1986}. A first-order phase
transition, which under equilibrium conditions occurs at
0.60~$T_\mathrm{c}$ at 29~bar pressure \cite{core_trans_diag},
separates these two core structures. At low pressures $P\lesssim
15$~bar only the nonaxisymmetric core exists.  In both cases the core
radius $r_{\rm c}$ is approximately equal to the coherence length $\xi
\gtrsim$~10~nm. An interesting curiosity to note is that this
transition was the first phase transition ever observed within a
defect, when it was discovered in 1981
\cite{CoreTransitionDiscovery}. A third vortex structure in bulk
$^3$He-B is the spin-mass vortex, a combination of a linear and a
planar defect with both spin and mass flow currents around its
core \cite{spinmass_prl,spinmass,neutronreview}. It will not be
discussed in this review.

\begin{center}
\begin{figure}[t]
\centerline{\includegraphics[width=0.7\linewidth]{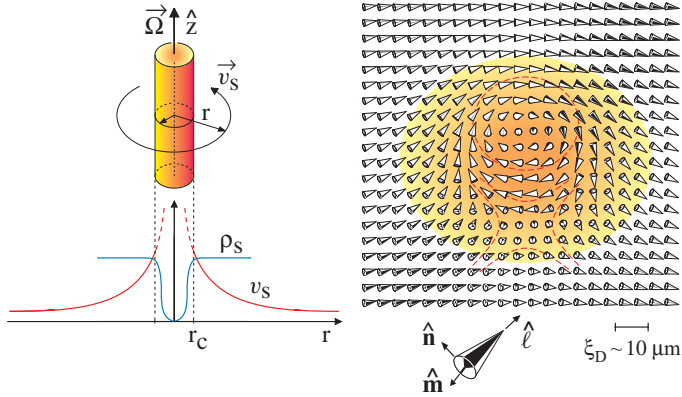}}
\caption{{\it (Left)} Macroscopic structure of quantized vortex
line in He superfluids. The core radius $r_{\rm c}$ is on the
order of the superfluid coherence length in $^4$He-II $(\xi \sim
0.1\,$nm) and $^3$He-B $(\xi \gtrsim 10\,$nm), but in $^3$He-A the
length scale is the healing length of the dipolar spin-orbit
interaction ($\xi_{\rm D} \gtrsim 10\,\mu$m). {\it (Right)}
Orbital order parameter texture of the soft core of the
double-quantum vortex in $^3$He-A in magnetic field. The cones
indicate the local direction and rotation of the orbital order
parameter triad of unit vectors ${\hat{\bf l}}, {\hat{\bf m}},
{\hat{\bf n}}$. The topological winding number of the ${\hat{\bf
l}}$ texture is $n=2$. The texture is nonaxisymmetric: it is
composed of a circular half, or {\it meron}, and a hyperbolic
meron, each with $2\pi$ circulation.} \label{ATCSymmetricFig}
\end{figure}
\end{center}

$^4$He-II and $^3$He-B are traditional Landau superfluids in that
their superflow is potential, $\nabla\times \mathbf{v}_{\rm s}=0$,
unless vortex line defects are present. In $^3$He-A, where the phase
and the orbital structure (represented by the orbital vector $\hat{\bf
l}$) of the order parameter are linked together, this condition is no
longer strictly satisfied. Instead, the so-called Mermin-Ho relation
holds \cite{MerminHo}:
\begin{equation}
\nabla \times \mathbf{v}_\mathrm{s}= {\hbar\over 4 m_3}
  \epsilon_{ijk} \hat{l}_i
(\nabla \hat{l}_j \times \nabla \hat{l}_k).
\label{MHEquation}
\end{equation}
This implies that rotational superfluid flow can be accomplished
via an inhomogeneous order-parameter texture $\hat{\bf l}({\bf
r})$. However, the energy cost of the necessary spatial
variations, resulting from the rigidity of the order parameter,
gives rise to a finite critical superflow velocity also in this
system. At this velocity, vorticity with continuously winding
structure of the order parameter orientation is formed so that in
most cases no hard vortex core is involved. In the simplest
   form the structure of an isolated continuous vortex has
the following spatial distribution of the orbital ${\hat{\bf
l}}$-field:
\begin{equation}
{\hat{\bf l}}(\rho,\phi)={\hat{\bf z}} \cos\eta(\rho) + {\hat{\brho}}
\sin\eta(\rho)~.
\label{lTextureContVortex}
\end{equation}
Here $\hat{\bf z}$, ${\hat{\brho}}$ and ${\hat{\bphi}}$ are the
unit vectors  of the cylindrical coordinate system; $\eta(\rho)$
changes from $\eta(0)=0$ to $\eta(\infty)=\pi$.  This winding
${\hat{\bf l}}$ texture  forms the so-called continuous soft core
of the  vortex \cite{dqvnature}, since it is in this
region where the non-zero vorticity of superfluid velocity is
concentrated:
\begin{equation}
\mathbf{v}_\mathrm{s}(\rho,\phi)=  {\hbar\over 2 m_3
\rho}[1-\cos\eta(\rho)]~{\hat {\bphi}}~~,~~\nabla\times
\mathbf{v}_\mathrm{s}=  {\hbar\over 2 m } \sin\eta ~\left(
{\partial \eta \over \partial \rho} \right) ~{\hat {\bf z}}~.
\label{v_sContVortex}
\end{equation}
The circulation of the superfluid velocity around a contour enclosing
the soft-core region is quantized, $\oint d{\bf r}\cdot
\mathbf{v}_\mathrm{s}=2\kappa$, corresponding to the quantization
number $n=2$. Thus the object described by
Eq.~(\ref{lTextureContVortex}) is a continuous double-quantum
vortex. By following the ${\hat{\bf l}}$ field across the cross
section of the soft-core texture, it is noted that the ${\hat{\bf l}}$
vector goes through all possible orientations on the unit sphere. Such
a topology of the vortex cross section in two spatial dimensions is
known as a {\it skyrmion}.
\index{vortex!vortex-skyrmion} \index{skyrmion!vortex-skyrmion}

In the magnetic field of the NMR measurements the continuous
vortex is deformed and its structure is nonaxisymmetric, see
Fig.~\ref{ATCSymmetricFig}. However, its topology is robust to
deformations, and the circulation remains the same: $\oint d{\bf
r}\cdot \mathbf{v}_\mathrm{s}=2\kappa$. It is important to note
that, since even in the soft-core region the order parameter
retains its bulk structure, the core size of the continuous \hea
vortex is not set by the coherence length $\xi \gtrsim 10\,$nm of
the superfluid state. Instead, the relevant length scale is the
three orders of magnitude larger dipolar healing length $\xi_{\rm
D} \gtrsim 10~\mu$m which originates from the spin-orbit coupling.

Using the two halves of the skyrmion texture, the circular and
hyperbolic {\it merons} (Fig.~\ref{ATCSymmetricFig}) as basic
building blocks, other structures of continuous vorticity can be
formed. An example are the various continuous periodic vortex textures
in zero or low magnetic field
\cite{VortexDiagram,VortexDiagramTheory}. Another important structure
is the vortex sheet
\cite{VorSheetPRL,VorSheetLong} which competes for living space
with the double-quantum vortex line. A concise lexicon of these
various structures can be found in Ref.~\cite{Lexicon}.

The concept of the quantized vortex line dates back to Lars
Onsager (1949) \cite{Onsager} and Richard Feynman (1955)
\cite{Feynman} who found that the Landau irrotationality
requirement $\nabla\times \mathbf{v}_{\rm s}=0$ has to be lifted
at singular lines where $\nabla\times \mathbf{v}_{\rm s} \neq 0$.
In the case of $^3$He-A these principles were put to a severe test
which they finally survived when, in the context of the work of
Mermin and Ho in 1976 \cite{MerminHo}, Chechetkin (1976)
\cite{Chechetkin} and Anderson and Toulouse (1977) \cite{AT} the
first example of a continuous vortex texture was proposed.

\subsection{Vortex states in rotating
superfluid}\label{RotSuperFluid}

The identification of the vortex structures of superfluid $^3$He,
and of the phase transitions separating these different
structures, is based to a large extent on NMR measurements on a
rotating sample. In rotation the vorticity $\nabla\times
\mathbf{v}_{\rm s}$ is aligned parallel to the rotation axis
$\mathbf{\Omega}$ and generally forms a regular array over the
cross section of the cylindrical sample. This is a particularly
simple situation where both the structural and dynamic properties
of these vortex structures can be analyzed.

\begin{figure}[t]
\begin{center}
\centerline{\includegraphics[width=0.5\linewidth]{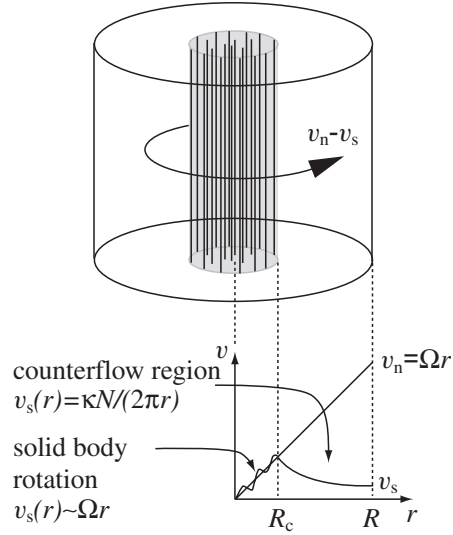}}
\caption{Schematic representation of a vortex cluster confined by
the azimuthally circulating counterflow of the normal and superfluid
components to the center of the rotating sample. The areal density of
rectilinear singly-quantized vortex lines within the cluster is
$2\Omega/\kappa$ and thus their number in a cluster of radius
$R_\mathrm{c}$ is $N=\pi R_{\mathrm{c}}^2\,
2\Omega/\kappa$.}\label{cluster}
\end{center}
\end{figure}

The minimum energy configuration in rotation is the state with the
equilibrium number of rectilinear vortex lines $N_{\mathrm{eq}}$,
which on average  mimics solid-body rotation of the superfluid,
{\it i.e.} $\langle \mathbf{v}_\mathrm{s}\rangle={\bf\Omega}\times
{\bf r}$, or $\langle\nabla\times
\mathbf{v}_\mathrm{s}\rangle=2{\bf\Omega}$. Since
$\langle\nabla\times \mathbf{v}_\mathrm{s}\rangle=n\kappa n_{\rm
v}$, the vortex density in the bulk is $n_{\rm v} =
2\Omega/(n\kappa)$. The formation of a new vortex is associated
with an energy barrier that has to be overcome before an
elementary vortex loop can be nucleated. At sufficiently low
applied flow velocities this is not possible, and metastable
states with a vortex number $N$ smaller than $N_{\mathrm{eq}}$ can
be formed.  These consist of a central vortex cluster
(Fig.~\ref{cluster}), with any number of vortex lines $0 < N \leq
N_{\rm eq}$. Within the cluster the rectilinear lines are packed
to their equilibrium density $n_{\rm v} = 2\Omega/(n\kappa)$,
confined by the counterflow of the normal and superfluid
components which circulates around the cluster with the velocity
$\mathbf{v} = \mathbf{v}_{\rm n} - \mathbf{v}_{\rm s} = [\Omega r-
n\kappa N/(2\pi r)]\,{\hat{\bphi}}$. The first term is the
velocity of the normal component, locked to co-rotation with the
cylindrical container (with radius $R$), while the second term
arises from the combined persistent superflow of the $N$
rectilinear vortex lines in the central cluster. An extreme case
is the Landau state -- the vortex-free state with $N=0$ and
$v_{\rm s} = 0$ (as expressed in the rest frame of the
laboratory). This is the state of maximum kinetic energy in the
rotating frame. In many of the rotating experiments described
below it is the initial state, the starting point for the
measurements. Independently of $N$, the maximum counterflow
velocity is at the cylindrical wall at $r = R$. This we call the
velocity of the externally applied flow or the rotation drive of
the cylindrical rotating container.

At constant rotation the stationary states are thus the
equilibrium vortex state and the various metastable states with a
depleted vortex cluster. In an ideal cylinder, which is exactly
aligned parallel to the rotation axis, it is possible to have more
than the equilibrium number of vortices $N_{\rm eq}$, owing to a
finite annihilation barrier \cite{AnnihilationBarrier}.
Experimentally the exact value of $N_{\rm eq}$ is important for
calibrating the measuring signals from a state with a well-defined
configuration and number of vortices. Transient time-dependent
rotating states are created in accelerating or decelerating
rotation \cite{VortexMotion}. In Secs.~\ref{TurbulenceKH} --
\ref{HelicalBundle} we describe measurements where rotation is
kept constant and the dynamics evolves from vortex seed loops
which have been introduced by external means into initially
vortex-free counterflow.

\subsection{Critical velocity of vortex formation}
\label{CriticalVelocity}

The lowest critical velocity in a rotating superfluid is that at
which the free energy of a rectilinear vortex line first becomes
negative in the container frame. The corresponding angular
velocity is known as the Feynman critical velocity,
$\Omega_\mathrm{c1}= \kappa/(2\pi R^2)\, \ln{(R/r_{\rm c})}$
\cite{Feynman}. It is analogous to the critical field $H_{\rm c1}$
for type II superconductors. For a rotating cylinder with a radius
of a few mm, $\Omega_\mathrm{c1} \sim 10^{-3}$ rad/s and is thus
very small. However, although at $\Omega > \Omega_\mathrm{c1}$ it
becomes energetically favorable to introduce vortices in the
rotating sample, some mechanism for their formation is required. A
number of such mechanisms exist, owing to sources both extrinsic
and intrinsic to the superfluid itself, each with its
characteristic critical velocity. For more discussion see
Ref.~\cite{VortexFormation}. Here we only summarize the basic
ideas important for the overall picture.

\begin{figure}[t]
\begin{center}
\centerline{\includegraphics[width=0.8\linewidth]{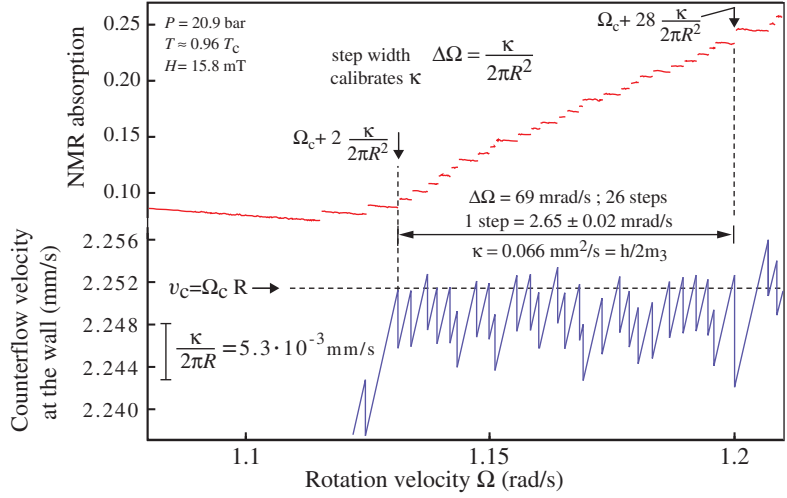}}
\caption{Measurement of single vortex formation as a function of
the applied rotation velocity $\Omega$ at high temperatures in
$^3$He-B \cite{SingleVortexSteps}. {\it (Top)} The vertical axis
shows the NMR absorption in the Larmor region of the $^3$He-B
spectrum. Vortex formation starts with the first step-like
increase, but the critical threshold at $\Omega_{\rm c}$ is
identified from the third step where the critical flow velocity in
the bottom plot reaches a more stable value. {\it (Bottom)}
Counterflow velocity at the cylinder wall $v = \Omega R - \kappa
N/(2\pi R)$, where N is the number of steps already formed. The
maximum possible value of counterflow for this sample container is
defined by the horizontal dashed line, $v = v_{\rm c} =
\Omega_{\rm c} R$.}\label{VortexSteps}
\end{center}
\end{figure}

In practice, in $^3$He-B the lowest critical velocity $\Omega_{\rm
c}$, which controls the formation of vortices, is found to depend
on the shape and size of the container and the roughness of its
surfaces. The simplest and most ideal case is a smooth-walled
cylinder which is mounted with its symmetry axis as parallel to
the rotation axis as possible. The surface quality is dependent on
the choice of material, the fabrication of the seams in the
corners, and the cleanliness of the walls. Even residual gases,
such as air or water, will condense on the wall during cool down,
form small crystallites, and may determine $\Omega_{\rm c}$. In a
good sample cylinder of typical radius 2 -- 3\,mm, $\Omega_{\rm
c}$ is relatively high, of order 1 -- 4\,rad/s, so that large
vortex-free counterflow can be achieved before the first vortex is
formed. In the worst case extrinsic sources govern $\Omega_{\rm
c}$. For instance, it can be determined by some pinning site, a
piece of dirt, at which a remnant vortex may remain pinned
indefinitely. If this site is occupied and $\Omega$ is increased
to the critical value associated with the site, the pinned remnant
loop will start to evolve.

In the most favorable case $\Omega_{\rm c}$ arises from a
combination of extrinsic and intrinsic reasons, if vortex
formation takes place at a sharp surface asperity in the form of a
pointed spike \cite{SingleVortexSteps,VortexFormation}. At a very
sharp spike the local velocity can exceed the average velocity at
the wall by one to two orders in magnitude. Thus superfluidity
will be broken first at this location when $\Omega$ is increased
to $\Omega_{\rm c}$, and a small vortex loop is formed
\cite{KopninSoininen}. The loop then evolves to a rectilinear
vortex line and reduces the counterflow velocity at the cylinder
wall to a sub-critical value $v = \Omega_{\rm c} R - \kappa/(2\pi
R)$.  If $\Omega$ is increased further by external means, vortex
formation occurs recurrently at the same site every time when the
counterflow reaches the critical value $v_{\rm c} = \Omega_{\rm c}
R$. Here $v_{\rm c}$ is therefore the limit for vortex-free flow
in this container. An example of such vortex formation in
single-quantum events as a function of $\Omega$ is illustrated by
the staircase pattern in Fig.~\ref{VortexSteps}. This measurement
has been performed in a container with $R=2\,$mm
\cite{SingleVortexSteps}. A similar measurement with $^4$He-II has
been demonstrated only with flow through orifices of sub-micron
size \cite{varoquaux,packard}.

An estimate of the intrinsic critical velocity, and of the energy
barrier which inhibits the formation of an elementary vortex
loop, can be obtained from the following simple consideration
\cite{VortexFormation}. The barrier is determined by the smallest
possible vortex ring. Since the radius of such a ring cannot be
smaller than the core radius (of the order of $\xi$), the energy of
the smallest ring can be estimated from Eq.~(\ref{VortexTension}) as
$E
\sim \rho_{\rm s} \kappa^2 \xi$. This gives $E/k_{\rm B} T \sim (\xi/a) (T_{\rm F}/T)$,
where we have used $\rho_{\rm s} \sim m/a^3$ for the superfluid
density and $T_{\rm F} = \hbar^2/2ma^2 k_{\rm B} \sim 1\,$K for the
degeneracy temperature of the quantum Fermi liquid, with $a$ as the
interatomic distance. For $^3$He-B we obtain $E/k_{\rm B}T >
10^5$. This should be compared to a similar estimate $E/k_{\rm B} T >
1$ for $^4$He-II, where the core size and the coherence length are
$\xi \sim a$.

How to overcome such an energy barrier \cite{nucleation,Langer}?
The rate for thermal activation over the barrier is $\propto
\exp(- E/k_{\rm B}T)$, and thus the barrier is practically
impenetrable by thermal activation or quantum tunneling
\cite{tunneling} at the appropriate temperatures for $^3$He
superfluids ($T\sim1$~mK)
\cite{VortexFormation,SingleVortexSteps}. Both mechanisms become
effective only when the local velocity at the asperity reaches a
value extremely close to the threshold where the energy barrier
disappears, and the hydrodynamic instability of the flow occurs.
This occurs at a velocity of order
  $v_{\rm c} \sim \kappa / \xi$. In $^3$He-B, this critical velocity
  for vortex formation is
comparable to the Landau critical velocity for quasiparticle
creation -- the pair-breaking velocity
$v_\mathrm{pb}=\Delta/p_{\rm F}\sim \kappa/\xi$, where $\Delta$ is
the superfluid energy gap.

In contrast, in $^3$He-A the smallest possible vortex loop is of
the order of the soft-core radius, the healing length of the
spin-orbital coupling  $\xi_{\rm D} \gtrsim 10~\mu$m. This is
several orders of magnitude larger than the coherence length
$\xi$. Consequently, the critical velocity for A-phase vortex
formation, $v_{\rm c} \sim \kappa / \xi_{\rm D}$, is considerably
smaller \cite{VortexFormation} while the energy barrier is higher
than in the B phase.

Therefore, in practical experimental conditions neither thermal
activation nor quantum tunneling are of importance in $^3$He
superfluids. Instead, vortex formation takes place when the
average counterflow velocity at the wall is increased to the point
where the local velocity at the sharpest asperity reaches the
critical value, the barrier vanishes, and the process thus becomes
an instability. In principle, pair breaking and quasiparticle
emission might occur already at a slightly lower velocity than
when the barrier actually disappears, and this might finally
trigger the hydrodynamic instability, which then results in vortex
formation. The process might happen in the following manner: near
the asperity the local velocity reaches the pair breaking value,
the creation and emission of quasiparticles increases the density
of the normal component, and as a result $\rho_{\rm s}$ decreases.
Due to the conservation of current $\rho_{\rm s} v_{\rm s}$ the
superfluid velocity then increases, enhancing the radiation of
quasiparticles, which increases $v_{\rm s}$ further. The final
result from the development of such a hydrodynamic instability
will be vortex formation. However, whatever is the real mechanism
of the instability generated by the flow in the vicinity of a
protuberance, it limits the maximum counterflow velocity that can
be achieved in a given sample container. With careful preparation
of the surfaces, the critical velocity $v_\mathrm{c}$ has been
raised up to about 0.1 -- 0.4\,$v_\mathrm{pb}$ in cylinders from
fused quartz with $R \sim 3$~mm.

The situation is quite different in $^4$He-II. Although the
maximum possible superfluid velocity, the Landau value, is three
orders of magnitude higher, the nucleation barrier height is  1 --
10\,K and comparable to the ambient temperature of $\sim 1\,$K. In
the flow through sub-micron-size orifices thermal activation has
been found to be an important mechanism in vortex nucleation
\cite{varoquaux2,packard} (at the lowest temperatures even quantum
tunneling has been argued to exist \cite{Davis,Ihas}). In
contrast, in applications of bulk volume \heii flow it is assumed
that there always exist an abundance of remnant vortex loops
pinned to walls \cite{awschalom} which start to expand in low
applied flow.

In \heb surface pinning is expected to be much less important than
in $^4$He-II because the vortex core radius is more than two
orders of magnitude larger. In the best conditions in a clean
quartz cylinder there are no {\it pinned remnant} vortices. In
such cases no other kind of information about surface pinning
exists at present time. However, {\it dynamic remnant} vortices
are present at low temperatures. When mutual friction dissipation
becomes exponentially small, it may take hours for the last vortex
to annihilate at the container wall in conditions of zero applied
flow. If a remnant vortex loop happens to be around when flow is
reapplied, it starts to evolve and may generate any number of new
independent vortex loops \cite{DynamicRemnant}. In fact, this
situation is expected to prevail also in bulk $^4$He-II over most
of the experimentally accessible temperature range and thus no
source with abundant pinned remnant vortices is needed to create
large numbers of vortices and turbulent flow.

In $^3$He-B we thus expect that a genuine intrinsic critical
velocity is determined by the most effective instability, since
the vortex formation barrier is impenetrable at all temperatures
and velocities below an upper limit $v_{\rm c}(T,P)$. This feature
has been utilized to study the different instabilities described
in the later sections. The upper limit is a container specific
critical velocity (which may change from one cool down to another,
depending on the container's preparation), of which an example is
seen in Fig.~\ref{VortexSteps}.  In such a measurement the
criterion for vortex formation is the lowest critical velocity, in
other words it is the vortex structure with the lowest $v_{\rm c}$
which is formed. If on the other hand one wants to establish the
true equilibrium vortex structure, one has to slowly cool the
sample at constant flow velocity below $T_{\rm c}$. At $T_{\rm c}$
critical velocities vanish and the criterion for the selection
becomes the lowest energy state. The equivalent of this procedure
in superconductivity is known as field cooling.

Cooling under rotation has to be used in $^3$He-A in order to
stabilize and identify the single-quantum vortex (with $n=1$)
which at low flow velocities has lower energy than the
double-quantum vortex which, in turn, has a much lower critical
velocity. Again, the large difference in critical velocities of
these two vortex structures arises because of their different core
structures. As distinct from the doubly quantized vortex, the
singly quantized $^3$He-A vortex has a hard vortex core with a
radius comparable to the superfluid coherence length $\xi$ (which
lies hidden and embedded within a three orders of magnitude larger
soft core of continuous structure \cite{vortex_phasediagram_prl}).
As a result, its critical velocity $v_{\rm c} \sim \kappa / \xi$
is close to the critical velocity for the formation of a B-phase
vortex. The formation of the purely continuous texture of a
doubly-quantized A-phase vortex does not involve breaking the
superfluid state anywhere; it merely requires reorienting the
degeneracy variables of the order parameter. That is why the
corresponding critical velocity is much smaller, $v_{\rm c} \sim
\kappa / \xi_{\rm D}$.

These theoretical predictions have been tested in numerous
different rotating experiments. Such measurements also indicate a
wide range of variation in the observed A-phase critical
velocities, proving that they depend on the prehistory of sample
preparation and thus on the quality of the global order parameter
texture \cite{aphasecritvel,GolovCritVel}. Entry into the texture
at unusually low critical velocity is provided by such regions
where the spin-orbit coupling is not minimized, if they extend to
the cylindrical boundary. Thus the largest reductions in critical
velocity are observed in the presence of different types of planar
domain-wall-like defects in the A-phase order-parameter field,
which are called solitons \cite{asolitons}. If the plane of the
soliton is oriented parallel to the rotation axis (called a splay
soliton), then the critical velocity approaches zero, although it
apparently always remains finite, and all vortex quanta will enter
from along one of the two connection lines of the soliton sheet
with the cylinder wall. This means that the emerging vortices are
not lines but a periodic chain of circular and hyperbolic vortex
quanta stacked within a soliton sheet
(Fig.~\ref{ATCSymmetricFig}). This structure is called a vortex
sheet \cite{VorSheetPRL}. It has the lowest critical velocity of
all A-phase vortex structures and also the fastest dynamic
response \cite{sheet_prl_ve}. Numerical calculations of the flow
instability of various different one-dimensional initial textures
\cite{critveltheory} show that semi-quantitative agreement with
the measured variation exists. A further peculiarity of A-phase
vortex textures is the existence of remnant vorticity in the form
of vortex lines pinned in soliton sheets in the bulk liquid
\cite{GolovPinning}.

The large difference between the critical velocities for the
formation of a doubly quantized A-phase vortex and a singly
quantized B-phase vortex makes it possible to create different
metastable flow states in a rotating two-phase sample.

\subsection{Vortices and AB interface in rotation}
\label{ABInterfaceInRotation}

\begin{figure}[t]
\centerline{\includegraphics[width=1.0\linewidth]{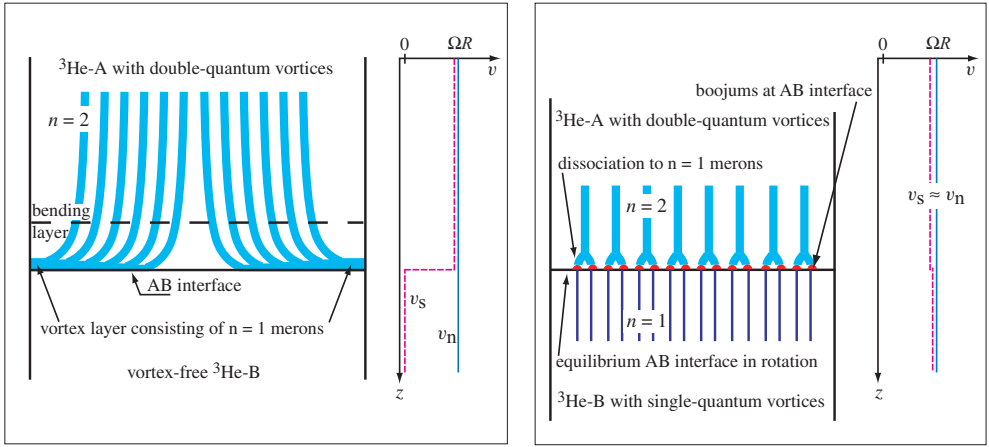}}
\medskip
\caption{Sketches of the AB interface in rotation. {\it Left:} In
the A phase double-quantum vortex lines are formed at low rotation
and we may assume it to be approximately in the equilibrium vortex
state. At the AB interface the double-quantum vortices curve over
to the cylinder wall and cover the AB interface as a vortex layer.
This layer supports the discontinuity in the tangential velocities
of the superfluid fractions at the AB interface. The width of the
AB interface is on the order of the superfluid coherence length
$\xi$ while that of the vortex layer is three orders of magnitude
larger, namely the dipolar healing length $\xi_{\rm D}$. Thus in
this metastable state below the critical velocity the continuous
A-phase vorticity does not penetrate through the AB interface, but
gives rise to the unusual axial distribution in the flow
velocities of the normal and superfluid components, $v_{\rm n}(r)
= \Omega r$ and $v_{\rm s}(r)$, as shown on the right for $r=R$.
{\it Right:} In equilibrium rotation, an A-phase double-quantum
vortex with winding number $n=2$, dissociates at the AB interface
into its $2\pi$ constituents, the circular and hyperbolic merons,
each with $n=1$. Each meron gives rise to a singular point defect,
a boojum, on the AB interface. The boojum is required as a
termination point of a singular $2\pi$ ($n=1$) B-phase vortex.
Thus in the equilibrium state the continuous vorticity crosses the
AB interface, transforming to singular vorticity. However, neither
point or line singularities are easily created in superfluid
$^3$He and therefore the vortex crossing takes place in a
Kelvin-Helmholtz instability event. } \label{VorAB-Interfaces}
\end{figure}


With two $^3$He superfluids which belong to the same order
parameter manifold it becomes possible to construct  a unique
situation which does not exist in other known coherent quantum
systems. With a profiled magnetic-field distribution it is
possible to achieve the coexistence of $^3$He-A and $^3$He-B, and
to stabilize an AB phase boundary in the sample. What happens to
such a superfluid two-phase sample in rotation? The substantial
mismatch in the vortex properties of the two phases -- their
critical velocity, quantization of circulation, and vortex
structure -- raises the question: How are the vortices going to
behave at the AB interface? In $^3$He-A the critical velocity is
low, while in $^3$He-B it is an order of magnitude higher. This
means that the A phase tends to be filled with essentially the
equilibrium number of double-quantum vortex lines, while in the B
phase there would at least initially be no vortices. Is such a
situation stable, how is it going to evolve, and how are the
single-quantum vortices of $^3$He-B going to fit in this picture
if they emerge later at higher velocities?

The left-hand side of Fig.~\ref{VorAB-Interfaces} depicts the
situation where the two-phase system is brought into rotation at
constant temperature. When the rotation is started, A-phase
double-quantum vortices are created at low critical velocity while
no vortices are formed in the B phase. This expectation, confirmed
by measurement \cite{KH_prl}, means that the A-phase vorticity is
not able to cross the AB interface and is accumulated on the
A-phase side of the interface such that it coats the interface
with a dense vortex layer. The layer is made up of a continuous
texture of vorticity \cite{risto_prl} and sustains the tangential
discontinuity in the flow velocities of the superfluid fractions
on the different sides of the AB interface. Thus we have
constructed a metastable state in which the two superfluids slide
with respect to each other with a large shear-flow discontinuity,
since the superfluid fraction in the A phase rotates
solid-body-like while that in the B phase remains stationary in
the inertial frame.

The minimum-energy state is shown on the right in
Fig.~\ref{VorAB-Interfaces}. Here the vorticity is conserved on
crossing the interface and both phases contain the equilibrium
number of vortices at any given angular velocity of rotation.
Accordingly, the number of double-quantum vortices in the A phase
is one half of the number of singular singly-quantized B-phase
vortices. On approaching the interface, the continuous A-phase
vortex splits into its two $2\pi$ constituents or merons. Each of
the merons ends in a boojum on the AB interface. The boojum is a
point-like topological singularity of the orbital $\hat{\bf l}$
vector at the AB interface, the termination point of a singular
B-phase vortex on the AB interface.

In practice singularities are not easily created in superfluid
$^3$He: like in the case of vortices the energy barrier is too
high compared to ambient temperature (typically by 6 -- 9 orders
of magnitude). Therefore the equilibrium state at the AB interface
is not obtained by increasing rotation at constant temperature and
pressure \cite{KH_prl}. Nor is it formed by cooling through
$T_{\rm c}$ at constant rotation and pressure, which is the usual
method to create the equilibrium vortex state below a second-order
phase transition. The reason is that here the first order
A$\rightarrow$B transition also has to be traversed, to form the
AB interface within the sample
\cite{vorlayer_prl,vorlayer_physica}. The closest approximation to
the equilibrium state is obtained by starting with the equilibrium
number of B-phase vortices in a single-phase sample at high
rotation, with the barrier field at zero or at sufficiently low
value. Next the barrier field is swept up (at constant $\Omega$,
$T$, and $P$), until A phase and the AB interface is formed. In
this case the equilibrium superfluid circulation is already
trapped in the sample and cannot all escape. Finally, by reducing
rotation to the point where B-phase vortices start to annihilate,
one has reached the equilibrium vortex state.

In contrast, if one simply starts increasing rotation of a
two-phase sample with an AB interface, very different behavior is
observed because of the energy barriers preventing the nucleation
of point and line singularities. This is one of the ways to
demonstrate the superfluid Kelvin-Helmholtz instability, where the
AB interface becomes unstable in the presence of an excessively
large counterflow velocity which is oriented parallel to the AB
interface (Sec. \ref{KHSection}). A complex chain of events is
then started in which also vortices escape across the AB interface
from the A to the B-phase side.

\subsection{Vortex dynamics and mutual friction}
\label{sect-VortexDynamics}

In $^3$He-B and  $^4$He-II, in the absence of vorticity, the
superflow is potential and the superflow velocity ${\bf v}_{\rm s}
= {\bm \nabla}\Phi $ is expressed in terms of the ``flow
potential'' which is proportional to the phase of the superfluid
wave function, $\Phi=(\kappa/2\pi)\varphi$. The superflow velocity
$ \mathbf{v}_{\rm s}$ obeys the Euler equation
\cite{Khalatnikov_book}:
\begin{equation}
\frac{\partial \mathbf{v}_{\rm s}}{ \partial t}+ \bm{\nabla}
\tilde\mu =0, \label{Euler-superIdeal}
\end{equation}
where $\tilde \mu =\mu +v_{\rm s}^2/2$ and $\mu$ is the chemical
potential. When quantized vortices (or continuous vorticity in
$^3$He-A) are present, the superfluid velocity is no longer
potential. The motion of a vortex leads to a phase-slip effect and
modifies the r.h.s. of Eq.~(\ref{Euler-superIdeal}).

Vortex lines form a part of the superfluid component, but the normal
component influences their movement through mutual friction which
arises from the scattering of normal excitations from the vortex
cores. In the zero-temperature limit, where the normal excitations
vanish, the motion of a vortex line is governed by the Magnus force
only, so that the vortex velocity coincides with the local superfluid
velocity at the position of the vortex element. At non-zero
temperatures the friction between the vortex and the normal component
-- the so-called mutual friction -- causes a drag force on the vortex
line and, as a result, the velocity $\mathbf{v}_\mathrm{L}=d{\bf r}_{\rm
L}/dt$ of a vortex segment in the flow deviates from
$\mathbf{v}_\mathrm{s}$.

In the presence of vortices, the flow potential is not uniquely
defined along the contours encircling the singular vortex lines.
If the vortices do not overlap, the flow potential can be written
as
\[
\Phi=\sum_{\beta =1}^N \Phi_\beta \left({\bf r}-{\bf r}_\beta
,t\right).
\]
Here ${\bf r}_\beta (s_\beta)$ are the coordinates of singular
lines specified by a parameter $s_\beta$. If the positions of
these lines also depend on time, ${\bf r}_\beta ={\bf r}_\beta
(t)$, the time derivative of the superflow velocity becomes
\begin{eqnarray*}
\frac{\partial {\bf v}_\mathrm{s}}{\partial t}&=&\frac{\partial }{\partial
t}{\bm \nabla}\Phi ={\bm \nabla}\frac{\partial^\prime \Phi
}{\partial t}-\sum_\beta \left(\frac{\partial {\bf
r}_\beta}{\partial t} \cdot {\bm \nabla}
\right){\bm \nabla}\Phi_\beta \\
&=&{\bm \nabla}\left[\frac{\partial^\prime \Phi }{\partial t}-
\sum_\beta \left({\bf v}_\beta \cdot {\bm \nabla} \Phi_\beta
\right)\right] +\sum_\beta \left[{\bf v}_\beta \times {\bm
\omega}_\beta \right]\,.
\end{eqnarray*}
Here $\partial ^\prime /\partial t $ is the derivative only of the
explicit $t$ dependence of $\Phi$ and we put ${\bf v}_\beta
=\partial {\bf r}_\beta /\partial t$. The vorticity of a single
vortex is
\[
{\bm \omega}_\beta = {\rm curl}\, {\bf v}_{{\rm s}\, \beta} ={\rm
curl}\, {\bm \nabla \Phi_\beta} \,.
\]
In $^3$He-B and $^4$He-II the vorticity from singular vortex lines
is expressed as
\begin{equation}
{\bm \omega} _\beta =\kappa_\beta \int \delta\left({\bf r}-{\bf
r}_\beta\right)\,d{\bf r}_\beta \,. \label{vorticity-single}
\end{equation}
Here ${\bf r}_\beta$ is the coordinate of the $\beta$-th vortex
line and $\delta \left({\bf r}-{\bf r}_\beta\right)$ is the
three-dimensional $\delta$-function, $d{\bf r}_\beta =\hat{\bf
s}_\beta \, ds_\beta$, $\hat{\bf s}_\beta$ is the unit vector in
the direction of the vortex line at the point ${\bf r}_\beta$, and
$ds_\beta$ is the arc length of the vortex line. The circulation
of each vortex $\kappa_\beta =n_\beta\kappa$ may have $n_\beta$
circulation quanta $\kappa$. Since the derivative
\[
\frac{\partial \Phi }{\partial t}=\frac{\partial^\prime \Phi
}{\partial t}- \sum_\beta \left({\bf v}_\beta \cdot {\bm \nabla}
\Phi_\beta\right)
\]
is usually defined as the ``superfluid chemical potential'' $\mu
_\mathrm{s} =-\partial \Phi /\partial t$ we have
\begin{equation}
\frac{\partial {\bf v}_\mathrm{s}}{\partial t}+{\bm \nabla }\mu _\mathrm{s}=
\sum_\beta {\bf v}_\beta \times {\bm \omega }_\beta \,.
   \label{Euler-super}
\end{equation}
One can write here $\mu_\mathrm{s} =\tilde\mu +\tilde \mu_\mathrm{s} $, where
$\tilde \mu_\mathrm{s} = \mu_\mathrm{s}-\tilde \mu$ is the deviation of the superfluid
chemical potential from the total chemical potential of the fluid,
which is the counterpart of the gauge-invariant scalar potential
in the theory of nonstationary superconductivity
\cite{Kopnin_book}. In equilibrium $\tilde \mu_\mathrm{s} =0$, so that
$\mu_\mathrm{s}=\tilde \mu$, which in fact is the famous Josephson
relation.

The velocity of each vortex is determined up to its component
perpendicular to the vortex line \cite{donnelly}:
\begin{equation}
{\bf v}_\beta =\hat{\bf s}_\beta\times ({\bf v}_\mathrm{s}\times\hat{\bf
s}_\beta) +\alpha \hat {\bf s}_\beta\times ({\bf v}_\mathrm{n}-{\bf v}_\mathrm{s})
-\alpha^\prime \hat{\bf s}_\beta\times \left[ \hat {\bf
s}_\beta\times({\bf v}_\mathrm{n}-{\bf v}_\mathrm{s})\right] \label{vl}
\end{equation}
Here $\alpha (T,P)>0$ and $\alpha' (T,P)<1$ are the temperature
and pressure dependent dissipative and reactive mutual-friction
parameters.

Inserting Eq. (\ref{vl}) into Eq. (\ref{Euler-super}) we find
\begin{eqnarray*}
\sum_\beta {\bf v}_\beta \times {\bm \omega }_\beta &=& \sum_\beta
\kappa_\beta \int \delta({\bf r}-{\bf r}_\beta) \, ds_\beta \left[
\hat {\bf s}_\beta \times [\hat{\bf s}_\beta \times (\hat{\bf
s}_\beta \times {\bf v}_\mathrm{s})] \right] \\
&+&\alpha^\prime \sum_\beta \kappa_\beta \int \delta({\bf r}-{\bf
r}_\beta) \, ds_\beta \left[ \hat {\bf s}_\beta \times [\hat{\bf
s}_\beta \times (\hat{\bf s}_\beta \times ({\bf v}_\mathrm{n}-{\bf
v}_\mathrm{s}))]
\right]\\
&-&\alpha \sum_\beta \kappa_\beta \int \delta\left({\bf r}-{\bf
r}_\beta\right)d s_\beta \left[\hat{\bf s}_\beta\times \left[\hat
{\bf s}_\beta\times ({\bf v}_\mathrm{n}-{\bf
v}_\mathrm{s})\right]\right]\,.
\end{eqnarray*}
The first line gives
\begin{eqnarray*}
\sum_\beta \kappa_\beta \int \delta({\bf r}-{\bf r}_\beta) \,
ds_\beta \left[ \hat {\bf s}_\beta \times [\hat{\bf s}_\beta
\times (\hat{\bf s}_\beta \times {\bf v}_\mathrm{s})] \right] \\
=-\sum_\beta \kappa_\beta  \int \delta({\bf r}-{\bf r}_\beta) \,
ds_\beta \left[ \hat {\bf s}_\beta \times {\bf v}_\mathrm{s} \right] ={\bf
v}_\mathrm{s} \times {\bm \omega}_\mathrm{s}
\end{eqnarray*}
where
\begin{equation}
{\bm\omega}_\mathrm{s}=\sum_\beta {\bm \omega}_\beta \label{omega_s}
\end{equation}
is the total vorticity of the superfluid. The second line can be
transformed in the same way. As a result,
\begin{equation}
\frac{\partial {\bf v}_\mathrm{s}}{\partial t}+{\bm \nabla }\tilde \mu
={\bf v}_\mathrm{s}\times {\bm \omega}_{\rm s} +{\bf f}_\mathrm{mf}\,.
\label{vs-dynam-micro}
\end{equation}
Here ${\bf f}_\mathrm{mf}$ is the mutual-friction force \cite{HallVinen}
\begin{eqnarray}
{\bf f}_\mathrm{mf}&=&-\alpha \sum_\beta \kappa_\beta \int
\delta\left({\bf r}-{\bf r}_\beta\right)d{\bf r}_\beta\times
\left[\hat {\bf s}_\beta\times ({\bf v}_\mathrm{n}-{\bf v}_\mathrm{s})\right]
\nonumber \\
&&+ \alpha^\prime \left[({\bf v}_\mathrm{n}-{\bf v}_\mathrm{s})\times
{\bm \omega}_{\rm s} \right] \label{mutfricforce-micro}
\end{eqnarray}
exerted by the normal component on a unit mass of superfluid via the
vortex lines. The first term in ${\bf f}_\mathrm{mf}$ is the viscous
component, with a negative projection on the relative velocity
${\bf v}_\mathrm{s}-{\bf v}_\mathrm{n}$,
\begin{equation}
{\bf f}_\mathrm{mf}\cdot ({\bf v}_\mathrm{s}-{\bf v}_\mathrm{n})=-({\bf v}_\mathrm{s}-{\bf
v}_\mathrm{n})^2\alpha \sum_\beta \kappa_\beta \int \delta\left({\bf
r}-{\bf r}_\beta\right)[1-\cos^2 \gamma_\beta]\, ds_\beta.
\label{Fmf-vsprojection}
\end{equation}
Here $\gamma_\beta$ is the angle between ${\bf s}_\beta$ and ${\bf
v}_\mathrm{s}-{\bf v}_\mathrm{n}$. Without mutual friction force, Eq.
(\ref{vs-dynam-micro}) coincides with the Euler equation of the
classical hydrodynamics of an ideal inviscid fluid
\begin{equation}
\frac{\partial {\bf v} }{\partial t}+{\bm \nabla }\tilde \mu ={\bf
v} \times {\bm \omega}. \label{nonvisc-dynam}
\end{equation}
In what follows we consider the case where all vortices have the
same circulation with $n=1$.

\begin{figure}[t]
\begin{center}
\centerline{\includegraphics[width=0.9\linewidth]{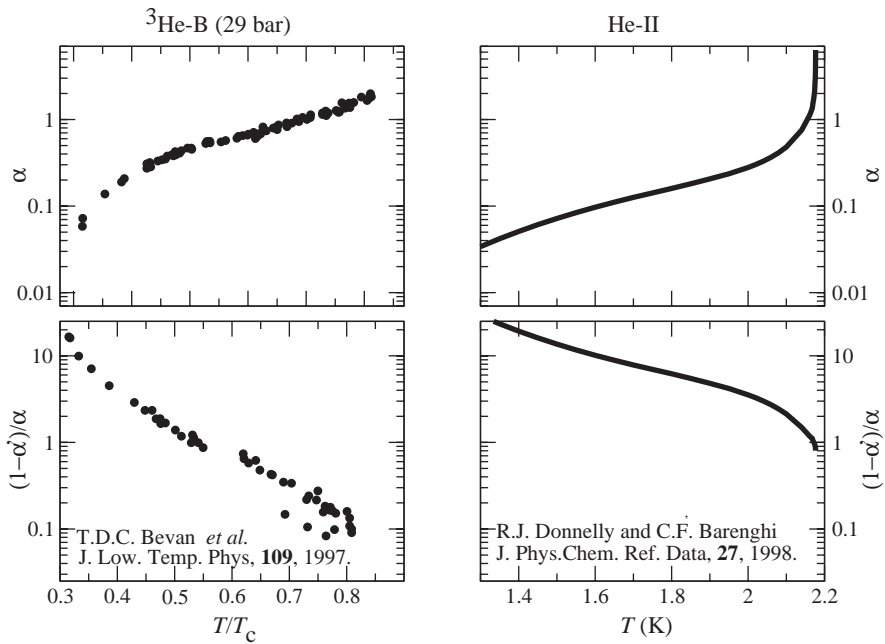}}
\caption{Mutual friction covers a different range of values in
\heb {\it (left column of panels)} and $^4$He-II {\it (right
column)}. The {\it top} row of panels shows the dissipative mutual
friction coefficient $\alpha (T)$ and the {\it bottom} row the
dynamic parameter $q^{-1} = (1-\alpha')/\alpha$. A transition in
vortex dynamics is expected at $q \sim 1$. The data for \heb is
from Refs.~\cite{bevan, bevanprl} and for $^4$He-II from
Ref.~\cite{he_data}.}\label{mf_data}
\end{center}
\end{figure}

For $^3$He-B, the mutual-friction parameters were measured in
Refs.~\cite{bevan,bevanprl}. This remarkable hydrodynamic
measurement was performed by examining capacitively changes in
antisymmetrically driven normal modes of a thin membrane which
lies in the plane perpendicular to the rotation axis. As will be
seen in Sec.~\ref{HelicalBundle} (Fig.~\ref{alpha}), these results
agree with later measurements on the longitudinal propagation
velocity of vortices in the rotating cylinder which also yield the
dissipative parameter $\alpha$. These experimental data agree with
calculations of the parameter values in
Refs.~\cite{kopnin_mfhe3B,kopnin_mf,KL2tau} (for a theory review
see \cite{kopnin_rep}). The mutual friction parameters are
discussed also in terms of the chiral anomaly and the
Callan-Harvey effect in relativistic quantum field theory in
Ref.~\cite{volovik_droplet}. Both $\alpha$ and $\alpha^{\prime}$
vanish at $T=0$. With increasing temperature the dissipative
mutual-friction parameter $\alpha$ increases so that above
0.6~$T_\mathrm{c}$ all dynamic processes are heavily overdamped in
$^3$He-B. In $^3$He-A $\alpha$ is expected to be in the overdamped
regime at all currently accessible temperatures
\cite{kopnin_mf,bevan,kopnin_mfhe3A}. In contrast, in $^4$He-II
$\alpha$ is much smaller at all experimentally relevant
temperatures, as seen in Fig.~\ref{mf_data}. The difference in the
magnitudes of the mutual-friction parameters for superfluid $^3$He
and $^4$He is so striking that the usual picture of $^4$He-II
vortices moving {\em with the superflow} fails for $^3$He vortices
which move {\em across the flow} in most of the accessible
temperature range. In this respect, motion of vortices in $^3$He
resembles that in most of type-II superconductors where they also
move perpendicular to the flow with practically no measurable Hall
angle \cite{kopnin_rep}.

Since the equation of motion (\ref{vl}) for the trajectory of a
vortex only depends on the two parameters $\alpha$ and
$\alpha^{\prime}$, its solutions can be classified in terms of the
dynamic parameter $q$ defined as
\begin{equation}
q\equiv \frac{\alpha}{1-\alpha' }~. \label{q}
\end{equation}
This parameter has already been introduced in Section
\ref{Sect-Helium-super}. According to both the theoretical
predictions \cite{KL2tau} and the experimental data for \heb the
parameter $q$ depends exponentially on temperature in the
low-temperature limit. The lower row of panels in
Fig.~\ref{mf_data} shows the inverse $q^{-1} = (1-\alpha')/\alpha$
for \heb and $^4$He-II. In superfluid dynamics $q$ acquires an
important function: it characterizes the relative influence of
dissipation and $1/q$ plays the role of the Reynolds number
(Sec.~\ref{TurbTheory}). A change in the characteristic solutions
and a corresponding transition in the dynamics can be expected in
the regime $q \sim 1$: this is located in the middle of the
temperature range for $^3$He-B, but only a few $\mu$K below
$T_{\lambda}$ for $^4$He-II.

\begin{figure}[t]
\begin{center}
\centerline{\includegraphics[width=\linewidth]{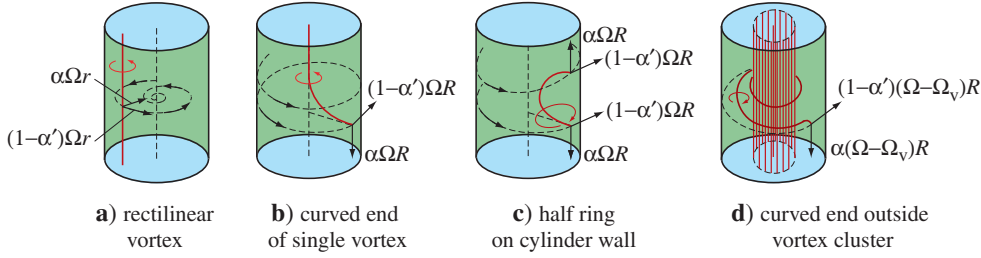}}
\caption{Sketches of vortex trajectories in an ideal rotating
cylindrical superfluid sample, as viewed in the rotating frame.
{\it (a)} A rectilinear vortex, which is released in the flow from
the cylinder wall, travels along a spiral to its equilibrium
location, to become aligned along the center axis of the cylinder.
{\it (b)} In a long cylinder the propagating end of the vortex
describes a spiral trajectory on the cylinder wall while it
expands into the vortex-free section of the cylinder. The trailing
end is aligned along the central axis, the equilibrium location
for the vortex.  {\it (c)} An elementary vortex usually forms as a
half ring on the cylinder wall. It expands both axially and
radially. Neglecting the contribution from its self-induced
velocity, the half ring positions itself perpendicular to the
azimuthally circulating applied flow and remains during its
expansion inside a radial plane which rotates around the cylinder
axis with angular velocity $(1-\alpha')\Omega$ with respect to the
cylinder walls. {\it (d)} If a central vortex cluster already
exists in the cylinder (with $N = \pi R_{\rm c}^2 \, 2\Omega
/\kappa$ rectilinear vortex lines), then the spiralling motion
leaves behind a helically wound vortex. At low temperatures, when
the axial motion is slow, a tightly wound helix forms around the
cluster. The end section of this vortex, where it approaches the
cylindrical wall, is most likely to suffer the Kelvin-wave
instability. Here the applied flow velocity is large
and the vortex is partly aligned azimuthally along the applied
flow. }\label{VortexTrajectories}
\end{center}
\end{figure}

It is instructive to inspect some solutions of Eq.~(\ref{vl}), as
sketched in Fig.~\ref{VortexTrajectories}, for an ideal rotating
cylinder where the rotation and cylinder axes coincide. We assume
the experimentally important situation where the applied
counterflow velocity dominates, {\it i.e.} we set in
Eq.~(\ref{vl}) $\mathbf{v}_{\rm n} = \mathbf{\Omega} \times
\mathbf{r}$ and $\mathbf{v}_{\rm s} \approx 0$, and consider a
large enough $q$ so that the vortex dynamics is regular (instead
of turbulent, see Sec.~\ref{TurbulenceKH}). The simplest case (a)
is a rectilinear vortex line parallel to the rotation axis which
is released in the flow from the cylinder wall. This might
correspond to the situation in which a double-quantum vortex forms
in $^3$He-A when the cylinder is filled with a perfect global
texture. In the cylindrically symmetric flow geometry the vortex
will retain its rectilinear shape while it travels inward along a
spiral which finally places the line in its equilibrium position
along the axis of the cylinder. Another important case is (b) a
vortex with a free end expanding in a long cylinder. The free end
terminates perpendicular on the cylinder wall and describes a
regular helix on the wall during its expanding motion. If we
neglect the influence from its own curvature on the motion, then
the curved section of the vortex remains within the same radial
plane which rotates at the angular velocity $(1-\alpha')\Omega$
around the cylinder axis, when viewed from the rotating frame. The
trailing section of the vortex becomes aligned along the central
axis and remains at rest there. An immediate extension of this
example is case (d) where the vortex expands in a cylinder in the
presence of a pre-existing cental vortex cluster. The rectilinear
trailing end of the vortex becomes now part of the vortex cluster
and is incorporated as one of the vortices in the peripheral
circle of lines. This section would prefer to be stationary in the
rotating frame, it resides in a region of the sample cross section
where the average counterflow velocity vanishes, and therefore it
moves only with difficulty from one lattice site to the next in
the outermost ring of vortex lines. The free end, on the other
hand, expands along a spiral trajectory and leaves behind a helix
which cannot relax instantaneously.

To explore the hydrodynamic transition between regular and
turbulent vortex dynamics at $q \sim 1$, we can ask the following
question: how is the dynamic equation (\ref{vs-dynam-micro})
modified for superfluids in the continuous limit, after averaging
locally over vortex lines? This can be done even for tangled
vortex states, if the lines are locally sufficiently parallel and
their radius of curvature is much larger than the vortex core
diameter. This would be the case, for instance, for rotating
states with transient time-dependent disorder, since the rotating
flow would rapidly polarize the component parallel to the rotation
axis while the order in the transverse plane would be restored
only later. In this case the averaging over nearby vortices on
the r.h.s. of Eq.~(\ref{vs-dynam-micro}) gives
\begin{equation}
\frac{\partial \mathbf{v}_{\rm s}}{ \partial t}+
\bm{\nabla}\tilde \mu = \mathbf{v}_{\rm s} \times \bm{\omega}_\mathrm{s} -
\alpha' (\mathbf{v}_{\rm s} -\mathbf{v}_{\rm n} )\times
\bm{\omega}_\mathrm{s}+ \alpha~\hat{\bm{\omega}}_\mathrm{s}
\times\left[\bm{\omega}_\mathrm{s} \times (\mathbf{v}_{\rm s}
-\mathbf{v}_{\rm n} )\right]. \label{SuperfluidHydrodynamics}
\end{equation}
This result is known as the coarse-grained hydrodynamic equation
for the superflow velocity \cite{soninhydro}. It will later be
used to develop an analogous superfluid interpretation to the
transition to turbulence, as can be derived in viscous
hydrodynamics from the Navier-Stokes equation \cite{nature}.

\subsection{Kelvin-wave instability of vortex lines}
\label{Sect-Kelvin-wave}

A quantized vortex can support helical Kelvin waves which become
important in turbulent vortex dynamics. The Kelvin waves may lead to
an instability which generally develops at large flows. This
instability can result in an increase in the number of individual
vortex lines under an applied flow if the mutual friction is
sufficiently small. It is the first step in the process by which more
vortices can be generated from one or a few existing vortices so that
bulk volume turbulence can switch on. The evolution towards the
instability starts when a vortex becomes sufficiently aligned along
the applied flow. A vortex line oriented parallel to the external flow
is an unstable configuration, as demonstrated by Donnelly {\it et al.}
by applying a thermal counterflow current parallel to rectilinear
vortex lines in rotating $^4$He-II \cite{barenghi_kwaves}. Above a
critical axial flow velocity rectilinear vortices became unstable and
formed a turbulent tangle with varying axial polarization, depending
on the axial counterflow velocity. The phenomenon was explained by
Glaberson {\it et al.} \cite{glaberson, ostermeyer} in terms of the
Kelvin-wave instability and was later reproduced also numerically by
Tsubota {\it et al.} \cite{RotTurbulence}.

Consider small transverse deformations to an isolated vortex line,
oriented along the $z$ axis, in externally applied counterflow ${\bf
v}=v~\hat{\bf z}$ parallel to it. Parametrizing the position vector
of an arbitrary element on the deformed vortex as
\begin{equation}
{\bf r}(z,t)=\zeta(z,t)\hat{\bf x}+\eta(z,t)\hat{\bf y}+z\hat{\bf z},
\end{equation}
we can write the unit tangent of the line as $\hat{\bf s}(z)
\approx (\partial\zeta/\partial z)\hat{\bf x}+(\partial\eta/\partial
z)\hat{\bf y}+\hat{\bf z}$ (to linear order in the small
quantities $\zeta$, $\eta$). The vortex curvature also gives rise to
a self-induced contribution to the superfluid velocity at the
vortex line. In the local approximation \cite{donnelly,soninhydro}
\begin{equation}
{\bf v}_{\rm s}^{\rm i}=\tilde \kappa ~\frac{\partial{\bf
r}}{\partial z} \times \frac{\partial^2{\bf r}} {\partial z^2}
\approx \tilde \kappa  \left( -\frac{\partial^2\eta}{\partial
z^2}~\hat{\bf x} +\frac{\partial^2\zeta}{\partial z^2}~\hat{\bf y}
\right),
\end{equation}
where $\tilde \kappa =(\kappa/4\pi)\ln(2\pi/kr_{\rm c})$. Inserting these
to the equation of motion for the vortex line, Eq.~(\ref{vl}), we
find
\begin{eqnarray}
\frac{\partial\zeta}{\partial t} &=& -\tilde \kappa
\frac{\partial^2\eta}{\partial z^2}+ \alpha\left( \tilde \kappa
\frac{\partial^2\zeta}{\partial z^2}+v\frac{\partial\eta}{\partial
z} \right) +\alpha' \left( \tilde \kappa
\frac{\partial^2\eta}{\partial z^2}-v\frac{\partial\zeta}{\partial
z} \right),
\nonumber \\
\frac{\partial\eta}{\partial t} &=& \tilde \kappa
\frac{\partial^2\zeta}{\partial z^2}+ \alpha\left( \tilde \kappa
\frac{\partial^2\eta}{\partial z^2}-v\frac{\partial\zeta}{\partial
z} \right) -\alpha' \left( \tilde \kappa
\frac{\partial^2\zeta}{\partial z^2}+v\frac{\partial\eta}{\partial
z} \right).
\end{eqnarray}
The dispersion relation for wavelike disturbances $\propto
\exp[-i(\omega t-kz)]$, {\it i.e.} Kelvin waves, can be found as
\begin{equation}\label{KWdispersion}
\omega=(1-\alpha')\tilde \kappa  k^2-\alpha' kv-i\alpha (\tilde
\kappa k^2-kv).
\end{equation}
With vanishing counterflow, $v=0$, the dispersion relation
simplifies to \cite{sonin_vordynamics}
\begin{equation}\label{kelvin_dispersion}
\omega=\tilde \kappa  k^2(1-\alpha'- i\alpha).
\end{equation}
Here we again encounter the dynamic parameter $q = \alpha/(1-\alpha')$
already introduced in Eq.~(\ref{q}) as the ratio of the imaginary
and real parts of the dispersion \cite{KelvinWaveDamping}. The
waves are always damped but at high temperatures, where $q>1$, the
waves are overdamped and do not propagate.

On the other hand, in the presence of externally applied flow the
long-wavelength modes with $k<v/\tilde \kappa $ have
$\mathrm{Im}(\omega)>0$, and exhibit exponential growth. In other
words, if an evolving vortex configuration at some time has long
enough vortex-line sections oriented parallel to external flow,
these will become unstable to exponentially growing helical
deformations. The expanding waves can then undergo reconnections,
either with the walls of the container or with other vortex lines.
This leads to a growing number and density of vortices and,
ultimately, if $q$ is small enough, to the onset of turbulence, as
discussed in Sec.~\ref{InjectionMethods}.

\section{Kelvin-Helmholtz instability in superfluids}
  \label{KHSection}

\subsection{Introduction}\label{KH-Intro}

The Kelvin-Helmholtz shear flow instability is a well-known
phenomenon of classical hydrodynamics which was first discussed by
Lord Kelvin already in the 1860's. It occurs at the interface
between two fluid layers which are in relative motion with respect
to each other. For instance, at low differential flow velocity the
interface between two stratified layers of different salinity or
temperature is smooth in the ocean, but at some critical velocity
waves are formed on the interface. Similarly, ripples do not form
on the water surface on a lake at infinitesimal wind velocities,
but form at some finite critical value. The Kelvin-Helmholtz
instability is thus a common phenomenon in nature around us. The
condition for instability is derived in many textbooks on
hydrodynamics \cite{Chandrasekhar,Lamb} in the limit of inviscid
and incompressible fluids. In superfluids it was first observed at
the rotating AB interface \cite{KH_prl}.

The AB interface instability occurs between the two $^3$He
superfluids, $^3$He-A and $^3$He-B, when their superfluid
fractions move with respect to the normal component tangential to
the AB interface (Fig.~\ref{VorAB-Interfaces} {\it left}). This
initial state is dissipationless while the state after the
instability is not, as surface waves or ripplons form on the
interface and their motion is highly damped. In conventional
liquids and gases the mathematical description of the interfacial
instability is inevitably only approximate: if viscosity is
neglected, the initial nondissipative states of the two liquid or
gas layers sliding with respect to each other are not exact. The
question then arises not only about the true value of the critical
velocity, but more generally about the existence and nature of the
instability. Superfluids are the only laboratory examples of cases
where viscosity is totally absent, and the mathematical
description of the instability can be presented analytically in a
simple form. The initial state of the AB interface with different
tangential superflow velocities across the interface is in
nondissipative thermodynamic equilibrium, until the critical
velocity of the hydrodynamic instability is reached. As a result
-- contrary to viscous normal fluids -- the instability threshold
is well defined. Experimentally it is manifested by the sudden
formation of vortices in the initially vortex-free $^3$He-B phase
(Fig.~\ref{VorAB-Interfaces} {\it right}) at a rotation velocity
at which no vortices would yet be formed without the presence of
the AB interface.

\subsection{Kelvin-Helmholtz instability in classical hydrodynamics}
\label{kh_classical}

The Kelvin--Helmholtz (KH) instability is one of the many
interfacial instabilities in the hydrodynamics of liquids, gases,
charged plasma, and even granular materials. It refers to the
dynamic instability of an interface with discontinuous tangential
flow velocities and can loosely be defined as the instability of a
vortex sheet. Many natural phenomena have been attributed to this
instability. The most familiar ones are the generation of
capillary waves on the surface of water, first analyzed by Lord
Kelvin \cite{kelvin_original}, and the flapping of sails and
flags, first discussed by Lord Rayleigh \cite{Rayleigh}.

Many of the leading ideas in the theory of interfacial
instabilities in hydro\-dy\-namics were originally inspired by
considerations about ideal inviscid flow. The corrugation
instability of the interface between two immiscible ideal liquids,
with different mass densities $\rho_1$ and $\rho_2$, occurs in the
gravitational field at the critical differential flow velocity
\cite{kelvin_original}
\begin{equation}
({\bf U}_1-{\bf U}_2)^4= 4\sigma g \frac{ |\rho_1-\rho_2|
(\rho_1+\rho_2)^2}{ \rho_1^2\rho_2^2}~, \label{KHClassical}
\end{equation}
where $\sigma$ is the surface tension of the interface, and $g$ is
the gravitational acceleration. To separate the gravitational and
inertial properties of the liquids, we rewrite the threshold
velocity in the form
\begin{equation}
  {\rho_1\rho_2\over \rho_1+\rho_2}({\bf U}_1-{\bf U}_2)^2=2\sqrt{\sigma
F}~. \label{InstabilityCondition1}
\end{equation}
Here  $F$ is the external field stabilizing the position of the
interface, which in the case of a gravitational field is
\begin{equation}
F=g(\rho_1-\rho_2)~, \label{GravityForce}
\end{equation}
but can in general originate from some other source. The surface
mode of ripplons, or capillary waves, which is first excited has
the wave number corresponding to the inverse `capillary length'
\begin{equation}
k_0=\sqrt{F/\sigma}~. \label{WaveVectorInstability}
\end{equation}
However, ordinary fluids are not ideal and the correspondence
between this theory and experiment is not satisfactory. One reason
for this is that the initial state cannot be properly prepared --
the shear-flow discontinuity is never an equilibrium state in a
viscous fluid, since it is not a solution of the Navier-Stokes
equation. As usual in hydrodynamics, it is not apparent whether
the notion of `instability' can be properly extended from  the
idealized inviscid model to the proper zero viscosity limit of the
Navier-Stokes equation and further to the case of finite viscosity
\cite{Birkhoff}.

\subsection{Experimental setup}

\begin{figure}[t]
\begin{center}
\centerline{\includegraphics[width=0.5\linewidth]{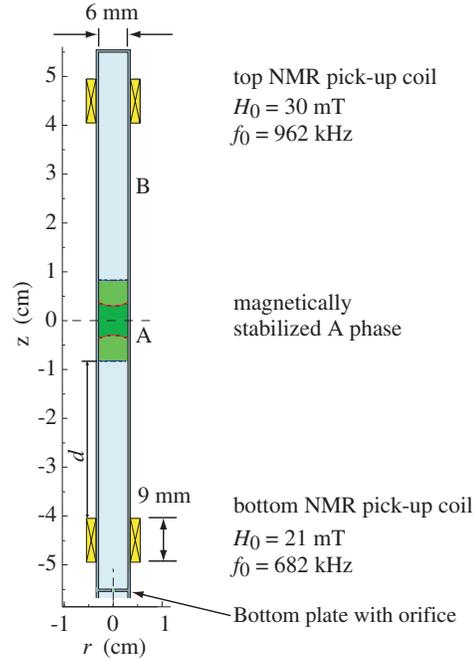}}
\caption{Two-phase superfluid $^3$He sample for measurements on
the Kelvin-Helmholtz instability. The sample volume is 6~mm in
diameter and 11~cm long. A small orifice of 0.5 -- 0.75~mm in
diameter in the center of the bottom end plate separates the
sample from the rest of the liquid $^3$He volume. The $^3$He below
the orifice is needed to establish thermal contact to the
refrigerator. NMR pick-up coils are located at both ends of the
sample tube. They are circular coils with their symmetry axis
transverse to the tube axis. Two solenoidal superconducting
magnets provide the homogeneous axially oriented polarizing fields
for NMR. A third barrier magnet creates the field to stabilize
$^3$He-A in the center section of the long sample. Two examples of
the A phase region (shaded) are shown at 0.55~$T_\mathrm{c}$: at
the current of $I_\mathrm{b}=8$~A in the barrier solenoid the A
phase extends further in the column and the two AB boundaries are
almost flat, while at $I_\mathrm{b}=4$~A the A phase region is
short and the AB boundaries are concave.}\label{setup}
\end{center}
\end{figure}

The superfluid Kelvin-Helmholtz experiments reported in
Ref.~\cite{KH_prl} were conducted in a rotating nuclear
demagnetization cryostat in which the liquid $^3$He sample can be
cooled below $0.2\,T_{\rm c}$ in rotation up to 4~rad/s. A
schematic illustration of the sample setup is shown in
Fig.~\ref{setup}. The sample is contained in a fused quartz tube
of $R=3$\,mm radius. An aperture of 0.5 -- 0.75\,mm diameter in
the bottom end plate restricts the flow of vortices into the
sample from the heat exchanger volume below. The quality of the
sample-tube surface is of great importance, since any larger
surface defects serve as nucleation and pinning sites. Before use
the container is treated with a HF solution, to etch away sharp
protuberances, and then cleaned with solvents. The same quartz
tube may display sizable variation in critical velocity and vortex
formation properties depending on how it is cleaned between
experiments, indicating that dust particles, dirt, frozen gas,
etc. serve as nucleation and pinning sites. Most measurements were
performed using a sample tube which allowed vortex-free rotation
up to the maximum rotation velocity of the cryostat.

\begin{figure}[t]
\centerline{\includegraphics[width=0.9\linewidth]{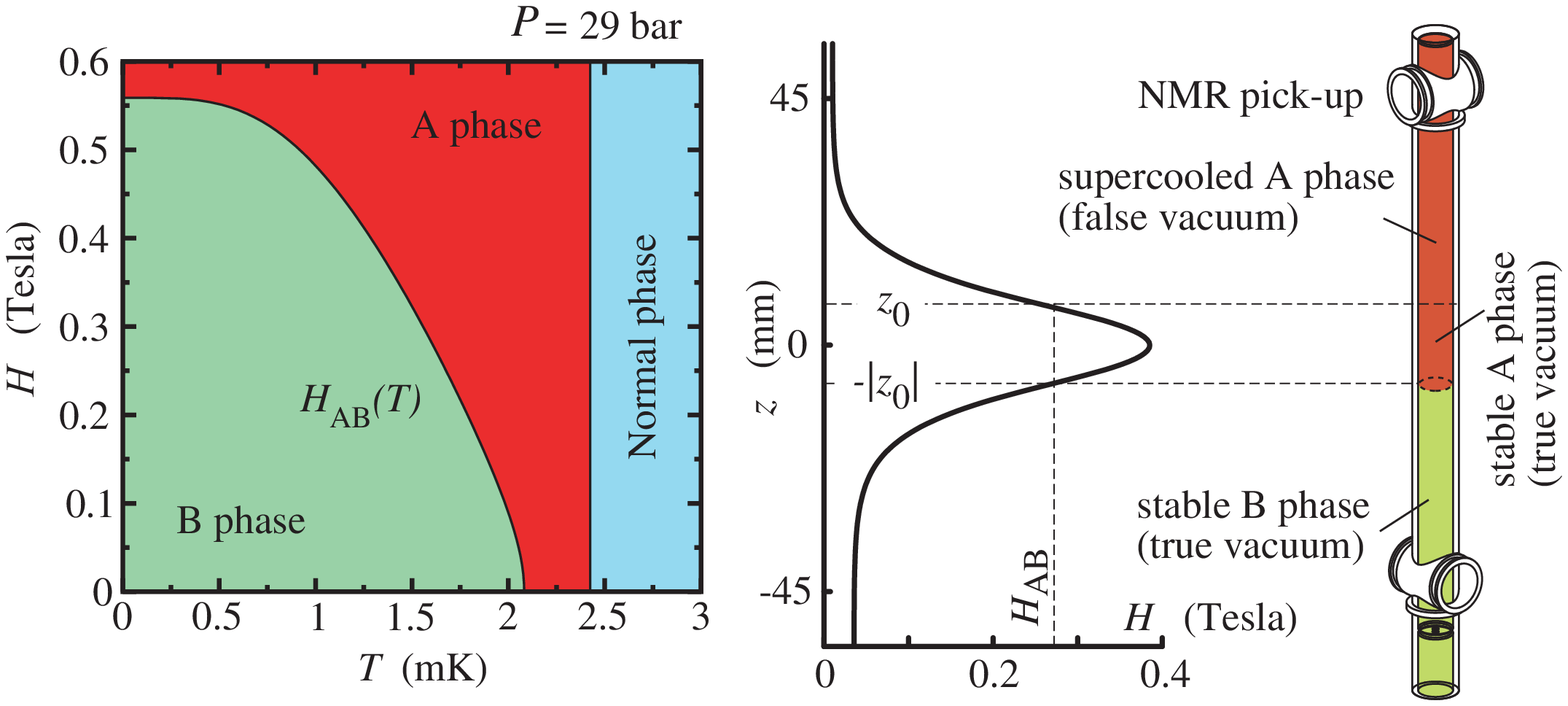}}
\medskip
\caption{Magnetic stabilization of $^3$He-A in the experimental
setup of Fig.~\ref{setup} \protect\cite{KH_prl}. {\it (Right)}
Sample container and the axially oriented magnetic field
distribution along the sample in the BA configuration (with
current $I_{\rm b} = 4.0\,$A in barrier magnet). The barrier field
maintains A phase in stable state in the section in the middle of
the sample, where $H > H_{\rm AB}(T)$. In the top and bottom sections
only the NMR polarization field is applied and $H < H_{\rm AB}(T)$. In
the top A phase may persist in meta-stable state (see text for
explanation). One AB interface exists here at the lower location
$H(-|z_0|)=H_{\rm AB}(T)$, where $z =-|z_0|$ is the average axial
location of the interface. When the temperature $T$ or the current
$I_{\rm b}$ is varied, changes occur in both  $z_0(T, I_{\rm b})$
and in the field gradient $\nabla H$ in which the interface
resides. The restoring force acting on the AB interface in
Eq.~(\protect\ref{InstabilityCondition2}) depends on the field
gradient $\nabla H$ and therefore on the field profile of the
barrier magnet. {\it (Left)} Magnetic phase diagram in the $(H,T)$
plane, where $H_{\rm AB}(T)$ marks the first-order phase transition
line between A and B phases (at $P=29.0\,$bar pressure)
\cite{hahn_prl}. The upper plane at $z_0(T, I_{\rm b})$ separates
the section in the middle, where $^3$He-A is the true vacuum
state, from the section in the top, where $^3$He-A is the false
vacuum. The false vacuum persists down to a container dependent
minimum temperature. } \label{KHInstabilitySetupFig}
\end{figure}

Two independent continuous-wave NMR spectrometers are used for
monitoring the sample. Their detector coils are located close to
both ends of the sample volume. In these regions of the sample a
low magnetic field is needed for NMR polarization which has to be
as homogeneous as possible.  Therefore three separate
superconducting solenoidal magnets with compensation sections, all
with their fields oriented along the rotation axis, are needed in
the experiment. Two magnets are required for the homogeneous NMR
fields and one as a "barrier magnet" around the central section of
the tube for stabilizing $^3$He-A. The barrier magnet provides
fields up to 0.6~T which is sufficient to stabilize the A phase at
all temperatures and pressures. In the inhomogeneous barrier field
$H(z)$ the equilibrium position $z_0$ of the AB interface is
determined by the equation $H(z_0)=H_{\rm AB}(T)$, where $H_{\rm AB}(T)$
is the first-order phase transition line between the A and B
phases in the $(H,T)$ plane (Fig.~\ref{KHInstabilitySetupFig} {\it
left}).

The two superfluids, $^3$He-A and $^3$He-B, have very different
magnetic properties, as seen in the phase diagram on the left of
Fig.~\ref{KHInstabilitySetupFig}. In the region where
$H>H_{\rm AB}(T)$ the A-phase has lower magnetic energy, while in the
neighboring region $H<H_{\rm AB}(T)$ the B-phase is favored. Thus the
gradient in the magnetic energy densities of the two liquids
provides a restoring force $F$ on the AB interface,
\begin{equation}
F= {1\over 2}~\nabla\left[(\chi_{\rm A}  -\chi_{\rm B} )
H^2\right] ~. \label{InstabilityCondition2}
\end{equation}
Here $\chi_{\rm A}>\chi_{\rm B}$ are the magnetic susceptibilities
of the A and B phases, respectively. Accordingly, this $\nabla H$
dependent restoring force on the AB interface has to be used
instead of gravity in the KH instability condition.

Various configurations of A and B phases can be stabilized and
trapped in the sample tube. (i) At zero or low barrier fields the
B phase might occupy the whole sample volume. In this {\it all-B
configuration} the two spectrometers probe the same volume. The
time difference between the two NMR readings can then be used to
study the propagation of vortices along the column. (ii) At
barrier fields above $H_{\mathrm{AB}} (T,P)$, the equilibrium
transition field between the A and B phases, the A phase is
stabilized in the center of the barrier magnet. In this {\it BAB
configuration} the top and bottom sections of the sample are
disconnected: vortex lines generated in one of the B-phase
sections do not pass to the other across the A phase region. (iii)
At pressures above 21~bar, $^3$He-A is stable also at zero
magnetic field from $T_\mathrm{c}$ down to $T_{\rm AB} (P)$. At
high temperatures the entire sample volume is then filled with A
phase {\it (all-A configuration)}. When the temperature is lowered
somewhat below the equilibrium A$\rightarrow$B transition, B phase
nucleates in the heat-exchanger volume and expands into the lower
section of the sample (at about $0.75\,T_{\rm c}$). If the barrier
field is sufficiently high to stabilize the A phase, the
advancement of the B phase is stopped by the stable A phase and an
AB interface is formed. Since the A phase in the top section can
supercool quite substantially \cite{osheroff_supercool}, the
sample remains in the {\it BA configuration}
(Fig.~\ref{KHInstabilitySetupFig} {\it right}). Thus the barrier
field isolates the top from an A$\rightarrow$B transition in the
lower section. Eventually, at a low enough temperature, B phase
also forms independently at the top. Since the ultimate
supercooling of $^3$He-A depends on the surface properties, the
A$\rightarrow$B transition also serves as a measure of the quality
of the quartz walls. At best the top was supercooled to
$0.52\,T_{\rm c}$ at 29~bar (when the equilibrium transition is at
$T_{\rm AB} = 0.85\,T_{\rm c}$). The BA configuration was most
important in the early stages of measurements on the AB interface
instability, since the top spectrometer is then recording the
vortex number in the A phase and the bottom spectrometer in the B
phase. To understand what happens in the two-phase sample when
rotation is started, and to correlate in real time the events on
both sides of the AB interface, simultaneous recordings of vortex
numbers in both phases are needed.

It turned out that in rotation $^3$He-A is in solid-body-like
rotational flow, practically locked to co-rotation with the
container, owing to its low critical velocity of vortex formation.
In contrast, in $^3$He-B the superfluid component is in the
vortex-free Landau state and stationary in the laboratory frame.
This nondissipative initial state becomes possible through the
formation of an A-phase vortex layer which covers the AB interface
and provides the discontinuity in the tangential flow velocities
(Fig.~\ref{VorAB-Interfaces} {\it left}). The tangential
discontinuity is ideal -- there is no viscosity in the motion of
the two superfluids so that this state can persist for ever. If
the rotation velocity is then incrementally increased to the
critical value, the Kelvin-Helmholtz instability occurs and some
vortices penetrate across the AB interface into the vortex-free B
phase (Fig.~\ref{VorAB-Interfaces} {\it right}). The first time,
when this happens in increasing rotation, is called the critical
rotation velocity $\Omega_{\rm c}(T,I_{\rm b})$. If rotation is
increased further, then the instability will occur repeatedly
every time when the counterflow velocity close to the cylinder
wall reaches the critical value. In each case, the signal for the
instability is the appearance of new vortices in the B-phase
section.

One might think that by substituting the interfacial restoring
force $F$ (\ref{InstabilityCondition2}) into
Eq.~(\ref{InstabilityCondition1}), and using the superfluid
densities of the A and B phases instead of the total density, the
critical velocity for the KH instability of the AB interface could
be obtained. However, it turns out that a proper extension of the
KH instability to superfluids incorporates the criterion in
Eq.~(\ref{InstabilityCondition1}) only as a particular limiting
case, namely when $T \rightarrow 0$, while in general the physics
of the instability is rather different from the ideal inviscid
model. A different but well-determined criterion is obtained for
the instability condition in terms of the velocities and densities
of the superfluid fractions. This criterion is also formulated in
the absence of viscosity, as one would expect in a perfect
superfluid environment. Therefore the tangential discontinuity at
the interface between $^3$He-A and $^3$He-B below the instability
is a stable state.

\subsection{Modification of Kelvin-Helmholtz instability in superfluids}
\label{kh_superfluid}

The criterion for the KH instability of ideal fluids in
Eq.~(\ref{InstabilityCondition1}) depends only on the relative
velocity across the interface. In practice there always exists a
preferred reference frame imposed by the environment. In the
superfluid case this is the frame fixed to the container. At
$T\neq 0$ the normal component provides the coupling to the
reference frame: in the state of thermodynamic equilibrium the
normal component moves together with the container walls, {\it
i.e.} $\mathbf{v}_\mathrm{n}=0$ in the frame of the container.
Owing to this interaction of the AB interface with its environment
the measured instability occurs at a lower differential flow
velocity, before the classical criterion in
Eq.~(\ref{InstabilityCondition1}) is reached. Moreover, it even
occurs in the case when the two superfluids move with the same
velocity. However, this does not imply that the renormalized
instability criterion would depend on the interaction with the
normal component -- in fact, it is still determined only by
thermodynamics. Waves are formed on the interface when the free
energy of a corrugation becomes negative in the reference frame of
the environment and the AB interface becomes thermodynamically
unstable, {\it i.e.} when it becomes possible to reduce the energy
via the normal component and its interaction with walls. The
instability condition Eq.~(\ref{InstabilityCondition1}) of the
ideal inviscid fluid is restored if the interaction with the
environment is not effective. This would occur, for example,
during rapid rotational acceleration at very low temperatures when
the instability caused by the interaction with the environment
does not have enough time to develop.

\begin{figure}[t]
\centerline{\includegraphics[width=0.8\linewidth]{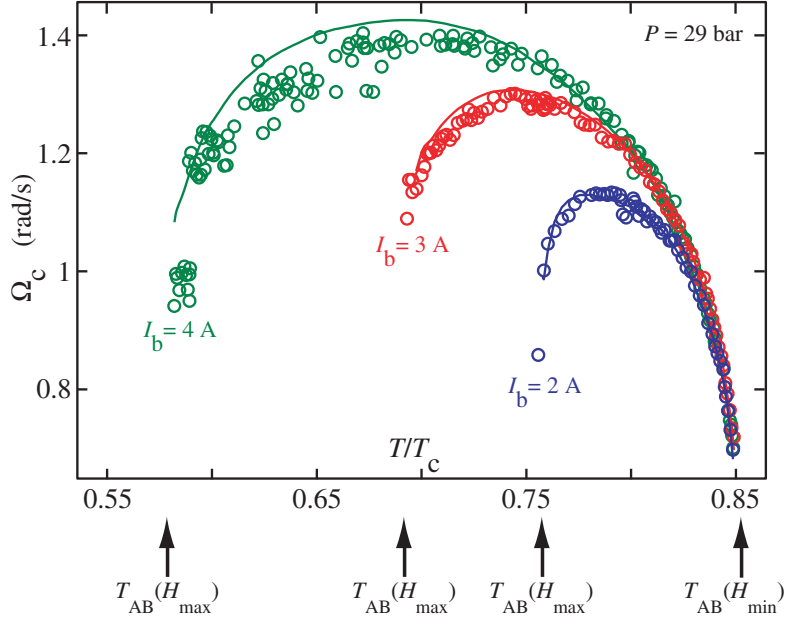}}
\medskip
\caption{AB interface instability as a function of temperature.
The measured critical velocity $\Omega_{\rm c} = U_{\rm B}/R$ is
plotted at three different fixed currents $I_{\rm b}$ in the
barrier magnet \protect\cite{KH_prl}. When $T$ is changed at
constant $I_{\rm b}$, the AB interface relocates roughly along the
contour where $H(z,r) = H_{\rm AB}(T)$. Thus its average position
$z_0$ and the field gradient $\nabla H$ change. These changes
alter the value of the restoring force in
Eq.~(\ref{InstabilityCondition2}). The solid curves represent the
instability criterion in
Eq.~(\protect\ref{InstabilityConditionNewnon-zeroT}), when
$U_{\rm B}=\Omega R$ and $U_{\rm A}= v_{{\rm nA}}= v_{{\rm nB}}=0$. No
fitting parameters are used. The vertical arrows below the figure
demarcate the temperature interval where the AB interface exists
within the barrier magnet at different currents $I_{\rm b}$: At
$T_{\rm AB}(H_{\rm min})$ the interface is stable in the minimum
field location. At $T_{\rm AB}(H(I_{\rm b})_{\rm max})$ the
critical field $H_{\rm AB}$ equals the maximum field inside the
barrier magnet.  } \label{KHInstabilityCurvesFig}
\end{figure}

The free energy of the disturbed AB interface $\zeta(x,y)$ in the
reference frame of the container contains the surface tension
energy, which corresponds to the potential energy in the `gravity'
field, and the kinetic energy of the two liquids
\cite{volovik_KH}:
\begin{eqnarray}
{\cal F}\{\zeta\}&=&{1\over 2}\int dx dy  \left[F\zeta^2 + \sigma
(\nabla\zeta)^2 \right]\nonumber \\
&&+ {1\over 2}\int dx dy \left[\int_{-h_{\rm B}}^\zeta dz
\rho_\mathrm{sB}\mathbf{v}_\mathrm{sB}^2 + \int_\zeta^{h_{\rm A}} dz
\rho_\mathrm{sA}\mathbf{v}_\mathrm{sA}^2\right].
\label{SurfaceFunctionalAnisotropic}
\end{eqnarray}
Here we take into account that in thermal equilibrium the normal
component is at rest in the container frame,
$\mathbf{v}_{\mathrm{nA}}=\mathbf{v}_{\mathrm{nB}}=0$, and only
the superflow contributes to the kinetic energy. The heights of
the A and B phase layers are denoted with $h_{\rm A}$ and $h_{\rm B}$. For
simplicity we ignore the anisotropy both in A phase and in B phase
at high fields and approximate the superfluid-density tensor with
a scalar. We write the superfluid velocities as
$\mathbf{v}_{\mathrm{sB}}({\bf r})={\bf U}_{\rm B}+ \delta
\mathbf{v}_{\mathrm{sB}}({\bf r})$ and
$\mathbf{v}_{\mathrm{sA}}({\bf r})={\bf U}_{\rm A}+ \delta
\mathbf{v}_{\mathrm{sA}}({\bf r})$, where ${\bf U}_{\rm B}$ and ${\bf
U}_{\rm A}$ are the velocities at an undisturbed flat interface. Using
$\nabla\times\mathbf{v}_\mathrm{s}=\nabla\cdot\mathbf{v}_\mathrm{s}=0$,
and the boundary conditions at the interface, $\hat{\bf
s}\cdot\mathbf{v}_{\mathrm{sA}}=\hat{\bf s}\cdot{\bf
v}_{\mathrm{sB}}=0$, one obtains the free energy of a surface mode
as (for details see Ref.~\cite{Ruokola} where also the anisotropy
of the tensor $\rho_{\mathrm{sA}}^{ij}$ is taken into account)
\begin{equation}
{\cal F}(\zeta_k)\propto  a^2 \left[F + k^2\sigma  -k \left(
\rho^{\rm eff}_{\rm A} U_{\rm A}^2 + \rho^{\rm eff}_{\rm B} U_{\rm B}^2 \right)\right]~.
\label{SurfaceFunctionalAnisotropic2}
\end{equation}
Here $k$ is the wave vector along $x$ for a surface corrugation
amplitude of the form $\zeta(x)=a\sin kx$, and
\begin{equation}
  \rho^{\rm eff}_{\rm A}=\frac{ \rho_{\mathrm{sA}} }{\tanh (kh_{\rm A})}~~,~~
  \rho^{\rm eff}_{\rm B}=\frac{ \rho_{\mathrm{sB}} }{\tanh (kh_{\rm B})} ~.
\label{EffectiveMass}
\end{equation}
In the relevant experimental conditions we are always in the `deep
water' limit, $kh_{\rm A}\gg 1$,  $kh_{\rm B}\gg 1$. The free energy becomes
negative for the first time for the critical ripplon wave number
$k_{0}=(F/\sigma)^{1/2}$  when
\begin{equation}
  \frac{1}{2}  \left( \rho_\mathrm{sB}U_{\rm B}^2 + \rho_\mathrm{sA}U_{\rm A}^2 \right)=\sqrt{\sigma F}~.
\label{InstabilityConditionNewnon-zeroT}
\end{equation}
A comparison of Eq.~(\ref{InstabilityConditionNewnon-zeroT}) to the
measured critical rotation velocity $\Omega_{\rm c}$ of the first KH
instability event is shown in Fig.~\ref{KHInstabilityCurvesFig}. The
curves have not been fitted; they have been drawn using accepted
values from the literature for the various quantities (for details see
Ref.~\cite{KH_prl}). Such a remarkable agreement for a complicated
phenomenon can only be achieved in superfluid $^3$He!

We thus conclude that, even under perfectly inviscid conditions,
in superfluids the critical velocity for the KH instability is not
given by the classical result for ideal fluids. The new criterion
for the instability in
Eq.~(\ref{InstabilityConditionNewnon-zeroT}) has at first glance
paradoxical consequences. The instability is not determined by the
relative velocity ${\bf v}_{\mathrm{sA}} -
\mathbf{v}_{\mathrm{sB}}$; in fact, the instability would occur
even if the two superfluid velocities were equal. The instability
would also occur for only a single superfluid with a free surface.
These new features arise from the two-fluid nature of the
superfluid. Therefore, the instability threshold is determined by
the velocities $\mathbf{v}_{\mathrm{s}i} - \mathbf{v}_\mathrm{n}$
of each superfluid $i$ with respect to the reference frame of the
walls and thus with respect to the normal fractions which in
thermodynamic equilibrium move together with the walls.
Accordingly, the free surface of a superfluid -- the interface
between the superfluid and the vacuum -- becomes unstable when the
superfluid velocity exceeds the critical value in the reference
frame of the normal fraction \cite{Korshunov,Korshunov2002}. With
many ($i$) superfluid fractions in the same liquid, such as the
neutron and proton superfluids in a neutron star
\cite{KH-NeutronStar,KH-NeutronStarPRL}, the threshold is
determined by some combination of the superfluid velocities ${\bf
v}_{\mathrm{s}i} - \mathbf{v}_\mathrm{n}$ \cite{Abanin}.

\subsection{Kelvin-Helmholtz instability in the low-temperature limit}
\label{kh_T=0}

On approaching the zero-temperature limit the density of normal
excitations rarefies, the coupling with the container walls
becomes weaker, and the superfluid density becomes the total
density: $\rho_{\mathrm{sA}}\rightarrow \rho_{\rm A}$ and
$\rho_{\mathrm{sB}} \rightarrow \rho_{\rm B}$. How is the superfluid
going to react to the environment under these conditions? Let us
compare the result in Eq.~(\ref{InstabilityConditionNewnon-zeroT})
with the ideal classical condition in
Eq.~(\ref{InstabilityCondition1}) in the limit when $T\rightarrow
0$. The classical instability condition reads
\begin{equation}
  \frac{\rho_\mathrm{A}\rho_\mathrm{B}}{\rho_\mathrm{A}+\rho_\mathrm{B}}
  \left( {\bf U}_\mathrm{B}-{\bf U}_\mathrm{A} \right)^2=2\sqrt{\sigma F}~,
\label{ZeroT1}
\end{equation}
while the superfluid instability occurs when
\begin{equation}
  \frac{1}{2}  \left(\rho_\mathrm{A} U_\mathrm{A}^2 +\rho_\mathrm{B} U_\mathrm{B}^2
\right)=\sqrt{\sigma F}~.
\label{ZeroT2}
\end{equation}
In the experiment the A-phase superfluid component is
approximately stationary in the container frame, $U_\mathrm{A} \approx 0$,
and the densities of the two liquids are practically equal,
$\rho_\mathrm{A} \approx \rho_\mathrm{B}=\rho$. Then the B-phase critical velocity
at the instability is $U_\mathrm{B}^2=4\sqrt{\sigma F}/\rho$ according to
the classical criterion, while from
Eq.~(\ref{InstabilityConditionNewnon-zeroT}) we obtain a result which
is smaller by a factor of $\sqrt{2}$, {\it i.e.} $U_\mathrm{B}^2=2\sqrt{\sigma F}/\rho$.

\begin{figure}[t]
\centerline{\includegraphics[width=0.6\linewidth]{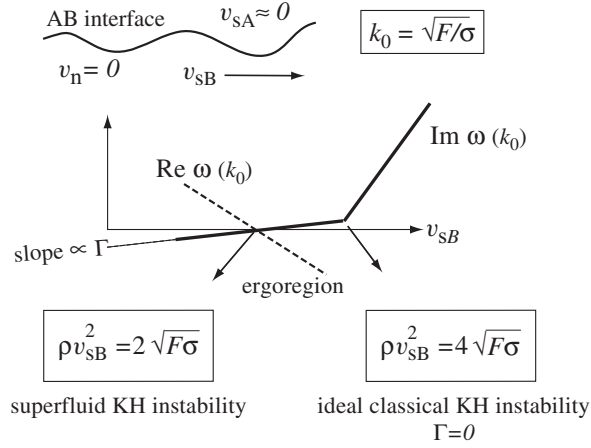}}
\medskip
\caption{Sketch of the imaginary and real parts of the frequency
$\omega (k)$ of the critical ripplon (when $k=k_0$) at the interface
between $^3$He-A and $^3$He-B. The diagram is constructed in the
rotating frame with $v_{{\rm sA}} \approx v_{{\rm nA}}= v_{{\rm
nB}}=0$ and $\rho_\mathrm{A} \approx \rho_\mathrm{B} = \rho$ on approaching the limit
$T\rightarrow 0$, and considering only incremental changes from the
critical conditons. At the superfluid instability, the imaginary part
${\rm Im}~\omega(k)$ crosses zero as a function of $v_{{\rm sB}}$ and
the attenuation of ripplons transforms to
amplification. Simultaneously the real part ${\rm Re}~\omega(k)$ also
crosses zero. The region where ${\rm Re}~\omega <0$ and where the
ripplon has negative energy is called ergoregion. The slope of the
imaginary part is proportional to the friction parameter $\Gamma$. If
$\Gamma$ is strictly zero, and thus the connection with the
environment is lost, the surface instability occurs at the value of
$v_{{\rm sB}}$ larger by a factor $\sqrt {2}$ --  at the corner
point which is part of the branch obeying the ideal classical KH
criterion of Eq.~(\ref{InstabilityCondition1}).  }
\label{KHvsLandauFig}
\end{figure}

The difference is imposed by the interaction with the environment
\cite{volovik_droplet}. To understand this, we repeat the
derivation of the classical KH instability
\cite{landau_fluid_dynamics} with one important modification. We
add to the equation of motion of the AB interface a friction force
which arises when the interface moves with respect to the
environment, {\it i.e.} with respect to the normal component. In
the reference frame of the container it has the form
\begin{equation}
F_{\rm friction}=-\Gamma \partial_t\zeta~.
\label{FrictionForceInterface1}
\end{equation}
In the low-$T$ limit, the friction between the AB interface and
the normal component arises from the Andreev scattering  of
ballistic quasiparticles from the interface. In this regime the
parameter $\Gamma \propto T^3$
\cite{KopninInterface,LeggettYip1990,YipLeggett1986}. For
simplicity, we choose a situation when both fluids move along the
$x$ direction, and consider surface waves (ripplons) of the form
$\zeta(x,t)=a\sin (kx-\omega t)$ in the container frame.
The classical dispersion relation for ripplon motion is then
modified by the presence of the friction term to
\begin{equation}
   \rho_\mathrm{A}\left({\omega\over
k}-U_\mathrm{A}\right)^2+\rho_\mathrm{B} \left({\omega\over k}-U_\mathrm{B}\right)^2 =
{F+k^2\sigma\over k} -i\Gamma{\omega\over k}~,
\label{Spectrumnon-zeroFrictionZeroT}
\end{equation}
modifying the nature of the instability. We rewrite the above
equation as follows:
\begin{eqnarray}
\frac{\omega}{k}=  \frac{\rho_\mathrm{A} U_\mathrm{A}+
\rho_\mathrm{B} U_\mathrm{B}}{\rho_\mathrm{A}+\rho_\mathrm{B}}
\nonumber
\\
\pm {1\over
\sqrt{\rho_\mathrm{A}+\rho_\mathrm{B}}}\sqrt{
{F+k^2\sigma\over k} -i\Gamma{\omega\over k} - {\rho_\mathrm{A} \rho_\mathrm{B} \over
\rho_\mathrm{A}+\rho_\mathrm{B}} (U_\mathrm{A}-U_\mathrm{B})^2} ~.
\label{Spectrumnon-zeroFrictionZeroT2}
\end{eqnarray}
If $\Gamma=0$, this reduces to the classical dispersion relation and
the instability occurs when the expression under the square root
becomes negative.  The ripplon spectrum then acquires an imaginary
part with both plus and minus signs. Thus at the threshold velocity of
Eq.~(\ref{ZeroT1}) the perturbation grows exponentially in time. A
sketch of the imaginary and real parts of the frequency of the
critical ripplon mode (with $k=k_0$) is shown in
Fig.~\ref{KHvsLandauFig}.

In the case of the superfluid instability we have to consider the
influence of the term with the friction parameter $\Gamma$ in
Eq.~\ref{Spectrumnon-zeroFrictionZeroT2}. When the imaginary part
${\rm Im}~\omega(k)$ crosses zero as a function of $U_\mathrm{B}$, the
attenuation of the ripplon modes is transformed to amplification.
This occurs first for ripplons with the wave vector given by
Eq.~(\ref{WaveVectorInstability}). While the instability condition
Eq.~(\ref{ZeroT2}) does not depend on the friction parameter
$\Gamma$, the slope of the imaginary part is proportional to
$\Gamma$. Therefore, when $\Gamma$ is strictly zero and the
connection with the reference frame vanishes, the interface
instability starts to develop when the classical KH criterion in
Eq.~(\ref{ZeroT1}) is reached. In the limit $T\rightarrow 0$ and
$\Gamma\rightarrow 0$, the experimental result is expected to
depend on how the observation time compares to the time needed for
the interface to become coupled to the environment, and for the
superfluid instability to develop at the lower critical velocity.
With sufficiently rapid rotational acceleration the classical KH
instability might then be reached. Unfortunately at present no
experimental verification exists on these predictions.

The real part of the ripplon frequency also crosses zero at the same
velocity than the imaginary part. Above the threshold the real part of
the ripplon spectrum, and thus its energy in the container frame,
becomes negative. This creates a connection to the physics of black holes
\cite{VolovikBH}. In general relativity the region where a
particle has negative energy is called the ergoregion. In the `shallow
water' limit $kh\ll 1$, when the spectrum of ripplons becomes
`relativistic', an exact analogy with the relativistic quantum field in the
presence of the black- or white-hole horizons emerges
\cite{SchutzholdUnruh2002}.

It is also possible here to identify a similarity with the
Kelvin-wave instability of an isolated vortex line in applied
flow. In the $T\rightarrow 0$ limit, when both $\alpha \rightarrow
0$ and $\alpha'\rightarrow 0$, the instability of a vortex line in
externally applied superflow towards the formation of Kelvin waves
resembles the AB interface instability. For the unstable modes in
Fig.~\ref{KHvsLandauFig} with $\mathrm{Im}~\omega > 0$, the real
part of the spectrum is negative, $\mathrm{Re}~\omega < 0$.
Similarly, for a vortex of finite length $L$ the wave-number is
constrained from below, $k > k_0=2\pi/L$, and the instability
forms at a critical velocity $v_{\rm c} \sim \tilde\kappa k_0$.
This $v_{\rm c}$ does not depend on the friction parameter
$\alpha$, whose role is to provide the dissipation from the
interaction between the vortex and the environment ({\it i.e.} the
normal component).

To conclude, we have found that the central property of the
superfluid KH instability is that the instability condition does
not depend on the relative velocity of the superfluids, but on the
velocity of each of the superfluids with respect to the
environment. The instability occurs even if the two fluids have
equal densities, $\rho_\mathrm{A}=\rho_\mathrm{B}$, and move with the same velocity,
$U_\mathrm{A}=U_\mathrm{B}$. This situation resembles that of a flag flapping in
wind. It was originally discussed with the view of the KH
instability of ideal fluids by Lord Rayleigh \cite{Rayleigh}.  One
might assume the instability to be that of a passive deformable
membrane between two distinct parallel streams having the same
density and velocity, as in the superfluid KH example: the flag
being represented by the AB interface, and the flagpole which pins
the flag serving as the reference frame fixed to the environment
so that the Galilean invariance is violated. However, the correct
explanation of the flapping flag requires the presence of friction
(for recent developments see {\it e.g.}
Ref.~\cite{FlexibleFilament}).

In the study of coherent quantum systems the special case of a free
surface is of great interest. Obviously the instability occurs even
if the two superfluids are on the same side of the interface, {\it
i.e.} with a free surface over a pool of two or more
interpenetrating superfluid components, such as a dilute solution of
superfluid $^3$He in superfluid $^4$He or the neutron and proton
superfluids in a neutron star \cite{Abanin}. The instability exists
even in the case of a single superfluid below the free surface under
the relative flow of the normal and superfluid components, as has been
pointed out by Korshunov
\cite{Korshunov2002}. He also derived two criteria for the
instability, depending on the coupling to the environment. In this
case the frame-fixing parameter which regulates the interaction with
the environment is the viscosity $\eta$ of the normal component of the
liquid. For $\eta\neq 0$ the critical counterflow velocity $v=v_{\rm
s} -v_{\rm n} $ for the onset of the surface instability is
independent of $\eta$, but does not coincide with the ideal classical
result which is obtained when viscosity is neglected ($\eta= 0$):
\begin{equation}
v^2(\eta\neq 0)=\frac {2}{\rho_ {\rm s} }  \sqrt{\sigma
F}=\frac{\rho_\mathrm{n}}{\rho}v^2(\eta=0)~. \label{InstabilitySurface}
\end{equation}
In laser-cooled rotating atom clouds in the
Bose-Einstein-condensed state the in\-sta\-bil\-ity of the free
surface is the generic vortex formation process
\cite{VorBEC,DynamicInstabilityBEC}. Another case of vortex
formation via surface instability is the interface between the
normal and superfluid states of the same fluid, which are flowing
at different velocities. Such a situation has been discussed
extensively for rotating $^3$He-B which is irradiated with thermal
neutrons \cite{neutronreview,Aranson}.

Finally, it is worth noting that these ideas might find
applications when studying the instability of quantum vacuum
beyond the event horizon, or the ergoregion of the black hole
\cite{volovik_droplet}. At the superfluid KH instability the
ergoregion develops on the AB interface, as the energy of the
surface waves becomes negative. Such ripplons are excitations of
the AB interface and provide a connection to the presently popular
idea in cosmology proposing that the matter in our Universe
resides on hypersurfaces (membranes or simply branes) in a
multidimensional space. Branes can be represented by topological
defects, such as domain walls or strings, and by interfaces
between different quantum vacua. In superfluid $^3$He, the brane
is represented by the AB interface between the two quantum vacua,
$^3$He-A and $^3$He-B. With the AB interface instability, it might
then be possible to model the instability of the quantum vacuum in
the brane world. The latter occurs in the ergoregion owing to the
interaction between the matter on the branes (represented by
ripplons) and the matter in higher-dimensional space (represented
by quasiparticles in the bulk superfluids).

\subsection{Measurement of AB interface instability}
\label{kh_injection}

\begin{figure}[t]
\begin{center}
\includegraphics[width=0.8\linewidth]{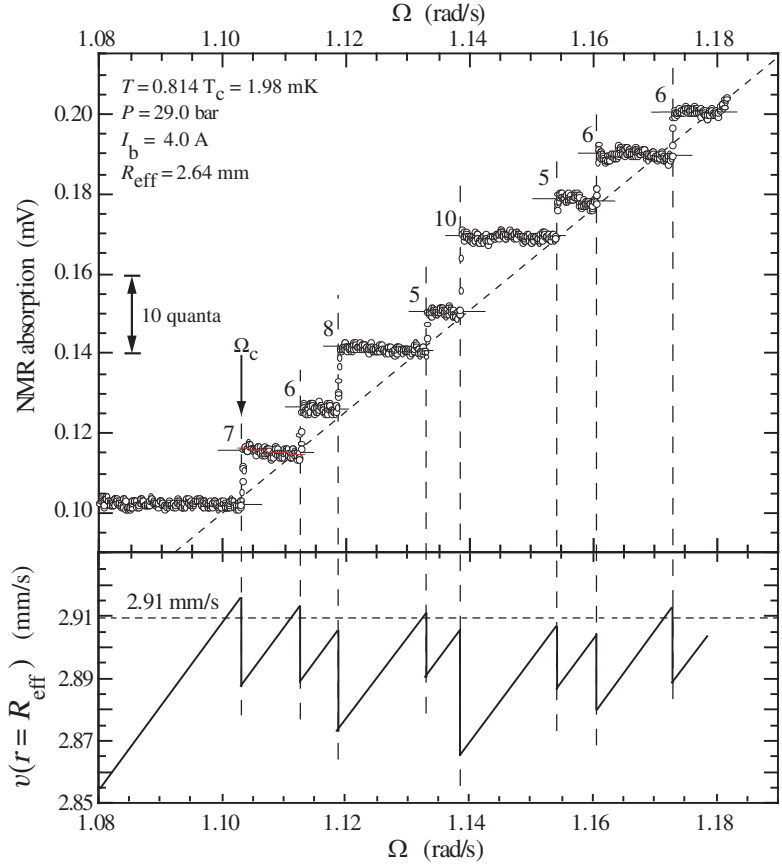}\\
\caption{Measurement of the superfluid KH instability of the AB
interface. {\it (Top)} A step-like increase in absorption is
observed in the Larmor region of the NMR line shape when the
rotation velocity $\Omega$ is slowly increased and the critical
value of counterflow velocity $v_{\rm c}$ is reached. The KH
instability occurs repeatedly in the form of different independent
events as a function of $\Omega$. A random number of $\Delta N$
rectilinear vortex lines are formed in each event in the B-phase
section of the sample (see Fig.~\protect\ref{setup}).
\emph{(Bottom)} At the instability the critical counterflow
velocity $v_{\rm c}(R_{\rm eff})$ is abruptly reduced by $\kappa
\Delta N/(2\pi R)$ to a sub-critical level. The constant critical
counterflow velocity $v_{\rm c}$ gives rise to the dashed lines in
each panel. $R_{\mathrm{eff}} \lesssim R$ is the effective radius
at which the instability occurs (see
Refs.~\cite{turb_meas_jltp,risto_prl} for details.)}\label{stairs}
\end{center}
\end{figure}

The superfluid KH instability of the AB interface is a new
physical phenomenon with wide-ranging ramifications, as discussed
in Secs.~\ref{KH-Intro} -- \ref{kh_T=0}. It also provides a whole
new set of tools for further measurements on vortex dynamics. The
standard KH measurement is that of the critical rotation velocity
$\Omega_{\rm c}$ shown in Fig.~\ref{stairs}. Here the NMR
absorption plotted on the vertical scale monitors the number of
rectilinear vortices in the B phase while the rotation velocity
$\Omega$ on the horizontal axis is slowly increased. Temperature
$T$ and barrier current $I_{\rm b}$ are kept constant during the
measurement. The first discontinuous jump in NMR absorption marks
$\Omega_{\rm c}$. This is the rotation velocity at which the first
vortices are formed in the initially vortex-free B phase. If
$\Omega$ is increased further, the instability occurs repeatedly
at the same value of critical counterflow velocity $\mathbf{v}(r =
R_{\rm eff}) = \mathbf{v}_{\rm c} \approx \mathbf{v}_{\rm n}(R) -
\mathbf{v}_{\rm sB}(R) =  \Omega_{\rm c} R $ (denoted with dashed
lines in both panels). Thus the staircase-like pattern of NMR
absorption is composed of steps of different height and can be
calibrated to provide the exact number of vortices which is
transferred across the AB interface in each single instability
event. Their number $\Delta N$ is denoted next to each instability
event. We see that $\Delta N$ is a small random number, which
after many similar measurements is found to be between 3 --- 30
with an average of $\sim 10$ \cite{turb_meas_jltp}. The
measurement in Fig.~\ref{stairs} is performed at high temperatures
where each of the $\Delta N$ vortex loops rapidly develops into a
rectilinear B-phase vortex line, after they have been tossed as a
tight bundle across the AB interface. From such measurements at
different temperatures, $\Omega_{\rm c}$ has been collected to
provide the curves at different values of $I_{\rm b}$ seen in
Fig.~\ref{KHInstabilityCurvesFig}.

\begin{figure}[t]
\includegraphics[width=0.8\linewidth]{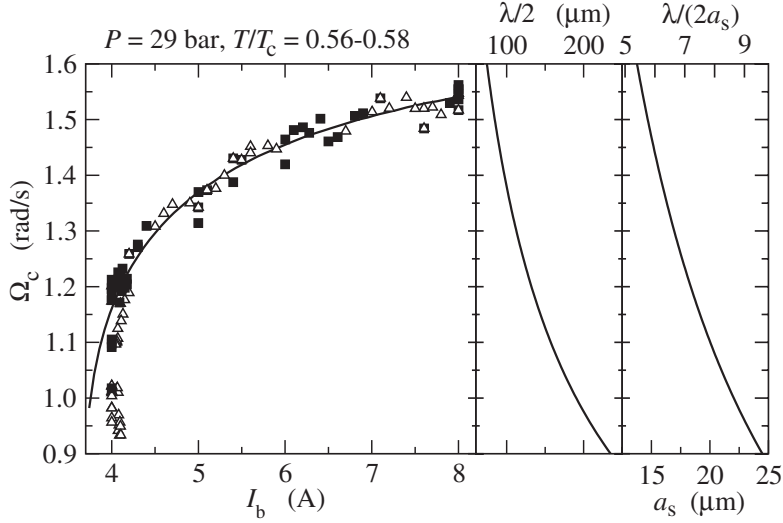}\\
\caption{AB interface instability as a function of the restoring
force at con\-stant temperature. (\emph{Left}) Measured critical
rotation velocity $\Omega_{\rm c}$ of the first KH instability
event as a function of the current $I_{\rm b}$ in the barrier
magnet. These measurements are conducted in the temperature regime
of the hydrodynamic transition between regular and turbulent
vortex dynamics: ($\vartriangle$) open symbols mark events in
which only a small number of $\Delta N$ new vortices are formed in
the B-phase section; ($\blacksquare$) filled symbols denote events
in which a turbulent burst increases the B-phase vortex number to
$N_{\rm eq}$. The solid curve represents
Eq.~(\ref{InstabilityConditionNewnon-zeroT}) without fitting
parameters. (\emph{Middle}) Half of the calculated ripplon
wavelength $\lambda = 2\pi/k_0$ at the instability according to
Eq.~(\ref{WaveVectorInstability}). (\emph{Right}) The bottom axis
gives the separation $a_{\rm s}$ between vortex quanta on the AB
phase boundary on the A-phase side, calculated from the number of
vortices in the A-phase section (when these are distributed evenly
in single-quantum units along the outer sample circumference
$r=R$). The top axis gives the number of single-quantum units
expected in one trough ($\lambda/2$)  of the surface wave. This
number $\lambda/(2a_{\rm s})$ agrees with the measured average for
$\Delta N \approx 10$ (see Ref.~\cite{turb_meas_jltp} for
details). }\label{kh fig}
\end{figure}

In the measurement of Fig.~\ref{stairs} the sample might be in
either the BA or BAB configurations. In the setup of
Fig.~\protect\ref{setup} the magnetic field distribution along the
vertical axis is almost symmetric with respect to the middle of
the sample. Also the end plates of the sample cylinder and the NMR
pick-up coils are at the same distance from the AB interfaces in
both B-phase sections. As a result the measured values of
$\Omega_{\rm c}$, the rotation velocity corresponding to the first
instability event, are identical for the top and bottom B-phase
sections in the BAB configuration. Nevertheless, the events at the
two AB interfaces in the BAB configuration are not correlated and
$\Delta N$ differs evidently randomly from one event to the next.
Therefore, as can be seen from Fig.~\ref{stairs}, subsequent
instability events at the two AB interfaces get soon out of step,
when $\Omega$ is increased above $\Omega_{\rm c}$, and occur in
the end randomly, but such that the critical counterflow velocity
remains fixed: $\mathbf{v}_{\rm c} \approx \Omega_{\rm c} R$. The
reason for this is that vortices play no direct role in the
instability condition of
Eq.~(\ref{InstabilityConditionNewnon-zeroT}): Even though the
vortices in the A-phase vortex layer covering the AB interface are
as closely packed next to each other as the soft core diameter
allows (see Fig.~\ref{kh fig} {\it right}), they do not determine
the instability. Instead it is the stability of the interface
itself in the tangential flow according to
Eq.~(\ref{InstabilityConditionNewnon-zeroT}).

\begin{figure}
\begin{center}
\includegraphics[width=0.86\linewidth]{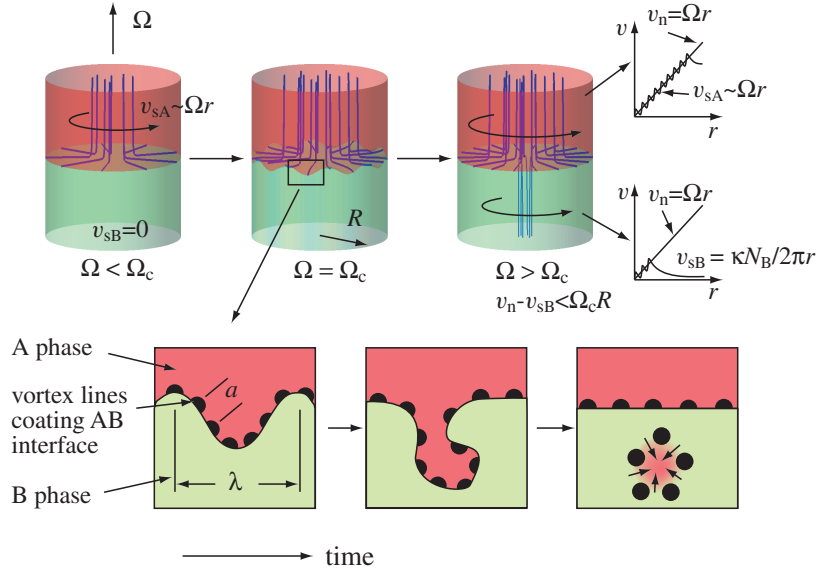}\\
\caption{Schematic illustration of how a KH instability event
might evolve on the AB interface. {\it (Top row)} Initially when
rotation is increased, vortices form at a low critical velocity in
the A phase section while none are formed in B phase.  At the
critical B-phase counterflow velocity the AB interface becomes
unstable towards wave formation, $\Delta N$ vortex loops end up on
the B-phase side of the interface, and then develop to $\Delta N$
rectilinear vortex lines. The end result from the instability is
that the $N_{\rm B} = \Delta N$ vortex lines, which now pass
through the AB interface after the first instability event at
$\Omega_{\rm c}$, wind up the B-phase superflow velocity and the
boundary settles down. In the upper right corner the velocities
are sketched of the normal ($v_\mathrm{n}$) and superfluid
($v_\mathrm{sA}$, $v_\mathrm{sB}$) components in the two phases.
They are shown here in the laboratory frame as a function of $r$
in the situation after the instability event (see also
Fig.~\protect\ref{cluster}). {\it (Bottom row)} A schematic
illustration of how  the vortex injection might happen. When the
boundary becomes unstable, waves form on the interface. A small
number of vortex lines becomes trapped in the deepest corrugation
which expands to the B-phase side (where $H < H_{\rm AB}(T)$ and A
phase is unstable). The corrugation becomes separated and the A
phase shrinks away but the circulation is left
behind.}\label{khseq_b}
\end{center}
\end{figure}

A schematic illustration is shown in Fig.~\ref{khseq_b} of the
process in which $\Delta N$ vortex loops might be transferred
across the AB interface. When the instability threshold is reached
while increasing $\Omega$, surface waves with wave vector
$k_0=\sqrt{F/\sigma}$ begin to form on the AB interface. At this
stage the $\Delta N$ vortices correspond to the A-phase
circulation quanta which reside in one of the corrugations of the
wave where it is about to become the deepest and most dominant
trough. In the lower row of diagrams in Fig.~\ref{khseq_b} the
possible sequence of events is depicted which might then take
place. The trough becomes a potential minimum for vortices, which
pushes the vortices even deeper in this well. Ultimately, as the
trough gets distorted, a Magnus force starts to act on the
vortices. The interface motion is highly damped and it moves
faster than the vortices. At some point the interface becomes
unstable below its equilibrium level in the region where $H <
H_{\rm AB}(T)$ and it springs back, the trough is smoothed out,
and a vortex bundle is left behind in B phase. As a result of such
a transfer of circulation across the AB interface, the counterflow
velocity at the phase boundary is now subcritical, and the
interface settles in its stationary state configuration.

A further illustration of the KH measurement and its agreement
with Eq.~(\ref{InstabilityConditionNewnon-zeroT}) is shown in
Fig.~\ref{kh fig}. This plot is generated by measuring
$\Omega_{\rm c}$ in similar manner as in Fig.~\ref{stairs}, by
increasing rotation to $\Omega_{\rm c}$ at different values of
$I_{\rm b}$, keeping temperature constant. Alternatively, the
measurement can be performed by decreasing $I_{\rm b}$ at constant
$\Omega$, so that the critical curve in Fig.~\ref{kh fig} is
traversed horizontally from right to left. Since the measurement
is at constant temperature, it is also performed at constant value
of $H_{\rm AB}(T,P)$. The result demonstrates the dependence of
the instability condition in
Eq.~(\ref{InstabilityConditionNewnon-zeroT}) on the restoring
force in Eq.~(\ref{InstabilityCondition2}), owing to the
dependence of $\nabla H$ on $I_{\rm b}$. When $I_{\rm b}$ is
changed, the AB interface moves within the barrier magnet to a
location where the field approximates the equilibrium transition
field, $H(r,z) \approx H_{\mathrm{AB}}$. Because of the large
curvature in the solenoidal field and the surface tension $\sigma$
of the AB interface, the equality is only approximately followed.
At large $I_{\rm b}$ the interface is flat, as seen in
Fig.~\ref{setup}, and $\Omega_{\rm c}$ changes only slowly  in
Fig.~\ref{kh fig}. In contrast, at low $I_{\rm b}$ the interface
is highly curved, its curvature changes rapidly with $I_{\rm b}$,
and accordingly $\Omega_{\rm c}$ is also a strong function of
$I_{\rm b}$. It is interesting to note that the $\Omega_{\rm c}$
data with the lowest $I_{\rm b}$ values in Fig.~\ref{kh fig}
correspond already to a configuration where the A-phase volume is
not a complete layer between two independent B-phase sections, but
where the AB interface separates out a torus-shaped A-phase volume
around the cylinder wall, with a B-phase channel connecting the
top and bottom sections of the sample \cite{AB-InterfaceShape}.
Surprisingly, even in this situation the measured data points lie
on a relatively smooth $\Omega_{\rm c} (I_{\rm b})$ dependence.

With a given barrier magnet the KH instability is restricted to
the range of velocities allowed by
Eq.~(\ref{InstabilityConditionNewnon-zeroT}). With the barrier
magnet of Fig.~\protect\ref{setup}  the critical velocities
$\Omega_{\rm c}$ are in the range 0.7 -- 1.6\,rad/s, as seen in
Figs.~\ref{KHInstabilityCurvesFig} and \ref{kh fig}. However,
vortex loop injection into vortex-free B-phase flow can be
performed at higher rotation velocities using the following
procedure: The barrier field is initially reduced below the
equilibrium value $H_{\rm AB}(T,P)$, so that there is no A phase
in the sample volume. The rotation velocity is then increased to
the desired value above the critical velocity $\Omega_{\rm c}$ of
KH instability. In this all-B-configuration the vortex free state
can be maintained up to the velocity at which vortex formation
from other sources starts. Next the barrier field is ramped up. A
phase then forms in a sudden hysteretic transition at a magnetic
field which somewhat exceeds the equilibrium value
$H_\mathrm{AB}$. Because of this superheated transition, the A
phase forms simultaneously within a larger volume, which is
unstable until the A-phase vortices have been formed. The AB
interface is also unstable, until a large number of vortices is
transferred into the B phase. The AB interface finally settles
down, when the B-phase superflow velocity has been reduced,
usually well below the critical value. This injection technique is
useful for studying the propagation of a large number of vortices
at velocities above $\Omega_{\rm c}$ at any temperature where the
AB interface exist.

The KH shear-flow instability provides a convenient mechanism for
precise vortex injection into initially vortex-free applied flow.
This is its principal application in the dynamical measurements
which are described in Sec.~\ref{TurbulenceKH}. Its critical
velocity is predictable and can be tuned externally. With the KH
instability it becomes possible to inject a bundle of vortex loops
as an externally triggered event: the rotation velocity can be
stabilized as close as within $\Delta\Omega\sim 0.01\,$rad/s below
the threshold $\Omega_{\rm c}$ and then suddenly increased by
$0.02\,$rad/s to start the KH event. Such reliability in vortex
injection allows new types of measurements on the dynamic
evolution of vortex lines in applied flow. The prime example is
the determination of the vortex flight time, a measurement of the
velocity at which a vortex propagates into vortex-free applied
counterflow in a rotating column (Fig.~\ref{VortexTrajectories}b).
Even more importantly, the KH instability provided the first firm
identification of the hydrodynamic transition from regular to
turbulent vortex dynamics which occurs on cooling below $\sim
0.6\,T_{\rm c}$; in other words, it allowed to distinguish a new
phenomenon from other sources of uncontrolled vortex formation.
\nopagebreak[2]

\section{Transition from regular to turbulent dynamics}
\label{TurbulenceKH}

\subsection{Introduction}

At high temperatures above $0.6\,T_{\rm c}$ the dynamical behavior
of vortices in $^3$He-B is regular, {\it i.e.} their number does
not increase during a time-dependent process. This is seen from
the fact that single-vortex steps are observed in
Figs.~\ref{VortexSteps} and in the KH measurement of
Fig.~\ref{stairs} the number of vortices $\Delta N$ formed in each
instability event fits within the distribution expected on the
basis of Fig.~\ref{kh fig} {\it (right)}. Some time ago it was
recognized that at lower temperatures single-vortex processes are
not observed in large applied flow \cite{VortexFormation};
instead, a large number of vortices is suddenly formed so that the
final state in rotation appears to include close to the
equilibrium number of vortex lines. More recently, a consistent
explanation of this phenomenon has been presented \cite{nature}.

On the basis of extensive measurements on the Kelvin-Helmholtz
instability as a function of temperature, it is now understood
that mutual friction divides the dynamics in $^3$He-B into a
high-temperature regime with regular vortex-number conserving
motion, and a low-temperature regime where superfluid turbulence
becomes possible. This hydrodynamic transition is unusually abrupt
as a function of temperature, as seen in Fig.~\ref{KH-8AvsT}. This
is the reason why it was first mistakenly interpreted to signal a
strongly temperature-dependent new critical velocity in
Ref.~\cite{VortexFormation}. The applied flow velocity, the
counterflow velocity $\mathbf{v}_{\rm n} - \mathbf{v}_{\rm s}$, is
an important factor in this transition: at low velocity a single
vortex has been observed to be in stable precessing motion for an
entire day at temperatures down to below $0.2\,T_{\rm c}$
\cite{Zieve3HeB}. A similar result has been verified for $^4$He-II
\cite{Zieve4He}. With increasing flow velocity the vortex
undergoes an instability towards the formation of Kelvin waves,
which may or may not lead to a rapid multiplication in the number
of vortex lines. The outcome from the instability depends on the
mutual friction parameter $q$.

The transition from regular to turbulent dynamics as a function of
mutual-friction dissipation is a new phenomenon. It has not been
observed in $^4$He-II; in this case the transition is expected
only a few tens of $\mu$K below $T_{\lambda}$, where $\rho_{\rm
s}$ is vanishingly small and the superfluid state very different
from that further below $T_{\lambda}$. In $^3$He-B the transition
is in the middle of the experimentally accessible temperature
range where the superfluid properties are continuous and well
developed.  Here the transition can be observed in one experiment
by scanning temperature from the superconductor-like dynamics at
high vortex damping to superfluid $^4$He-like turbulence at low
damping. This shows that superfluid dynamics can be varied and
that the traditional $^4$He-like superfluidity is just one
limiting case in this spectrum. The opposite extreme is superfluid
$^3$He-A where sufficiently low temperatures, to reach turbulence,
are probably not experimentally realistic.

In this section we first describe how the transition appears in
the KH meas\-ure\-ments. We then proceed with some models for its
explanation.

\subsection{Regular {\it vs} turbulent dynamics in
Kelvin-Helmholtz measurements}
\label{khturbulence}

\begin{figure}[t]
\begin{center}
\includegraphics[width=0.9\linewidth]{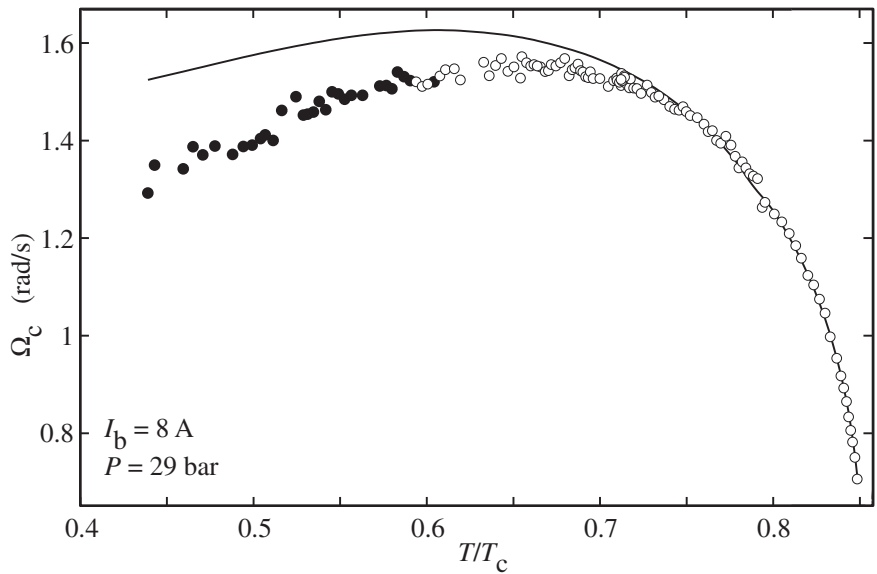}
\caption{Critical rotation velocity $\Omega_{\rm c}$ of the first
KH instability event at high barrier field, when the A-phase
section is stable down to $T \rightarrow 0$, or $H_{\rm b}(r,z)
|_{\rm max} > H_{\rm AB}(T = 0)$. The data points have been
classified according to the nature of the dynamic evolution after
injection: $(\circ)$, regular vortex number conserving;
$(\bullet)$, turbulent burst. A transition in the dynamics is seen
to take place at about 0.59\,$T_\mathrm{c}$, with little overlap
of open and filled symbols in the transition regime. The solid
curve denotes the calculated dependence from
Eq.~(\ref{InstabilityConditionNewnon-zeroT}) without fitting
parameters. In the absence of more appropriate parameter values,
mainly isotropic zero-field values have been used which do not
seem to provide perfect agreement at high fields and low
temperatures. } \label{KH-8AvsT}
\end{center}
\end{figure}

Fig.~\ref{KH-8AvsT} shows the KH critical velocity measured as a
function of temperature. Here the data points have been classified
according to whether the final state after the first critical
event only includes the vortices generated in the KH event itself
(open symbols), or if it incorporates almost the equilibrium
number of vortices (filled symbols). A sharp division line, with
little overlap of open and filled data points, is seen to divide
the plot into two regimes as a function of temperature: at high
temperatures only the KH vortices are generated while at low
temperatures the equilibrium vortex state is obtained after the
instability event. The KH critical velocity as a function of
temperature is a continuous smooth curve across this division line
and continues to follow the calculated dependence. The same
features are illustrated in Fig.~\ref{kh fig} which is measured at
a temperature close to the division line. A surprising and
characteristic property of these plots is that there are
essentially no data points with an arbitrary intermediate number
of vortices, not even in the transition regime.

The conclusion from these measurements is that at temperatures
below the division line a short turbulent burst follows the KH
event, after the closely packed bundle of vortex loops has arrived
across the AB interface into the vortex-free B-phase flow
(Fig.~\ref{khseq_b}). The turbulent burst generates the vortices
needed to reach almost the equilibrium number of vortices for the
B-phase flow. The propagation of these vortices along the long
rotating sample as a vortex front followed by a twisted vortex
bundle is discussed in Sec.~\ref{HelicalBundle}. A compelling
argument for the interpretation in terms of a turbulent burst is
the continuity of the KH critical velocity across the division
line -- it is unrealistic to assume that the nature of the
instability would change so suddenly. Since these early
observations \cite{nature} it has been understood that the KH
instability is just one example of vortex seed loop injection in
externally applied flow. Other examples are examined in
Sec.~\ref{InjectionMethods}. However, since the KH instability was
originally the most convincing case of vortex injection and of the
turbulent burst, with very particular properties, we describe this
example here in more detail.

\begin{figure}[t]
\begin{center}
\includegraphics[width=0.6\linewidth]{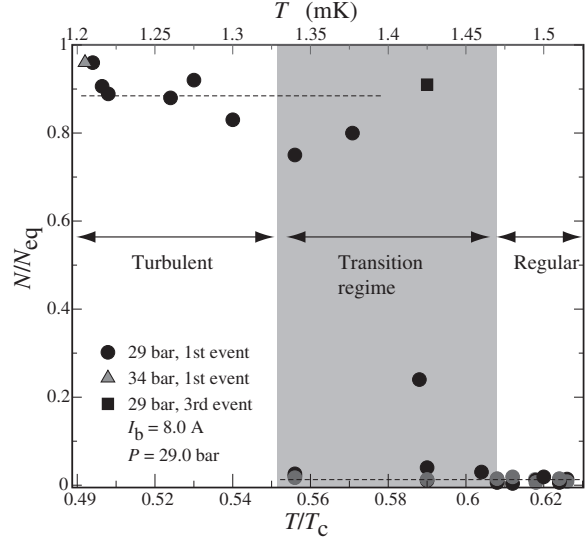}
\caption{The number of rectilinear vortex lines $N$ after KH
injection, normalized to the equilibrium number of lines $N_{\rm
eq}$, plotted as a function of temperature. At around
0.58\,$T_\mathrm{c}$, a sharp transition in the number of lines is
observed. At high temperatures the injection results in only a few
lines, but at low temperatures the equilibrium number of lines is
observed \cite{turb_meas_jltp}. } \label{fillingratio}
\end{center}
\end{figure}

Figure~\ref{fillingratio} shows a close-up of the transition
region as a function of temperature. The number of vortex lines
$N$ is plotted, normalized by the equilibrium number
$N_{\mathrm{eq}}$, after a single event of KH injection. At high
temperatures the injection leads always to a small number of
rectilinear lines, as expected for the KH instability
(Fig.~\ref{stairs}). Below 0.6\,$T_\mathrm{c}$ the number of
lines, after the system has settled down, is very different: the
injection results in the almost complete removal of counterflow.
The two regimes are separated by an abrupt transition at $T_{\rm
on}$~$\approx 0.58\,$$T_\mathrm{c}$ which has a narrow width of
$\Delta T_{\rm on} \approx 0.04\,T_{\rm c}$. We attribute the
final state at temperatures below $T_{\rm on}$ as the fingerprint
from a brief burst of turbulence which multiplies the vortex
number close to $N_\mathrm{eq}$. Here with KH injection the
turbulent burst occurs at the injection site at a distance of
30\,mm from the closest end of a detection coil. Thus there is no
direct NMR signal which would identify the burst itself, only the
propagating vortex state after it has traveled from the AB
interface to the detection coil. This means that the turbulent
burst is short in duration and localized to some section of sample
length adjacent to the AB interface.

In the transition regime the turbulent burst may not be triggered
by the first vortex injection event. Instead, a turbulent burst
can be preceded by one or more KH injection events which do not
lead to vortex multiplication. However, once turbulence sets in,
it generates the equilibrium number of lines in almost all the
cases.  Such events where turbulent vortex multiplication
terminates before the number of lines reaches the equilibrium
value are rare.

\begin{figure}[t]
\begin{center}
\includegraphics[width=0.8\linewidth]{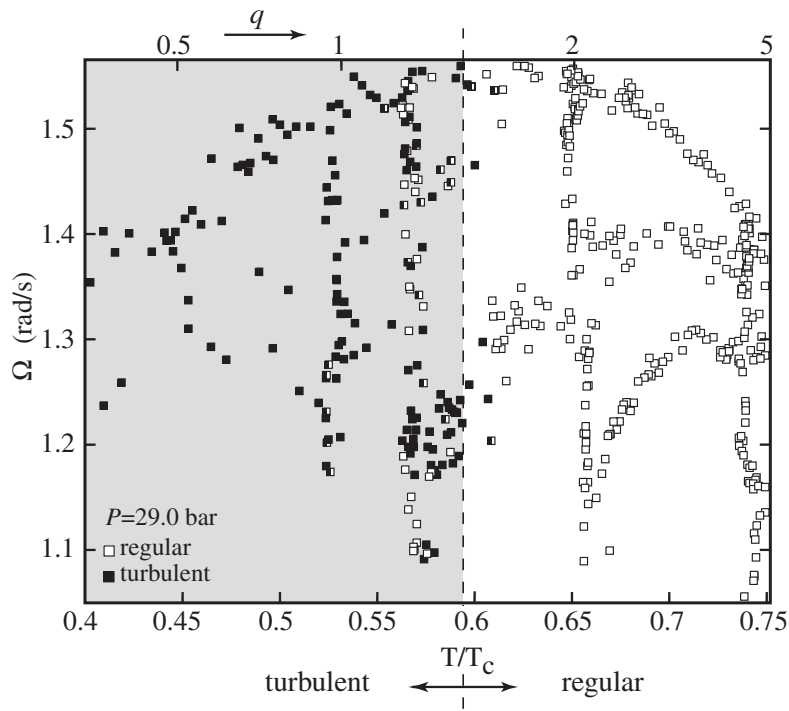}\\
\caption{Temperature-velocity phase diagram of turbulence. Each
marker in the plot represents the result from a measurement where
rotation is increased from zero to the critical velocity
$\Omega_{\rm c}$ of the AB interface instability. The cases where
turbulence is observed after the instability are marked with
filled symbols ($\blacksquare$) and regular cases with open
symbols ($\square$) \cite{nature}.  The horizontal top axis gives
the temperature dependent and velocity independent dynamic
parameter $q = \alpha / (1-\alpha^{\prime})$.  We conclude that
the phase boundary is primarily determined by its mutual friction
dependence. } \label{turb_29bar}
\end{center}
\end{figure}

The transition from Fig.~\ref{fillingratio} is shown as a phase
diagram in Fig.~\ref{turb_29bar}, plotted in terms of the B-phase
counterflow velocity $v=\Omega R$ and temperature $T$. In this
diagram each data point represents a KH injection measurement,
accumulated with different settings of the externally controlled
parameters, so that as wide a variation as possible is obtained
for the critical rotation velocity $\Omega_{\rm c}$ and
temperature $T$. Each marker in Fig.~\ref{turb_29bar} thus
indicates a KH injection event into vortex-free counterflow with
some parameter values which are not of interest in this context.
What we are interested in here is the division in filled and open
symbols: events followed by a turbulent burst are again marked
with filled symbols $(\blacksquare)$ while events which only lead
to a few rectilinear lines are marked with open symbols
$(\square)$. The transition occurs at about 0.59~$T_\mathrm{c}$;
at higher temperatures no injection events lead to vortex
multiplication while at lower temperatures all injections lead to
the equilibrium vortex state. The striking conclusion from this
plot is that the phase boundary between turbulence at low
temperatures and regular dynamics at high temperatures is vertical
and thus independent of the counterflow velocity.

\begin{figure}[t]
\begin{center}
\includegraphics[width=0.7\linewidth]{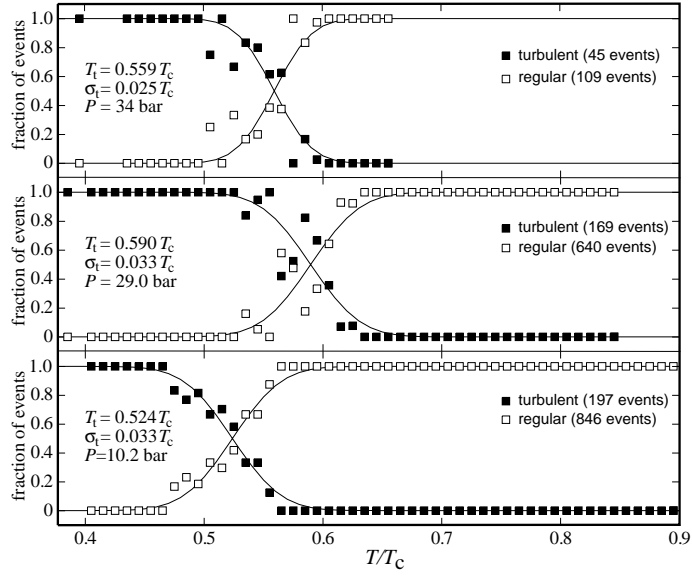}\\
\caption{Transition between regular and turbulent vortex dynamics
at different pressures. The transition is assumed velocity
independent and the events are categorized according to their
temperature \cite{turb_phase_d_jltp}. The transition temperature
increases with pressure, but the half width of its distribution is
approximately  $\sigma\approx 0.03$ $T_\mathrm{c}$ at all
pressures. }\label{transwidth}
\end{center}
\end{figure}

Plots similar to Fig.~\ref{turb_29bar} were also measured at 34
and 10\,bar pressures \cite{turb_phase_d_jltp}. In
Fig.~\ref{transwidth} the results are summarized. We assume that
the transition is velocity independent in the measured range of
velocities, and compile the data to show the probability of the
transition between regular and turbulent cases as a function of
temperature. The fitted Gaussian distributions give a narrow
half-width of $\sigma\approx0.03$~$T_\mathrm{c}$ at all pressures,
centered around a transition temperature in the range 0.52 --
0.59\,$T_{\rm c}$, depending on pressure. Mutual friction data is
available at 10 and 29\,bar pressures \cite{bevan}. At 10\,bar the
transition occurs at $q=0.6$ and at 29\,bar $q=1.3$. Thus the
transition appears to move to higher $q$ value with increasing
pressure. Measurements were also carried out at zero pressure
where the transition was found to be below $0.45\,T_{\rm c}$ at
low rotation velocities of 0.5 -- 0.7\,rad/s.

To conclude, KH injection measurements indicate that the phase
boundary between turbulent and regular vortex dynamics is foremost
a function of temperature and independent of the applied flow
velocity at velocities above $| \mathbf{v}| = |\mathbf{v}_{\rm n}
- \mathbf{v}_{\rm s}| \gtrsim 2.5\,$mm/s. Thus the transition
occurs as a function of mutual friction, such that the dynamic
parameter $q$ is of order unity.

\subsection{Classical and superfluid turbulence}\label{TurbTheory}

Since classical and quantum turbulence share many common features,
we begin with the basic concepts from classical turbulence by
inspecting the properties of the different terms in the
Navier-Stokes equation \cite{landau_fluid_dynamics}
\begin{equation}
\frac{\partial \mathbf{v}}{\partial t}+ \bm{\nabla }\tilde \mu =
   \mathbf{v}\times
   \bm{\omega}+  \nu\nabla^2 \mathbf{v} \ .
   \label{NormalHydrodynamics}
\end{equation}
Here $\nu$ is the kinematic viscosity (viscosity $\eta$/density
$\rho$), $\bm{\omega}=\bm{\nabla}\times \mathbf{v}$ is the vorticity
in the inertial (laboratory) reference frame. Turbulence is governed
by the interplay of the two terms on the r.h.s. of this equation, the
inertial (first) term and the viscous (second) term.

The transition to turbulence is determined by the Reynolds number
${\rm Re}=LU/\nu$, formed from the characteristic values for the
three quantities describing the flow: its velocity $U$, the
geometric size of the system $L$, and the kinematic viscosity
$\nu$. For small Reynolds numbers, the viscous term, $-\nu k^2
\mathbf{v}_{\bf k}$ for a perturbation with wave vector ${\bf k}$,
stabilizes laminar flow. In contrast, at large ${\rm Re}\gg 1$ the
inertial term in Eq.~(\ref{NormalHydrodynamics}) dominates, and
laminar flow becomes increasingly unstable against the formation
of a disorganized flow of eddies. In the most carefully prepared
experiments laminar flow has been maintained in a circular pipe up
to ${\rm Re}\sim 10^5$ \cite{faberbook}. However, the higher the
Reynolds number the smaller the disturbance needed to trigger
turbulence \cite{mullin_prl}. The evolution of turbulence is
governed by the Kolmogorov energy cascade: the kinetic energy of
the flow is transferred to smaller and smaller length scales, via
the decay and break up into ever smaller vortex loops along the
so-called Richardson cascade \cite{Frisch}, until a length scale
is reached where the energy can be dissipated by viscosity.

In superfluids turbulence acquires new features. First, the
superfluid consists of two inter-penetrating components: the
frictionless superfluid and the viscous normal fractions. The
total density $\rho=\rho_\mathrm{n}+\rho_\mathrm{s}$ is the sum of
the densities of these two components. The normal component
behaves like a regular viscous fluid while the superfluid
component is an ideal superfluid. At the superfluid transition the
density of the superfluid component vanishes, but increases with
decreasing temperature, until in the $T\rightarrow 0$ limit the
normal component vanishes. Secondly, the vorticity of the
superfluid component is quantized as discussed in
Sec.~\ref{sect-quantvortices}. If both the normal and superfluid
components are able to move, the turbulent state bears more
resemblance to the turbulence of classical viscous liquids. This
is often the case in superfluid $^4$He-II
\cite{vinen_turb_review}. In superfluid $^3$He, however, the
normal component is so viscous that it is essentially immobile.
The flow is then carried by the superfluid component which
contains a large number of quantized vortices.  Here a new class
of turbulent flow becomes possible: {\it one-component superfluid
turbulence}. It is this state of turbulent flow that we consider
in what follows. A window for this regime of superfluid turbulence
appears because of the existence of one more large parameter in
superfluid $^3$He, namely the ratio of its normal kinematic
viscosity to the superfluid circulation quantum, $\nu /\kappa$. In
$^3$He at $T\sim 0.5\, T_\mathrm{c}$ this ratio is $\nu /\kappa \sim 10^3$
in contrast to $^4$He-II where it is of the order of unity. A more
detailed discussion is given later in this section.

Because of its large viscosity, the normal component of $^3$He-B
moves together with the container. As a result,
Eq.~(\ref{vs-dynam-micro}) takes the following form in the frame
where the normal component is locally at rest ($\mathbf{v}_{\rm
n}=0$):
\begin{equation}
\frac{\partial \mathbf{v}_{\rm s}}{ \partial t}+ \bm{\nabla}\tilde
\mu  = (1-\alpha')\mathbf{v}_{\rm s} \times \bm{\omega}_\mathrm{s} + {\bf
f}_{\rm visc} \, , \label{SuperfluidHydrodynamics2}
\end{equation}
where
\[
{\bf f}_{\rm visc}=\kappa \alpha \sum_\beta \int \delta\left({\bf
r}-{\bf r}_\beta\right)d{\bf r}_\beta\times \left(\hat {\bf
s}_\beta\times {\bf v}_\mathrm{s}\right)
\]
is the viscous part of the mutual friction force.

The first inertial term on the r.h.s. drives the flow instability
towards turbulence in the same way as the inertial term in the
Navier-Stokes equation does for potential flow in classical
hydrodynamics \cite{landau_fluid_dynamics}. The second dissipative
term on the r.h.s. of Eq.~(\ref{SuperfluidHydrodynamics2}) is the
counterpart of the dissipative term in the Navier-Stokes equation
(\ref{NormalHydrodynamics}). It tends to stabilize the flow, since
it leads to energy dissipation
\begin{equation}
\frac{\partial }{\partial t}\frac{v_s^2}{2}={\bf v}_\mathrm{s}\cdot
{\bf f}_{\rm visc}, \label{Dissipation}
\end{equation}
where ${\bf v}_\mathrm{s}\cdot {\bf f}_{\rm visc}<0$ according to
Eq.~(\ref{Fmf-vsprojection}). The fundamental difference between
the dissipative terms in classical and superfluid dynamics is that
in superfluids this term has the same scaling dependence on
velocity and vorticity $f_{\rm visc}\sim \alpha \omega_s v_s$ as the
inertial term. This is a consequence of the two-fluid dynamics,
where the vortices provide the mechanism of momentum and energy
transfer between the two components of the fluid. Thus the
effective Reynolds number -- defined as the ratio of the inertial
and dissipative terms in the relevant hydrodynamic equation -- has
to be changed. In superfluids, it is the ratio of the two terms on
the r.h.s. of Eq.~(\ref{SuperfluidHydrodynamics2}):
\begin{equation}
{\Re}= (1-\alpha')/\alpha =1/q \ . \label{SuperfluidReynolds}
\end{equation}
According to this definition, and in analogy with classical
hydrodynamics, turbulence in superfluids is expected when
$1/q\gtrsim 1$ and laminar (regular) flow when $1/q\lesssim 1$. As
distinct from viscous liquids, this condition is independent of
extrinsic quantities such as the counterflow velocity  or the
characteristic dimension $R$ of the sample. The temperature
dependence of $\Re$ for \heb and He-II is explicitly shown in
Fig.~\ref{mf_data}.

The above dimensional arguments on the transition between regular
and turbulent vortex dynamics are in agreement with the results in
Fig.~\ref{turb_29bar}, where the boundary between the laminar and
turbulent regimes was also found to be at $q\sim 1$. Thus the
velocity independent parameter $\Re=1/q$, which controls the
transition to turbulent flow in superfluids, plays the same role
as the velocity dependent Reynolds number ${\rm Re}=UR/\nu$ in
classical hydrodynamics.

These considerations are valid provided that the applied
counterflow velocity is high enough to sustain vortices, {\it
i.e.} it exceeds the Feynman critical velocity
(Sec.~\ref{CriticalVelocity}), which is easily satisfied in any
experiment. From the Feynman criterion, one can define another
dimensionless parameter,  the ``superfluid Reynolds number''
\begin{equation}
   {\rm Re}_\mathrm{s}=U_\mathrm{s}R/\kappa,
\label{Superfluid2Reynolds}
\end{equation}
where $U_\mathrm{s}$ is the mean superfluid velocity with respect to the
normal component, {\it i.e.} the counterflow velocity. If the
condition ${\rm Re}_\mathrm{s}\gtrsim 1$ is fulfilled up to a logarithmic
prefactor $\ln (R/\xi)$, it becomes energetically favorable to add
a vortex line in the bulk flow.

\subsection{Onset of turbulent burst}
\label{TransitionTubulence}

In KH injection a tight bundle of many small vortex loops
(Fig.~\ref{KH-InjectVorConfig}) is transferred across the AB
interface into vortex-free B-phase flow. How is a turbulent burst
started from such a vortex bundle? In this section we discuss the
initial phase of vortex multiplication in applied flow using a
simple phenomenological model from Ref.~\cite{kopnin_turbulence}
which is constructed in the spirit of the Vinen equation for
superfluid turbulence \cite{vinen_eq}.

To characterize the initial conditions we need two numbers: the
intrinsic velocity-independent ${\Re}=1/q$ in
Eq.~(\ref{SuperfluidReynolds}) and the ``superfluid Reynolds
number'' ${\rm Re}_\mathrm{s}$ in Eq.~(\ref{Superfluid2Reynolds}). We
assume ${\rm Re}_\mathrm{s}\gg 1$, which corresponds to the typical
experimental situation in $^3$He-B that the presence of many more
vortices would be energetically possible. If a large energy
barrier prevents vortex nucleation, then vortices are not
necessarily created even at high ${\rm Re}_\mathrm{s}\gg 1$ and the
superfluid remains in a metastable state of counterflow. At
velocities well below the intrinsic critical velocity of vortex
formation (Sec.~\ref{CriticalVelocity}), $v_{\rm c}\sim \kappa /\xi$
\cite{VortexFormation}, superfluid turbulence can be initiated if
quantized vortices are injected by some extrinsic means into
vortex-free flow. In the rotating sample, provided that $1/q
\gtrsim 1$, turbulence then develops in a sudden burst where the
initial seed vortices start to multiply and form a vortex tangle
in the bulk volume.

We start from an initial configuration containing many
(essentially more than one) randomly oriented vortex loops, which
might have been either injected or created by some other
precursory process from a set of a few seed vortices. An example
of the precursor can be the Kelvin-wave instability discussed in
Sec.~\ref{Sect-Kelvin-wave}. We show that the initial array of
entangled vortices is unstable towards a burst-like multiplication
of vorticity provided the mutual friction is low enough. We assume
that vortex multiplication occurs within a certain region in the
fluid where the number of vortices is large. The region where
multiplication takes place can exist near the location of the seed
loops and/or near the walls of the rotating container, where the
rotating counterflow velocity reaches its maximum value and the
number of vortices in effect is increased by the presence of image
vortices. The vortex tangle created in such a region of
multiplication penetrates next into the rest of the fluid volume.

Since superfluid vorticity is quantized, the formation of new
vortices during the onset becomes the key issue. We consider this
process, taking mutual friction into account, and derive an
equation for the evolution of the density of entangled vortex
loops during the initial stages of the transition to turbulence.
The multiplication of seed vortices can be studied with Eq.\
(\ref{SuperfluidHydrodynamics2}). Taking the curl of both sides of
Eq.~(\ref{SuperfluidHydrodynamics2}) we obtain the equation for
the superfluid vorticity
\begin{equation}
\frac{\partial \bm{\omega}_\mathrm{s}}{\partial t}=(1-\alpha ^\prime
)\left[\bm{\nabla}\times (\mathbf{v}_\mathrm{s}\times \bm{\omega}_\mathrm{s})\right]+
\bm{\nabla}\times {\bf f}_{\rm visc} \ . \label{vorticity}
\end{equation}
Here we assume that the vorticity produced by turbulence is much
larger than the angular velocity, $\omega_\mathrm{s} \gg \Omega$, {\it
i.e.} the vortex density is much higher than that in equilibrium.
This allows us to neglect $\bm{\nabla}\times {\bf v}_\mathrm{n}=2{\bf
\Omega}$ in comparison to ${\bm \omega}_\mathrm{s}$.

Let us now average Eq.~(\ref{vorticity}) over randomly oriented
vortex loops with dimensions spread over an interval around a
characteristic size $\ell$, using the ideas of
Ref.~\cite{vinen_eq}. In a state of entangled vortex loops, their
three-dimensional density is $ n_{\rm v} \sim \ell ^{-3} $ while
the vortex-loop length per unit volume (two-dimensional vortex
density) is $ L=\ell n_{\rm v} =\ell ^{-2}= n_{\rm v} ^{2/3} $.
After averaging Eq. (\ref{vorticity}) only its scalar value is
meaningful, as any of one its components to a specific direction
vanishes. Let us express the two terms on the r.h.s. of Eq.\
(\ref{vorticity}) in terms of the vortex density $L$, keeping in
mind that the velocity produced by the vortex tangle is of order
$\tilde\kappa /\ell$, while its vorticity is $\omega _s\sim
\tilde\kappa /\ell^2 \sim \tilde\kappa L$, where $\tilde\kappa
=(\kappa /4\pi)\ln (\ell /\xi) $.

Regarding the onset of turbulence, the reactive coefficient
$\alpha' $ in Eq.\ (\ref{vorticity}) simply renormalizes the
inertial term of conventional hydrodynamics [the first term on the
r.h.s. of Eq.\ (\ref{NormalHydrodynamics})] that drives the
instability towards turbulent vortex formation. Therefore, the
vortex density {\em increases} owing to the first term in Eq.\
(\ref{vorticity}) according to
\begin{equation}
\dot L_+ =A (1-\alpha ^\prime )v_\mathrm{s} L^{3/2} =A (1-\alpha ^\prime
)(U-v_0) L^{3/2}  \label{creation}
\end{equation}
where $A\sim 1$ is a constant. The superfluid velocity in Eq.\
(\ref{creation}) is assumed to be ${v}_\mathrm{s}=U-{v}_0$ where $U$ is the
counterflow velocity, and $ v_0\sim \tilde\kappa/\ell $ is the
self-induced velocity for a vortex loop of length $\ell$ and core
radius $\xi$. The kinetic energy of the superfluid grows due to
the increase in the loop density. The energy is taken from the
external source at the length scale $R$ with the rate $dE/dt \sim
E_\mathrm{L} \dot L_+$, where $E_\mathrm{L}$ is the energy of the vortex per unit
length.

Vortex multiplication saturates when the loop density reaches a
value such that $U =v_0\sim \tilde\kappa L^{1/2}$. If the density
happens to become larger, it will decrease towards $L_{\rm sat}\sim
(U/\tilde\kappa)^2$ while the kinetic energy is returned back to
the external source. In other words, saturation is reached when
the ``turbulent superfluid Reynolds number'' ${\rm Re}_{\rm s}^{({\rm
turb})}=U\ell /\tilde\kappa$ becomes of order unity. The condition
Re$_\mathrm{s} \gg 1$ ensures the separation of scales, $\ell \ll R$, that
is required for the formation of a vortex tangle. The vorticity at
saturation, $\omega_{\rm sat}\sim \tilde\kappa /\ell^2_{\rm sat} \sim
\Omega \,{\rm Re}_\mathrm{s}$, is much larger than the equilibrium
vorticity $2\Omega$.


The second (dissipative) term in Eq. (\ref{vorticity}) acts to
stabilize vortex-free flow, thus {\em reducing} vortex density in
a way similar to that in normal fluids. Estimating the dissipative
term as $f_{\rm visc}\sim \alpha \omega_\mathrm{s} v_\mathrm{s}$ we
find for the rate of decrease in vortex density
\begin{equation}
\dot L_- = -B\alpha (U-v_0) L ^{3/2}   \label{inflation}
\end{equation}
where $B\sim 1$ is a constant. In contrast to viscous fluids, the
dissipative term has here the same scaling dependence as the
driving term in Eq~(\ref{creation}).

The multiplication of vortex loops, as described by Eq.\
(\ref{creation}), can be understood in terms of vortex collisions
and interconnections. Such processes were indeed seen in numerical
simulations on quantized vortices
\cite{nature,schwarz_88,RotTurbulence}. Reconnections of vortices
accompanied by the formation of a vortex tangle in normal fluids
were considered recently in Refs.\ \cite{Kivotides1,Kivotides2}.
Each reconnection of quantized vortices takes place over a
microscopic time of the order of the quasiparticle collision time,
which is much shorter than the characteristic times involved in
hydrodynamic processes. It is accompanied by some small amount of
dissipation within a volume of the order of $\xi ^3$. We consider
these reconnection processes as instantaneous and neglect the
corresponding dissipation. The rate of increase in vortex loop
density should be quadratic in $n_{\rm v}$, and thus $ \dot n_+ =
A v_{\rm r} n_{\rm v}^2 \ell ^2 $. Here $v_{\rm r}$ is the
relative velocity of the vortex loops, $\ell ^2$ is the loop
``cross section'', and the constant $A\sim 1$ describes the
``efficiency'' of the vortex multiplication due to pair
collisions. Using the definition of $L$ the vortex multiplication
rate becomes $\dot L_+ \sim v_{\rm r} L^{3/2}$. The vortex
velocity is determined through the mutual friction parameters
$\alpha$ and $\alpha ^\prime$ such that $ \mathbf{v}_\mathrm{L}=(1
-\alpha ^\prime ){\bf v}_\mathrm{s} -\alpha \,
\hat{\bm{\omega}}_\mathrm{s}\times \mathbf{v}_\mathrm{s} $. After
averaging over randomly oriented vortex loops the last term
vanishes, resulting in the average relative velocity of loops
$v_{\rm r}$ proportional to the longitudinal component of
$\mathbf{v}_\mathrm{L}$, $v_{\rm r}\sim (1-\alpha ^\prime
)v_\mathrm{s}$. The rate $\dot L_+$ thus agrees with Eq.\
(\ref{creation}).

The effect of the (second) viscous term on the r.h.s. of Eq.\
(\ref{vorticity}) is to decrease the loop density in the region of
multiplication, by inflating the loops in the applied counterflow,
and by extracting them from the region of multiplication into the
bulk where the counterflow is smaller. The viscous component of
the mutual-friction force leads to variations in the vortex loop
length $ \dot \ell \sim 2\pi v_\mathrm{L} \sim \alpha v_\mathrm{s}
\sim \alpha (U-v_0) $. Finally, for the rate of variation in the
vortex-loop density from the viscous component one obtains Eq.
(\ref{inflation}). The length increases while the density
decreases as long as saturation is not reached, $U -v_0>0$. If
saturation is exceeded, $v_0>U$, the density increases since the
loops shrink owing to the friction. Accordingly, the vortex loops
are extracted from the region of multiplication into the bulk
before saturation $U>v_0$, and they are extracted out of the bulk
fluid if $U<v_0$.

As we see, both inertial and viscous mutual friction terms, Eqs.\
(\ref{creation}) and (\ref{inflation}), have the same dependence
on vortex density, {\it i.e.} on the vortex length scale. The
total variation in loop density in the region of multiplication is
the sum of the two processes, $\dot L =\dot L_+ +\dot L_-$.
Putting $v_0=\tilde\kappa /\ell =\tilde\kappa L^{1/2} $ we obtain
\begin{equation}
(1-\alpha^\prime)^{-1}\dot L=(q_{\rm c}-q) \left( U L^{3/2}-\tilde\kappa
L^2\right) \ . \label{eq-Vinen1}
\end{equation}
Here the critical value $q_{\rm c}=A/B$ is generally of order unity.

Equation (\ref{eq-Vinen1}) resembles the Vinen equation
\cite{vinen_eq} for turbulence in superfluid $^4$He-II. However,
the difference is that the coefficient $q_{\rm c}-q$ can now have either
positive or negative sign, depending on the mutual-friction
parameters. As a result, two limits can be distinguished.

In the low-viscosity regime, which is typical for $^4$He-II,
$q_{\rm c}-q >0$. In this regime, the rate of multiplication is faster,
and the number of created vortex loops is large: each new vortex loop
serves as a source for producing more vortices. As a result,
avalanche-like multiplication takes place, which leads to the
formation of a turbulent vortex tangle. As the number of vortex
loops grows, the self-induced velocity increases and finally the
saturated density $L_{\rm sat}$ is reached.

In $^3$He-B the opposite regime is possible, with $q_{\rm c}-q
<0$. In this viscosity-domi\-nated regime the rate of extraction
of vortex loops exceeds the rate of multiplication; there is no
time for vortices to multiply since all seed vortices and newly
created vortices are immediately wiped away into the bulk fluid.
Provided initially created vortex loops have the density $L$
smaller than the saturation value $L_{\rm sat}\sim (U/\tilde
\kappa)^2 $, the number of vortices in the final state is
essentially equal to the number of initial vortices, and the
turbulent state is not formed. The corresponding stable solution
to Eq.\ (\ref{eq-Vinen1}) is $L\rightarrow 0$. Equation
(\ref{eq-Vinen1}) in this limit describes also one more regime of
the vortex tangle evolution which is realized when the initially
created vortex loops have the density $L$, exceeding $L_{\rm
sat}\sim (U/\tilde \kappa)^2 $: The vortex loops will shrink down
to very small sizes due to the viscous mutual-friction force and
collapse, $\ell \rightarrow 0$ while $L\rightarrow \infty$, since
the counterflow $U$ is no longer sufficient to support loops of
such small size. In this regime, the solution of
Eq.~(\ref{eq-Vinen1}) approaches another point of attraction,
$L\rightarrow \infty$.

One can see that the condition of instability for an entangled
vortex array $q\lesssim 1$ is essentially the same as the
condition for the propagation of underdamped Kelvin waves along an
isolated vortex line, established in Sec.~\ref{Sect-Kelvin-wave}.
This indicates that the threshold $q\lesssim 1$ in the
mutual-friction controlled dynamics is not just an accidental
combination of parameter values, but may be of more general and
fundamental importance for superfluid dynamics; however, its role
remains yet to be fully investigated.

To justify our assumption that the normal component does not
participate in superfluid turbulence we compare the viscous
force $ \eta_\mathrm{n} k^2 v_\mathrm{n} $ in the Navier-Stokes equation
(\ref{NormalHydrodynamics}) and the mutual friction force  of
Eq.~(\ref{mutfricforce-micro}) exerted on the normal component
$F_{\rm mf}\sim \alpha \rho_\mathrm{s} \omega_\mathrm{s}
(v_\mathrm{s}-v_\mathrm{n})$. Here $\eta_\mathrm{n} = \rho_\mathrm{n}
\nu$ is the normal dynamic viscosity, and $k$ is the wave vector of the
velocity field. Variations of $v_\mathrm{n}$ are smaller than those of
$v_\mathrm{s}$ when $ \nu k^2 \gg (\rho_\mathrm{s} /\rho_\mathrm{n}) \alpha
\omega_\mathrm{s} $. We estimate $k^2\sim \ell ^{-2}$ and $\omega_\mathrm{s}
\sim \kappa /\ell ^2$ in terms of the vortex-line density. The condition
becomes
\[
\nu /\kappa \gg (\rho_\mathrm{s} /\rho_\mathrm{n})\alpha \ .
\]
On the left-hand side of this inequality we find a new parameter,
the ratio of the kinematic viscosity and the circulation quantum
$\nu /\kappa$, which is a characteristic of the particular
superfluid.  For $^3$He at $T=0.5 \, T_{\rm c}$ we have $\nu
=\eta_\mathrm{n} /\rho_\mathrm{n} \sim 0.4$\,cm$^2$/s and $\kappa
$=0.066 mm$^2$/s, so that $\nu /\kappa \sim 10^3$. The inequality is
then well satisfied since $\rho_\mathrm{s}/\rho_\mathrm{n} \sim 1$ and
$\alpha \sim 0.5$ for $T=0.5 \, T_{\rm c}$. Therefore, the normal component
remains at rest in the container frame. Note that in $^4$He the situation
can be qualitatively different, since the normal component can be
involved in the fluid flow owing to its much smaller viscosity.

The overall evolution of the vortex density can be seen as an
interplay of two processes. The first is the turbulent instability
in the region of multiplication, governed by Eq.\
(\ref{creation}). The second process is the expansion of vortex
loops from the region of multiplication into the bulk due to the
dissipative component of the mutual friction force.
Eq.~(\ref{inflation}) taken with the {\it opposite} sign gives
thus the rate of vortex-loop-density flow {\it into the bulk}. In
this form, it exactly coincides with the Vinen equation
\cite{vinen_eq} as derived by Schwarz \cite{schwarz_88} (see also
\cite{Nemirovski06}), and applies to bulk superfluid turbulence
which is continuously sustained by an external source that acts in
the region of multiplication. A typical example is the grid
turbulence considered in Refs.~\cite{vinen_eq,schwarz_88}; the
region of multiplication is here assumed to be the vicinity of the
grid. Therefore, the rate at which vortices are supplied into the
bulk from the region where they are generated is
\begin{equation}
\dot L_{\rm bulk} = +B\alpha (U-v_0) L ^{3/2}=B\alpha (UL
^{3/2}-\tilde \kappa L^2) \ .
\end{equation}
If the supply continues long enough, the vortex density in the
bulk also saturates at $L_{\rm sat}=(U/\tilde\kappa )^2$. The solution
\cite{schwarz_91} describing the relaxation towards saturation for
constant $U$ has the characteristic rate $ \tau_\mathrm{b}^{-1}= \alpha
U^2/2\kappa $. This relaxation time increases with decreasing
temperature as the parameter $\alpha$ decreases.


\subsection{Energy cascades in developed superfluid
turbulence}
   \label{DevelopedTurbulence}


In the previous section \ref{TransitionTubulence} we discussed the
onset of superfluid turbulence and the resulting state which is
characterized by a single scale $\ell =\tilde\kappa/U$ at which
saturation occurs. We call such a single-parameter state the Vinen
regime of superfluid turbulence. This state is very different from
the turbulence in viscous liquids, where the energy spectrum obeys
the celebrated Kolmogorov-Obukhov 5/3-law
\begin{equation}
E_k=C \varepsilon^{2/3}_k k^{-5/3}\ .
\label{V5}
\end{equation}
Here $E_k$ is the one-dimensional density  of turbulent kinetic
energy in $k$-space,  defined such that the total energy density
$E$ (in physical space)  is given by
\begin{equation}
\label{V4}
E\equiv \frac 12\, \langle |\mathbf{v}|^2  \rangle =\int dk\, E_k\ ,
\end{equation}
and $\varepsilon_k$ is the energy flux in $k$-space,  which is
constant in the inertial range of $k$, where viscous dissipation
can be neglected, $\varepsilon_k=\varepsilon$.

\begin{figure}[!!!t]
     \centerline{\includegraphics[width=0.6\linewidth]{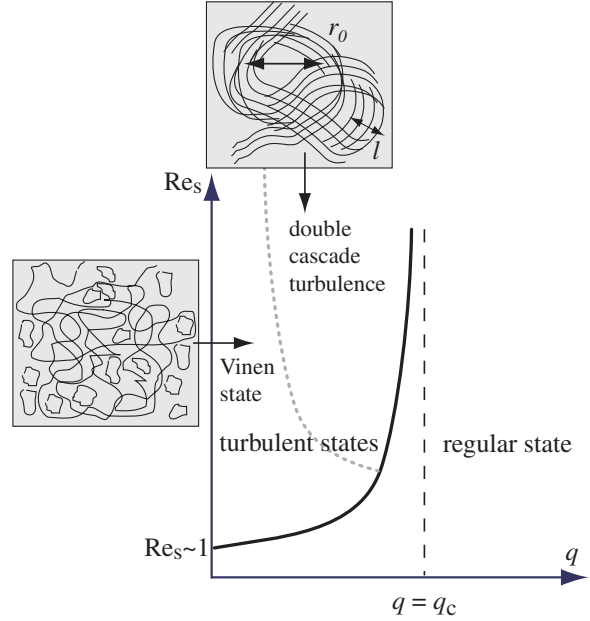}}
\caption{Possible phase diagram of dynamical  vortex states in the
$(q,{\rm Re}_\mathrm{s})$ plane. At large flow velocity ${\rm Re}_\mathrm{s}\gg 1$
well above the Feynman critical velocity, the boundary between
turbulent and `regular' vortex flow approaches a vertical line
$q=q_{\rm c}\sim 1$. The dashed curve marks the crossover between two
regimes of superfluid turbulence occurring at small $q$:  (i)
Developed turbulence of classical type, which is characterized by
two Richardson-type cascades owing to  mutual friction
dissipation. The Kolmogorov-Obukhov law $E_k\propto k^{-5/3}$
coexists with the $E_k\propto k^{-3}$ law. (ii) Quantum turbulence
of the Vinen type at even smaller $q$, which is characterized by a
single length scale $\ell=\kappa/U$. }
     \label{PhaseDiagram}
\end{figure}

Here, we study what is the outcome of the turbulent instability in
a superfluid in different regimes of Reynolds parameters $\Re
=1/q$ and Re$_s$. We show that in the intermediate range of
Reynolds numbers $1/q$ (Fig.~\ref{PhaseDiagram}) the turbulent
instability leads to a state of developed turbulence which is
closer to its classical analogue: It exhibits a Richardson--like
cascade \cite{volovik_spectrum} and is different from the Vinen
state of turbulence.

As a starting point we  utilize the coarse-grained hydrodynamic
equation for the dynamics of the superfluid with distributed
vortices, Eq.~(\ref{SuperfluidHydrodynamics}), or, after taking
the curl,
\begin{equation}
\frac{\partial \bm{\omega}_\mathrm{s}}{\partial t}=(1-\alpha ^\prime
)\bm{\nabla}\times (\mathbf{v}_\mathrm{s}\times \bm{\omega}_\mathrm{s})
+\alpha \bm{\nabla}\times [\hat{\bm{\omega}}_\mathrm{s} \times
(\bm{\omega}_\mathrm{s}\times \mathbf{v}_\mathrm{s})] \ . \label{vorticity-coarse}
\end{equation}
As we have seen, turbulence develops only if friction is
relatively small compared to the inertial term, $1/q \gtrsim 1$.
Here, we discuss the regime of well developed turbulence which
occurs at $\Re =1/q\gg 1$ when the inertial term is strongly
dominating. In this limit, $q\approx\alpha$, while both
$\alpha^\prime \ll 1$ and $\alpha \ll 1$. We show that well
developed turbulence can have an analog of the
Richardson-Kolmogorov cascade, which becomes modified by the
nonlinear mutual-friction dissipation.

The main difference of this cascade in superfluids from that in
viscous liquids is that the dissipation of energy from the mutual
friction in Eq.~(\ref{Dissipation})  occurs at all scales, and
thus the energy flux $\varepsilon_k$ must be essentially
$k$-dependent. From Eq.~(\ref{Dissipation}) it follows that the
energy losses from dissipation are
\begin{equation}
\frac{\partial E_k}{\partial t}= - \Gamma\, E_k  ~~~,~~\Gamma\sim
q\omega_0~~,
\label{DissipationEnergy2}
\end{equation}
where $\omega_0=\left< |{\bm{\omega}}_s|\right>$ is the average
vorticity. In steady-state turbulence, these energy losses must be
compensated by the energy exchange between different $k$ in the
cascade, $\partial \varepsilon _k/\partial k= - \Gamma\, E_k$.
Using the relation between the energy $E_k$ and the energy flux in
momentum space $\varepsilon_k$  in Eq.~(\ref{V5}), which follows
from general dimensional reasoning in the spirit of Kolmogorov,
one obtains the following balance equation for the flux
$\varepsilon_k$:
\begin{equation}
{\partial \varepsilon _k \over \partial k}  =
- \, \Gamma \,\varepsilon_k^{2/3} k^{-5/3}\,.
\label{V10}
\end{equation}
Such an equation for the energy budget has been used in
Refs.~\cite{Kov,Leith1, Leith2}; the more complicated  version
with a second derivative
\cite{ConnaughtonNazarenko1,LavalDubrulleNazarenko}  was  used for
superfluid turbulence by Vinen \cite{vinen_grid}. In the absence
of dissipation, {\it i.e.} when $\Gamma=0$, Eq.\ (\ref{V10})
immediately produces the  solution  with constant energy flux,
$\varepsilon_k=\varepsilon$.  Then  Eq.~(\ref{V5}) turns into the
Kolmogorov-Obukhov $5/3$-law for $E_k\sim \varepsilon^{2/3}
k^{-5/3}$.

Eq.~(\ref{V10}) must be supplemented by a boundary condition: a
fixed energy influx into the turbulent system from large length
scales of order of container size, $\epsilon_{k=1/R}=U^3/R$, where
$U$ is the counterflow velocity at this scale. A general solution
of Eq.\ (\ref{V10}) gives the following energy spectrum:
\begin{equation}
E_k = {U^{2} \over k (kR)^{2/3}} \left[1 + {\gamma \over
(kR)^{2/3}} - \gamma \right]^2~,
\label{E}
\end{equation}
where the dimensionless parameter
\begin{equation}
\gamma=\Gamma R/U=q\omega_0 R/U~,
\label{gammasmall}
\end{equation}
and the mean vorticity is expressed through  the energy spectrum,
\begin{equation}
   \omega^2_0=\left< |\omega|\right>^2 \simeq
\int\limits_{1/R}^{k_*}dk \, k^{2}\,  E_k\,.
\label{MeanVorticity}
\end{equation}
The ultraviolet cut-off $k_*$ in Eq.~(\ref{MeanVorticity}) is
determined by the microscopic scale at which the circulation in
the $k_*$-eddy reaches the circulation quantum $\kappa$: $v_{k_*}/
k_*=\kappa$, and thus the coarse-grained dynamics is no longer
applicable. Since $k^2v^2_k=E_k$, the cut-off is determined by the
spectrum and, as a result, we obtain a closed system of equations
which can be analyzed for different regions of the Reynolds
parameters, $\Re=1/q$ and Re$_\mathrm{s}=RU/\kappa$.


Let us consider the turbulent state that corresponds to $q\ll 1$
and $q^2{\rm Re}_\mathrm{s}\gg 1$. In this case the parameter $\gamma$ is
close to unity, $1-\gamma\sim (q^2{\rm Re}_\mathrm{s})^{-1/3}\ll 1$, and
the solution (\ref{E}) has the form
\begin{equation}
E_k \simeq  {U^{2} \over R^{2} k^{5/3}} \left[{1 \over k^{2/3}} +
{1 \over k\sb{\rm cr}^{2/3}} \right]^2 \, , \label{E3}
\end{equation}
where $k\sb{\rm cr}$ marks the crossover between the Kolmogorov law at
$ k>k\sb{\rm cr}$ to the steeper law $E_k\propto k^{-3}$ at
$k<k\sb{\rm cr}$. Three important scales -- the scale $k\sim 1/R$ at
which pumping occurs, the crossover scale $k\sb{\rm cr}$, and the
microscopic (quantum) cut-off scale $k_*$ -- are well separated in
this regime:
\begin{equation}
k_*\sim k\sb{\rm cr} q^{-3/2} \gg  k\sb{\rm cr} \sim  \frac{1}{R}
   ({\rm Re}_\mathrm{s} q^2)^{-1/2} \gg  \frac{1}{R} \ .
\label{separation}
\end{equation}

At $q^2{\rm Re}_\mathrm{s}\sim 1$ one has $k\sb{\rm cr}=1/R$, {\it i.e.}
the region of the $k^{-3}$ spectrum shrinks. Here two scenarios are
possible. In the first mutual friction is unessential and thus is
unable to compensate the Kolmogorov cascade. When the intervortex
distance scale is reached the Kolmogorov energy cascade is then
transformed to the Kelvin-wave cascade \cite{vinen_turb_review} of
isolated vortices. In the second scenario the turbulent state is
completely reconstructed and the Vinen state discussed in
Sec.~\ref{TransitionTubulence} and
Refs.~\cite{vinen_eq,schwarz_88} emerges. This state contains a
single scale $\ell=\kappa/U$ and thus no cascade. A possible phase
diagram of the turbulent states is shown in
Fig.~\ref{PhaseDiagram}. The connection of this phase diagram with
the flow states observed in various experiments on superfluid
$^4$He-II and $^3$He-B is discussed in Ref.~\cite{Skrbek}.

These phenomena found in $^3$He-B added a new twist to the general
theory of turbulence in superfluids which was developed earlier by
Vinen \cite{vinen_classical,vinen_turb_review} and which was based
on numerous experiments in superfluid $^4$He-II where the
first signs of turbulence were observed already in the 1950's
\cite{vinen_eq}. The new theory based on $^3$He-B experiments
incorporates two Reynolds parameters (the velocity-dependent
$UR/\kappa$, and the velocity-independent $q$). It suggests different
types of developed superfluid turbulence in different regions of
Reynolds parameters, and allows to derive deviations from the
classical Kolmogorov-Obukhov scaling law $E(k)\sim k^{-5/3}$.

The extension of these ideas to the more general case when both
the normal and the superfluid components may become turbulent has
been performed in Ref.~\cite{LvovNazarenkoSkrbek} on the basis of
two-fluid hydrodynamics. The results of this analysis are
applicable to superfluid $^4$He-II, where the viscosity of the
normal component is many orders of magnitude smaller than in
$^3$He-B, or in $^3$He-$^4$He mixtures where, owing to the
presence of $^3$He quasiparticles, mutual friction at low
temperature is also expected to be significantly higher than in
pure superfluid $^4$He-II.

\subsection{Injection of seed vortex loops in applied counterflow}
\label {InjectionMethods}

\begin{figure}[t]
\begin{center}
\includegraphics[width=1\linewidth]{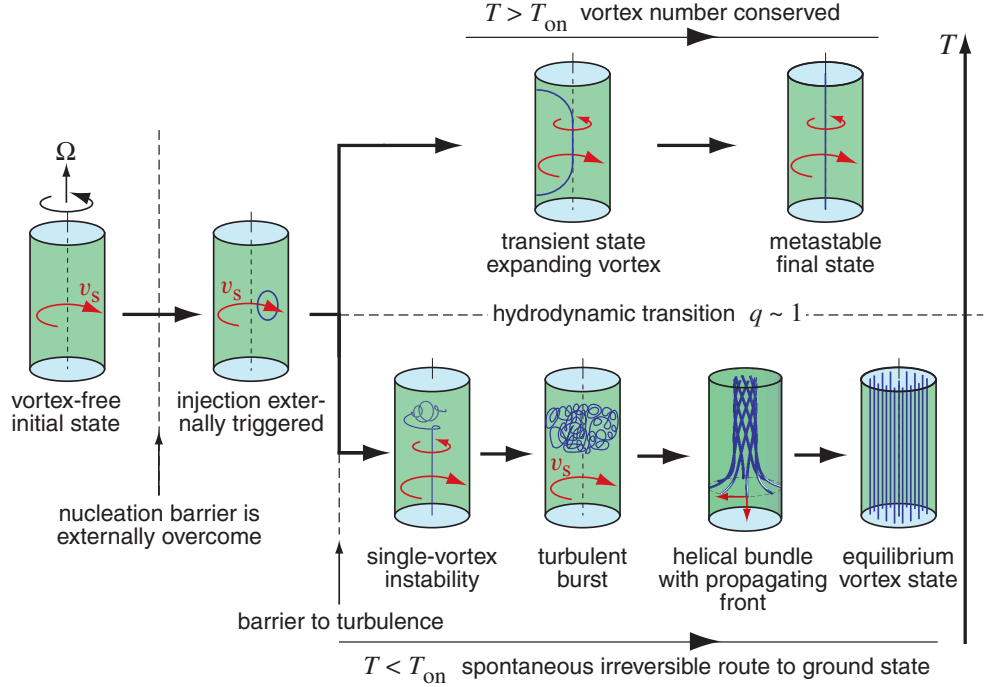}\\
\caption{Principle of vortex injection measurements. Different
techniques have been explored to inject vortex seed loops in
rotating vortex-free flow. One can then study the evolution of the
seeds as a function of temperature. Two regimes are found: (1) at
high temperatures the vortex number stays constant while (2) at
low temperatures a turbulent burst increases the vortex number
close to that in the equilibrium vortex state. The exact onset
temperature $T_{\rm on}$ of turbulence has been found to be
sensitive to the details of the injection process. This indicates
that the energy barriers of the different processes, which
destabilize the seed loops and lead to their turbulent
multiplication, depend in addition to temperature also on the
number, configuration, and density of the
loops.}\label{VorInjectPrinciple}
\end{center}
\end{figure}

Why do seed vortices injected in applied counterflow undergo a
transition to turbulence? A superficial answer can be given by
inspecting the diagram  in Fig.~\ref{VorInjectPrinciple}. Here
vortex-free flow is generated by subjecting the superfluid sample
to uniform rotation. This is a metastable state of high energy
where the superfluid fraction does not participate in the
rotation. As long as the resulting maximum counterflow velocity
remains below the limit of spontaneous vortex formation of the
particular sample setup, the vortex-free state persists and no
vortices are formed. By injecting vortex seed loops in the applied
flow the energy barrier for the nucleation of vortices can be
externally bypassed. We can then watch what happens to the
injected seeds. At high temperatures damping is large, the seed
vortices quickly evolve to rectilinear lines, the number of
vortices remains conserved, and the final state remains highly
metastable.

However, below a sudden onset temperature $T_{\rm on}$ the
situation changes radically: dissipation in vortex motion has
dropped to a level where Kelvin wave excitations are not
over-damped and helical waves with large amplitude can be formed
in such sections of the seed vortex where the flow has a component
oriented parallel to the vortex core. The formation of new
vortices in dynamic processes becomes then possible and suddenly
the superfluid can reduce its overall energy state and reach
equilibrium. The bottleneck in this chain of instabilities is the
single-vortex instability in applied counterflow. However,
eventually at sufficiently low temperature even this will
inevitably happen. If the density of the injected seed loops is
high enough from the beginning, as appears to be the case in KH
injection, then inter-vortex interactions allow turbulence to
switch on immediately and at a higher temperature than the
single-vortex instability. The evolving turbulence becomes a
collective phenomenon, which can be described either in terms of
the Vinen-type theory in Sec.~\ref{TransitionTubulence} or the
developing superfluid turbulence in
Sec.~\ref{DevelopedTurbulence}. Unfortunately, in rotating flow of
\heb above $0.4\,T_{\rm c}$ turbulence is limited to a very short
burst so that a measurement of its evolution with time becomes a
challenging effort. The propagation of the vorticity in the
rotating cylinder after the turbulent burst is described in
Sec.~\ref{HelicalBundle}.

Measurements employing injection of seed loops in applied flow are
a new feature in turbulence studies. They have become possible in
$^3$He-B where vortex-free flow can be achieved at sufficiently
high flow velocity.  With the aid of seed loop injection one can
study different instabilities in the serial chain of dynamic
processes which lead to vortex multiplication and ultimately to
turbulence. One may expect that the closer the injected seed loop
configuration is to a situation where the loops can interact and
create a turbulent burst the higher will be the transition
temperature to turbulence.  Injections with a small number of
loops which are far apart require a lower temperature to develop
to turbulence, {\it i.e.} the transition becomes more and more
irregular being dependent on such details as the number of
injected loops, their shape, size and density, their mutual
interaction and interaction with the container walls, where also
surface roughness and pinning may matter. These dependences can be
investigated using different injection techniques \cite{LT24}.

{\it Magnetically driven B$\rightarrow$A transition:}  Large
numbers of seed vortices can be created by sweeping up the barrier
field $H_{\rm b}$ in the setup of Fig.~\ref{setup}. The field
sweep is conducted in rotation at constant temperature, when the
entire sample is in B phase and vortex free. When the field
reaches some slightly hysteretic value above $H_{\rm AB}(T,P)$
over a short section in the middle of the long sample, then the A
phase, its vortices, and two AB interfaces all start to form
essentially simultaneously \cite{AB-InterfaceShape}. A detailed
explanation how this happens (in a situation which is unstable
from the beginning) has not been worked out, but if the rotation
velocity $\Omega$ significantly exceeds the critical KH velocity
$\Omega_{\rm c}(T,P)$ a large number of vortices is suddenly
injected in the B-phase sections.

\begin{figure}[t]
\begin{center}
\includegraphics[width=0.75\linewidth]{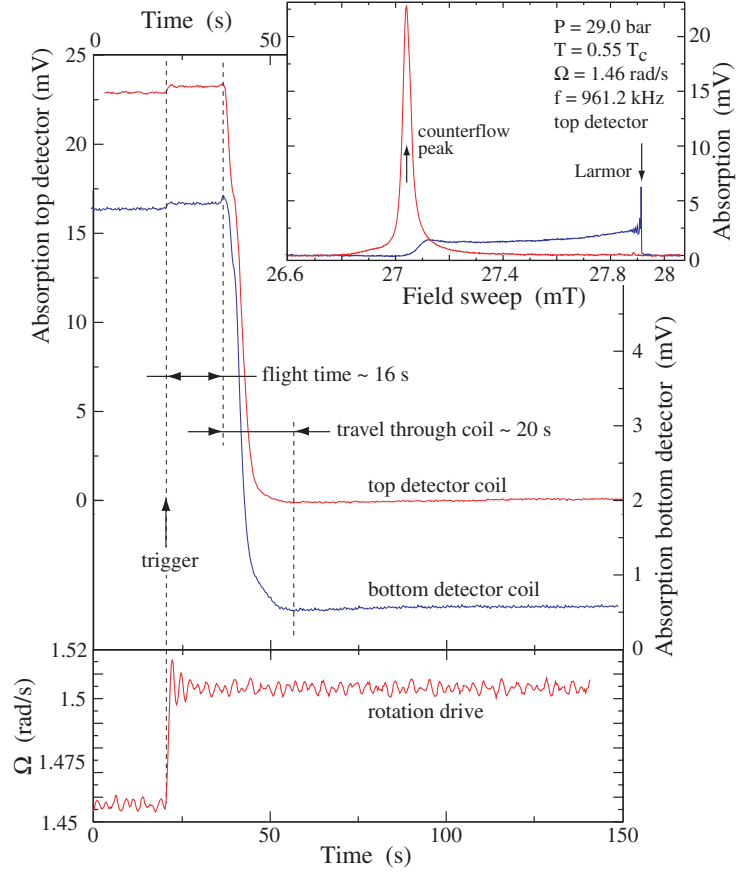}\\
\caption{NMR measurement of KH instability in the turbulent
temperature regime. {\it (Top right panel)} NMR absorption line
shapes in $^3$He-B. The initial vortex-free state displays a large
maximum which is shifted from the Larmor position by a temperature
dependent amount. The height of this counterflow peak is roughly
proportional to the counterflow velocity. The final equilibrium
vortex state has a flat distribution of absorption. At fixed
temperature the integrated absorption under the two line shapes is
equal. {\it (Main panel)} Counterflow peak heights measured as a
function of time in the BAB configuration, when the KH instability
is triggered in the setup of Fig.~\protect\ref{setup} (with
$I_{\rm b} = 8.0\,$A). Seed loop injection at the AB interface is
followed by a turbulent burst, rapid polarization, and the forming
of a propagating vortex state consisting of a front followed by a
twisted bundle. The counterflow peak height starts decreasing when
the first fastest vortices in the front reach the closer end of a
detector coil. The peak height vanishes when the last vortices in
the front pass through the far end of a detector coil. In this
example at $0.55\,T_{\rm c}$  no stable front exists: the fastest
vortices move at $v_z \sim 2\,$mm/s while the slowest vortices, on
an average, are only half as fast by the time they have passed
through the coil. Since the measuring setup in
Fig.~\protect\ref{setup} is symmetric, the top and bottom detector
coils display identical responses. {\it (Bottom panel)} Externally
controlled trigger for KH injection. The rotation drive is
suddenly increased by $\Delta \Omega \approx 0.05\,$rad/s so that
$\Omega$ jumps above the critical KH velocity of $\Omega_{\rm c} =
1.48\,$rad/s. This is signalled as an incremental increase in the
counterflow peak heights of both signals in the main panel.
}\label{KH-Measurement}
\end{center}
\end{figure}

{\it Kelvin-Helmholtz instability:} The Kelvin-Helmholtz
instability of the AB inter\-face, discussed in
Secs.~\ref{kh_injection} and \ref{khturbulence}, is so far the
most reproducible of the available vortex-injection mechanisms.
The injection is carried out at constant external con\-di\-tions
while the rotation velocity $\Omega$ is suddenly incrementally
increased by $\Delta \Omega \approx 0.05\,$rad/s across the
critical KH value $\Omega_{\rm c}(T,P)$. At temperatures above
$T_{\rm on}$ such injection generates a limited number of vortices
in the B phase and consequently the staircase pattern in
Fig.~\ref{stairs} can be displayed, by triggering multiple
instability events one after the other. Below $T_{\rm on}$ the
staircase pattern is not obtained, since already the very first
injection sends the sample in the equilibrium vortex state and
removes essentially all applied counterflow. The measured
signature from the injection in the turbulent temperature regime
is shown in Fig.~\ref{KH-Measurement}. The radical difference from
Fig.~\ref{stairs} is evident: The dynamic formation of new
vortices in a turbulent burst increases the vortex number
immediately to $N_{\rm eq}$.

The evolution following a triggered injection event can be
monitored by recording the NMR absorption height as a function of
time either at the location of the counterflow peak, as is done in
Fig.~\ref{KH-Measurement} to measure the reduction in the
macroscopic counterflow, or in the region of the Larmor edge, as
is done in Fig.~\ref{stairs} to exhibit a signal which is
generated more directly by vortices. In both regions the NMR line
shape arises from the interplay of vortex-free counterflow and of
vortices via their orientational influence on the order parameter
texture, as explained in the Appendix.

\begin{figure}[t]
\begin{center}
\includegraphics[width=0.8\linewidth]{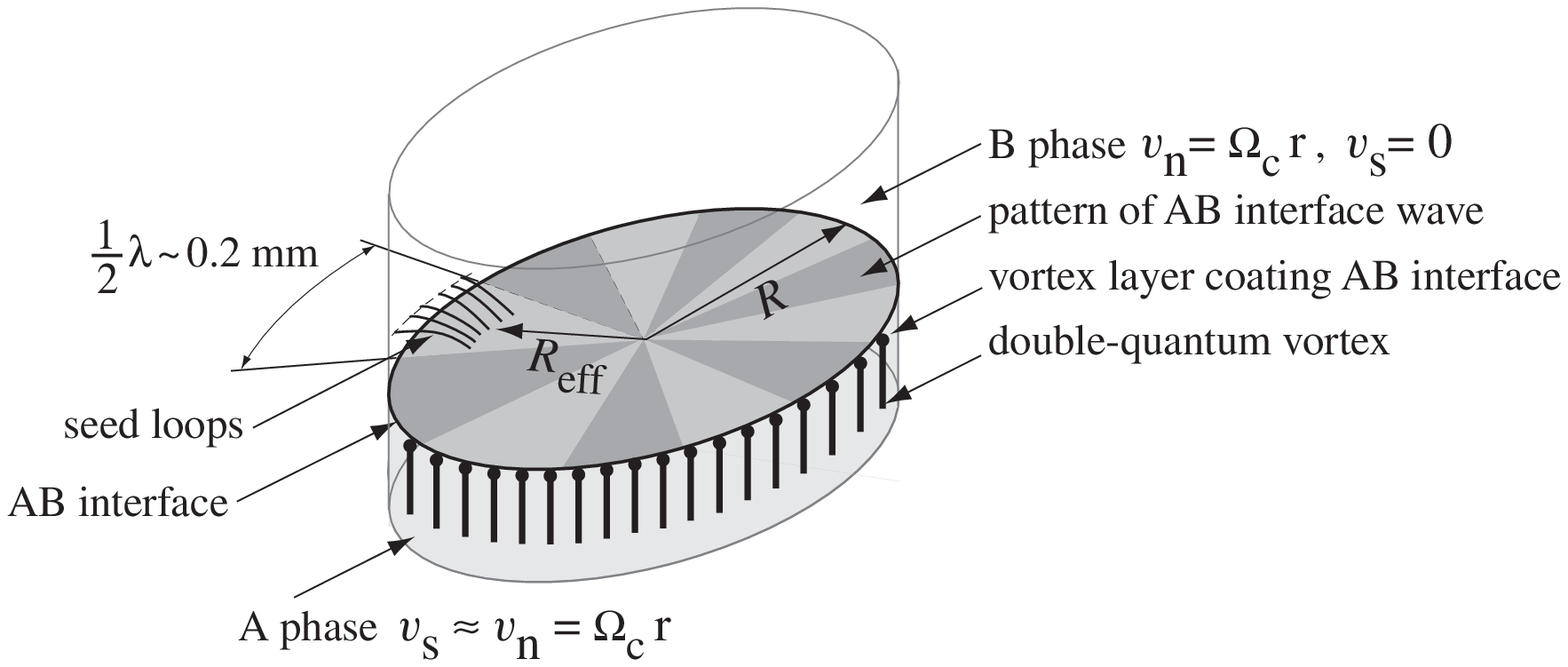}
\caption{Sketch of the initial vortex loop configuration in KH
injection. A section of the cylindrical two-phase sample is shown.
Roughly the equilibrium number of double-quantum vortices exists
on the A phase side. At the AB phase boundary the A-phase vortices
curve into a surface layer which coats the interface on the
A-phase side. In this layer the vortices run radially out to the
cylinder wall \protect\cite{risto_prl}. The B phase is vortex
free, except for the seed loops which have escaped across the AB
interface in a KH instability event. Initially they are contained
in a tight bundle which is oriented radially in the cylinder,
starting from $R_{\rm eff}$ at the AB interface and ending at the
cylindrical wall. The bundle has roughly a diameter $\sim 0.1$ --
0.2\,mm (corresponding to $\lambda/2$ in Fig.~\protect\ref{kh fig}
{\it center}), a length $\sim 0.4\,$mm (corresponding to $R -
R_{\rm eff}$ in Fig.~\protect\ref{stairs}), and contains on an
average 10 vortex loops with a spacing $a_{\rm s} \sim 15$ --
$25\,\mu$m (Fig.~\protect\ref{kh fig} {\it right}). In the
turbulent temperature regime the vortices within the bundle are
rapidly destabilized by Kelvin wave excitations, owing to the
applied flow $\mathbf{v}=\mathbf{v}_{\rm n}-\mathbf{v}_{\rm
s}=\mathbf{\Omega} \times \mathbf{r}$ and the interactions between
the loops.} \label{KH-InjectVorConfig}
\end{center}
\end{figure}

The principal new information in Fig.~\ref{KH-Measurement}  is
related to the motion of the vortices after the turbulent burst.
From the time interval between the trigger and the first response
from the propagating vortex front one can measure the longitudinal
velocity of the fastest vortices \protect\cite{flight_time_jltp}.
It is found to be approximately $v_{\rm z} \approx \alpha \Omega
R$. The subsequent time interval, during which the counterflow
peak decays to zero, is at low temperatures a measure of the
thickness of the vortex front: the convolution of the moving
vortex front with certain width through a detector coil of given
length. In Fig.~\ref{KH-Measurement} at relatively high
temperatures no stable front is formed. Here the time interval
between the trigger and the moment when the counterflow peak
vanishes measures the flight time of the slowest vortices from the
AB interface to the far end of the detector coil. These properties
of the front and the helical vortex bundle behind it will be
discussed in Sec.~\ref{HelicalBundle}. We note that the response
in Fig.~\ref{KH-Measurement} is deterministic and fully
reproducible from one measurement to the next, in spite of the
fact that it involves a short-lived stochastic turbulent burst.
The reproducibility is attributed to the well-behaved value of the
KH critical rotation velocity $\Omega_{\rm c}(T,P)$ plus to the
fact that the propagation of the vortices along the rotating
column controls the slow time scale in Fig.~\ref{KH-Measurement}
which masks all other faster processes.

KH injection produces the transition to turbulence at the highest
onset tem\-pera\-ture. The reason is interpreted to be the initial
configuration in which the seed vortices end up on the B-phase
side of the interface: As sketched in
Fig.~\ref{KH-InjectVorConfig}, they form a tightly packed bundle
of many roughly parallel vortex loops, as estimated in
Fig.~\protect\ref{kh fig}. Apparently, in the applied B-phase flow
they immediately start interacting turbulently, when the
amplitudes of Kelvin wave excitations are not over-damped. This
leads instantaneously to a turbulent burst and gives the highest
onset temperature $T_{\rm on}$ which, at least at higher flow
velocities, turns out to be independent of the flow velocity. In
contrast such transitions to turbulence, which have to be preceded
by the single-vortex instability, depend on the flow velocity and
occur at lower temperature.

{\it Neutron absorption:} The nuclear capture reaction of a $^3$He
nucleus with a thermal neutron provides an externally controllable
mechanism for vortex line injection in vortex-free flow of
$^3$He-B \cite{neutron_jltp,neutronreview}. A thermal neutron
incident on liquid \he has a short mean absorption length of only
$\sim 0.1\,$mm before it suffers the capture reaction n+$^3$He
$\rightarrow$ p+$^3$H+764\,keV. The reaction energy is released in
the form of kinetic energy of the two reaction products. It is
dissipated by them in ionizations and recoil, such that roughly a
volume of the fluid within a radius $\lesssim 50\, \mu$m from the
reaction site is locally heated above $T_\mathrm{c}$. Within
microseconds this ``neutron bubble'' cools back to the ambient
bath temperature, but a random vortex tangle is left behind
\cite{kibble_zurek_nature}. In the absence of an externally
applied counterflow the loops in the tangle shrink and disappear,
but in vortex-free rotation the largest loops with proper
orientation and polarization are extracted from the tangle and
expand into the bulk fluid, where they initially appear as
separated rings.

The extraction of rings from the tangle is governed by the
magnitude of the applied vortex-free flow velocity $v$ according
to the well-known formula for the equilibrium state of a vortex
ring: A ring of radius $r$ is in stable state at the flow velocity
\begin{equation}
v(r) = \frac{\kappa}{4\pi r}\; \ln{\frac{r}{\xi}}\;.
\label{VortexRing}
\end{equation}
A ring with larger radius than $r(v)$ will expand in the flow
while a smaller will contract. Therefore a minimum threshold
velocity $v_{\rm cn}$ exists at which the first vortex ring can be
extracted from the tangle. This velocity corresponds to the
maximum possible ring size, which has the radius $R_{\rm b}$ of
the neutron bubble: $r(v_{\rm cn}) \sim R_{\rm b}$. At larger flow
velocities smaller rings can be pulled from the tangle.
Simultaneously the number of such smaller rings can be larger than
one, since several smaller rings can fit within the neutron
bubble. Again there exists a minimum threshold velocity $v_{\rm
cni}$ which is required in order to extract $i$ rings of equal
size from the bubble. Their number $i$ is obtained from a volume
argument, {\it i.e.} according to how many spheres of radius
$r(v)$ can fit inside the neutron bubble without overlap: $i \sim
(R_{\rm b}/r(v))^3$. Using Eq.~(\ref{VortexRing}) we then obtain
$v_{\rm cni} \sim i^{1/3}\, v_{\rm cn}$. A better experimental and
theoretical justification of these features can be found in
Ref.~\cite{ruutu_neutron_prl,neutronreview}.

A neutron absorption event can be used to inject vortex rings in
the rotating flow. The number and size of the rings depends on the
local velocity of flow. Since the mean absorption length is only
$\sim 100\,\mu$m and the flow velocity increases with radius as
$v= \Omega r$, all rings will initially be located in the vicinity
of the outer wall at a well-defined flow velocity $v =\Omega R$.
In this situation neutron absorption becomes an externally
controllable injection method. However, because of the random
tangle from which the injected rings originate, vortex formation
from a neutron absorption event is by nature a stochastic process.
This means that at a flow velocity $\Omega R \sim v_{\rm cni}$ the
number of rings obtained from a given neutron absorption event can
be anything from zero up to the maximum limit $i$. At high
temperatures $T > T_{\rm on}$, each extracted ring evolves
independently to a rectilinear vortex line. These can be
individually counted in similar fashion as in Fig.~\ref{stairs}.


\begin{figure}[t]
\centerline{\includegraphics[width=0.8\linewidth]{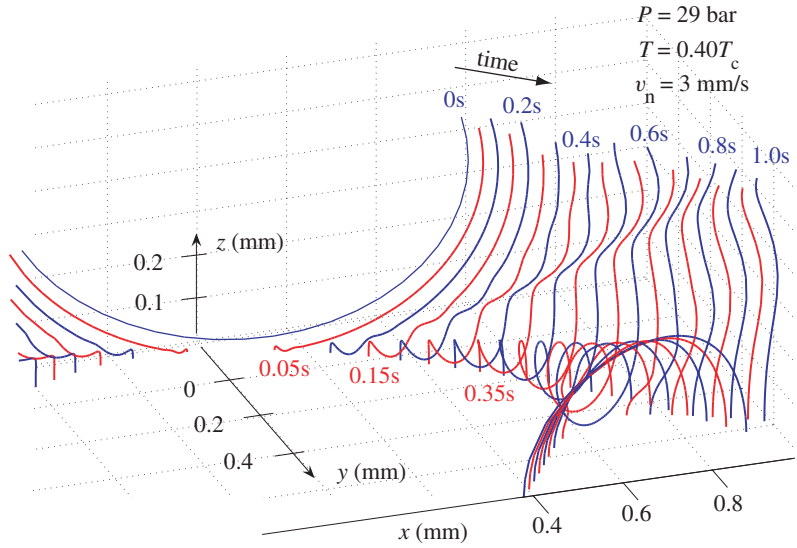}}
\caption{Principle of the boundary induced vortex instability. A
vortex ring of radius 0.5\,mm is placed in flow with constant
normal fluid velocity and its evolution in time is followed.
Initially the center of the ring is 0.5\,mm above a plane solid
boundary, which is the $xy$ plane at $z=0$, and the plane of the
ring is slightly tilted from the $y=0$ plane. The ring first
expands and reorients itself in the applied flow and then collides
with the boundary. The collision with the boundary creates two
reconnection kinks on the vortex. The kink on the right has the
correct orientation and helicity and starts to expand in the
applied flow. The kink grows to a loop which in turn reconnects at
the boundary creating thereby a new independent vortex. The
calculation uses the hydrodynamic parameters of $^3$He-B at
$0.40\,T_{\rm c}$ ($q$ = 0.21), with the normal fluid velocity
$v_\mathrm{n}$ = 3\,mm/s oriented along the positive y-direction.}
\label{simurecon}
\end{figure}


In the turbulent temperature regime either no vortices are
obtained or, if even one ring is extracted from the tangle, the
event leads to dynamic vortex formation and a turbulent burst. By
choosing the value of the applied flow velocity $\Omega R$ in
relation to the threshold velocity $v_{\rm cni}$ (which is
measured at $T > T_{\rm on}$), one can study the onset $T_{\rm
on}$ as a function of the number of rings $i$ which are extracted
from a single neutron absorption event. Of particular interest is
to check whether with $\Omega R \gtrsim v_{\rm cn}$ a single
vortex ring can lead to a turbulent burst.  In practice, because
of the small size of the neutron bubble, $R_{\rm b} \lesssim
50\,\mu$m, the applied flow velocity for vortex injection via a
neutron absorption event has to be relatively high: $\Omega_{\rm
cn} = v_{\rm cn}/R \gtrsim 1.4\,$rad/s. Such measurements confirm
that at sufficiently low temperatures even a single extracted ring
leads to a turbulent burst.

{\it Other injection techniques:} There exist also other types of
``injection'', in which a small number of curved seed vortices can
be introduced in externally applied rotating counterflow. Vortex
multiplication can, for example, be initiated by a dynamic remnant
vortex which is left behind from a preceding experimental run and
which has not yet had sufficient time to annihilate at the
container wall \cite{DynamicRemnant}. A further source of curved
seed vortex loops are those vortices of the rotating equilibrium
vortex state which connect to the cylindrical side wall owing to
the misalignment between the rotation axis and the symmetry axis
of the cylinder \cite{slow,EquilVorState}. The common feature of
all such experiments with different injection mechanisms and
varying initial conditions is that vortex multiplication and the
onset of turbulence become possible at the temperature
corresponding to $q \sim 1$, although the measured onset $T_{\rm
on}$ may be lower and depends on the initial conditions, {\it
i.e.} on the applied counterflow velocity and the configuration
and density of the initial seed vortices.

In general, it is found both from experiments and numerical
simulations that, at sufficiently low $q$, injection of even a
single vortex loop into externally applied flow always leads to
turbulence. Based on simulations in the single-vortex regime,
where inter-vortex interactions can be neglected, one can conclude
that the most important mechanism for generating a second
independent vortex from the seed loop involves a reconnection of
the seed loop at the solid boundary. The mechanism is illustrated
in Fig.~\ref{simurecon}, which shows a vortex ring drifting in
applied flow, colliding with a plane boundary, and the ensuing
formation of an expanding new loop adjacent to one of the
reconnection kinks. Reconnection with the boundary induces a
spectrum of helical Kelvin wave excitations on the vortex
\cite{svistunov,KelvinWaveCascade}. At finite temperature in the
absence of external flow these are damped, but in applied
counterflow a Kelvin wave with correct helicity and proper
orientation can start to expand \cite{slow}. This expanding loop
can then reconnect again with the boundary, leading to the
generation of further loops.

\section{Propagating vortex front and twisted vortex state}
\label{HelicalBundle}

\subsection{Introduction}

In the rotating container, the multiplication of the injected
vortices and the ensuing turbulence are transient phenomena which
ultimately lead to the establishment of the stable equilibrium
vortex state, with the equilibrium number of rectilinear vortex
lines. In this state the superfluid component imitates solid-body
rotation. Thus the superfluid and normal components are finally
locked to corotation. In the long cylinder, the transient phase
acquires new features, which consist of the spirally winding
propagation of the vorticity into the metastable Landau state and
the subsequent relaxation of twisted vortices to rectilinear
lines.

Consider a long rotating superfluid column initially in a
metastable state of vortex-free flow with high kinetic energy: the
superfluid component at rest, and the normal component in
solid-body rotation. If nearly the equilibrium number of vortices
is suddenly created locally in some part of the sample, how does
the vorticity spread over the rest of the sample to reach stable
equilibrium? In other words, how is the initially stationary
superfluid component dragged into corotation? To shed light on
this new hydrodynamic problem, we discuss here an experiment where
vortex propagation in superfluid $^3$He-B is studied by monitoring
the NMR signal as a function of time, at a location which is far
away from the injection site.

As discussed in the preceding chapter, the dynamics of quantized
vortex lines in $^3$He-B is strongly influenced by the strength of
the dissipative mutual-friction force, imposed by the normal
component. This force drives the longitudinal motion of vortices
along the rotating superfluid column, and is responsible for
dissipating the excess kinetic energy of the initial Landau state.
The reduction in this frictional damping with decreasing
temperature is responsible for the transition to turbulence and is
also expected to influence further the propagation on approaching
the lowest temperatures below $0.3\, T_{\rm c}$.

Initially it came as a surprise that both numerical simulations
and measurements did not bear any evidence of an expanding
turbulent vortex tangle in the rotating sample. In fact, even
under the conditions of fairly low damping (at temperatures of
$\sim 0.3 \, T_{\rm c}$), the injected vorticity does not spread
in the form of an incoherent tangle. Instead, vortices propagate
along the sample in a time-invariant configuration with a narrow
\emph{vortex front}. Behind the front the vortices are left in a
\emph{helically twisted} configuration \cite{twist}. This novel
dynamic vortex state arises as a consequence of the spiral motion
of the vortex front which, in addition to the axial propagation,
also has an azimuthal velocity component with respect to the frame
of the container. The latter is derived from the reactive
component of mutual friction. Experimentally the twisted vortex
state can be identified through its associated superfluid velocity
field which has a component along the rotation axis and leaves a
clear fingerprint on the NMR signal. Both of these new features,
the vortex front and the twisted vortex state, are also reproduced
in numerical simulations.

\subsection{NMR response from propagating vortices}

\begin{figure}[p]
\begin{center}
\includegraphics[width=1\linewidth]{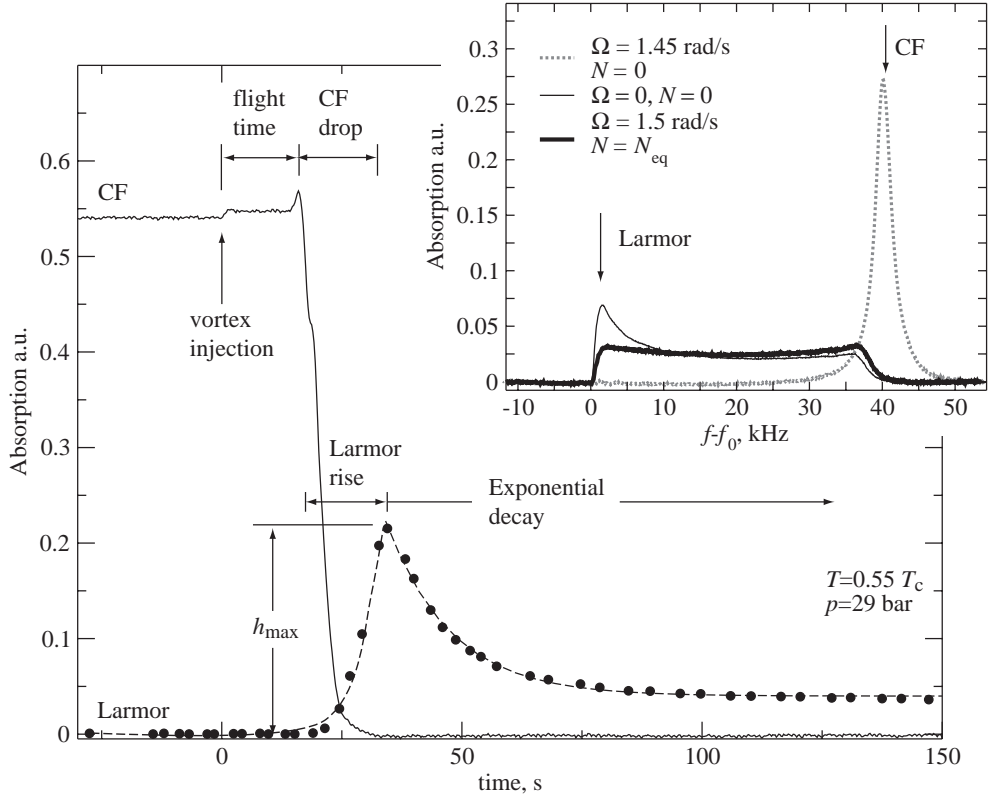}\\
\caption{Time sequence of NMR absorption signals. Similar to
Fig.~\protect\ref{KH-Measurement}, the signals show the time
evolution when vortex loops are injected into vortex-free flow
from the AB interface with the setup of Fig.~\protect\ref{setup}
in BAB configuration. The two traces represent the counterflow
peak height (CF) and a maximum close to the Larmor edge of the NMR
signal (Larmor, ($\bullet$)). The corresponding frequency shifts
are denoted with arrows in the NMR absorption spectra in the
insert. The Larmor signal is recorded with the bottom spectrometer
and the CF signal with the top spectrometer. They are not exactly
comparable in amplitudes, but are normalized such that the
integral over the absorption spectrum is unity. The spectra in the
insert are measured with the bottom spectrometer as a function of
the NMR field which has been converted to frequency shift. The
Larmor region is recorded by sweeping the NMR field linearly back
and forth in a narrow interval and by plotting the peak height.
The location of this maximum is not exactly constant during the
time when the Larmor absorption is time dependent. The CF signal
is recorded at fixed NMR field. The initial state is vortex free
rotation at 1.45\,rad/s. Vortex injection is triggered by a small
increase of 0.05\,rad/s which is seen as a small step increase in
the CF signal. At the moment of injection several vortex loops
cross the phase boundaries, undergo a burst of turbulence, and
start the propagation towards the detectors at both ends of the
sample. No change in the signals is seen until the first vortices
reach the closest end of the detector coil and the CF signal
starts decreasing owing to the removal of the azimuthal
counterflow. Following this, the Larmor signal rises to a maximum
$h_\mathrm{max}$ which provides a measure of the axial counterflow
velocity in the twisted vortex state. Subsequently the Larmor
signal decays exponentially in amplitude which corresponds to the
relaxation of the twist when the vortices are connected at one end
to a solid end plate of the sample cylinder. } \label{sequence}
\end{center}
\end{figure}

We begin by examining NMR signals from vortex-injection
measurements with the Kelvin-Helmholtz technique.   As described
in Sec.~\ref{InjectionMethods}, vortex injection into originally
vortex-free flow at low enough temperatures leads to a turbulent
burst, where the number of vortex lines momentarily increases
locally to $N_\mathrm{eq}$ in some cross section of the sample.
The newly created vortices begin to propagate along the rotating
column, gradually replacing the metastable vortex-free state of
large rotational counterflow with vorticity. Figure~\ref{sequence}
shows the measured NMR signals as a function of time, as recorded
with  spectrometers located at a distance of $\sim$ 4\,cm from the
injection site at the AB phase boundary.

The NMR signals in Fig.~\ref{sequence} represent traces of the
absorption as a function of time at two different locations of the
NMR spectrum, corresponding to two distinct maxima in the NMR line
shape. The relative amount of absorption concentrated at these
peaks is a useful measure of the azimuthal counterflow and,
consequently, of local superfluid vorticity: if the counterflow is
small, an absorption peak forms near the Larmor frequency, whereas
in situations with large counterflow most of the absorption
collects at another peak at slightly higher frequency, which
corresponds to the so-called counterflow peak (see insert of
Fig.~\ref{sequence}). In the main panel the heights of the Larmor
and counterflow peaks (marked as CF) are recorded as a function of
time. Actually, in the precise setup shown in Fig.~\ref{setup} the
two peaks are being monitored with two independent spectrometers
located near the opposite ends of the cylindrical sample. In KH
injection these two NMR traces refer to different, disconnected
sets of propagating vortices, but since the sample arrangement is
exactly symmetrical in the two halves, the two signals can be
viewed as representing NMR responses on a common time scale.

At the start of the experiment, the B-phase volume contains no
vortices and is in a state of high counterflow at
$\Omega=1.45\,$rad/s: a large CF signal is then visible, whereas
in the Larmor region there is very little absorption. At time
$t=0$, the angular velocity is suddenly increased to 1.50\,rad/s
(seen as a small increase in the CF peak height), above the KH
critical velocity which has been adjusted between 1.45 and
1.50\,rad/s with the barrier magnetic field. At this moment,
vortices are injected into the B phase sections, undergo the
turbulent burst, and begin to propagate towards the NMR coils.
After a temperature-dependent flight time (approximately 20\,s in
the conditions of Fig.~\ref{sequence}) the vortices arrive at the
two coils, which is signalled by a rapid decrease in the measured
azimuthal counterflow and eventually by a complete disappearance
of the CF signal. Simultaneously, the peak height of the Larmor
signal increases, reaches a maximum, and then slowly relaxes to
the value corresponding to the equilibrium NMR spectrum, with
roughly the equilibrium number of rectilinear vortex lines.

Two interesting conclusions can be immediately drawn from the
responses in Fig.~\ref{sequence}. First, the fact that after the
vorticity arrives at the NMR coil the counterflow signal drops
rapidly from its initial maximum value to zero indicates that the
propagating vortices form a front, which separates the vortex-free
region from the region occupied by vortices. Similar behavior is
observed also in numerical calculations, which will be discussed
below in Sec.~\ref{simu_front}.

The second peculiar feature has to do with the response of the
Larmor signal, its nonmonotonic time dependence. The transient
overshoot of the Larmor peak  appreciably above the
equilibrium-state value is especially noteworthy. Such a response
cannot be obtained with any configuration which would incorporate
only rectilinear lines. Instead, this is a fingerprint of the
helical twist acquired by the propagating vortices. In the
following, we discuss this aspect in more detail and demonstrate
how both numerically calculated NMR spectra for the twisted vortex
state and detailed simulations of vortex motion support this
interpretation.

\subsection{Helically twisted vortex state}\label{front_subsec}

\begin{figure}[t]
\begin{center}
\includegraphics[width=0.45\linewidth]{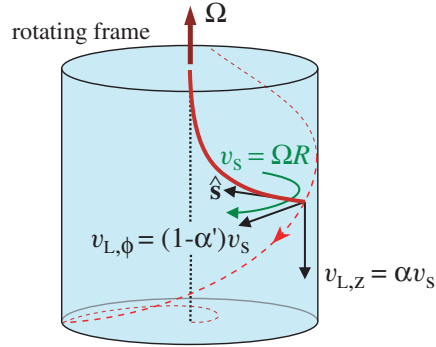}\\
\caption{The motion of a single vortex loop in a rotating
cylinder. If viewed in the frame rotating with the container, the
end of the loop that connects to the side wall moves with a
velocity that has both an azimuthal as well as an axial component.
Therefore, the plane of the loop rotates while the vortex expands
in the rotating flow. The rest of the vortex is aligned along the
central axis which is the equilibrium position of the
vortex.}\label{onevort}
\end{center}
\end{figure}

To understand the NMR features discussed above it is instructive
to begin by studying the motion of a single vortex loop in
rotating flow. Consider the instantaneous velocity of the
vortex-line element ${\hat \mathbf{s}}$ at the point where the
line connects to the cylindrical side wall in Fig.~\ref{onevort}.
In the rotating frame, the velocity of the normal component
vanishes, while the local superfluid velocity is ${\bf v}_{\rm
s}=-\Omega R~\hat{\bphi}$ (here we ignore the small self-induced
contribution from the curvature of the loop). From Eq.~(\ref{vl}),
one then finds the velocity of the element ${\hat \mathbf{s}}$ as
\begin{equation}
{\bf v}_{\rm L}=-(1-\alpha')\Omega R~\hat{\bphi}-\alpha\Omega
R~\hat{\bf z},
\end{equation}
{\it i.e.} the vortex end has both azimuthal and vertical velocity
components, as shown in Fig.~\ref{onevort}.  From this, we
immediately find that the expansion time, or the ``time of
flight'' which is required for the single curved vortex to advance
an axial distance $d$ along the rotating column, is
$d/(\alpha\Omega R)$. Combined with this, the azimuthal velocity
gives rise to a spiralling motion where the plane of the loop is
constantly turning during its expansion.

\begin{figure}[t]
\begin{center}
\includegraphics[width=0.6\linewidth]{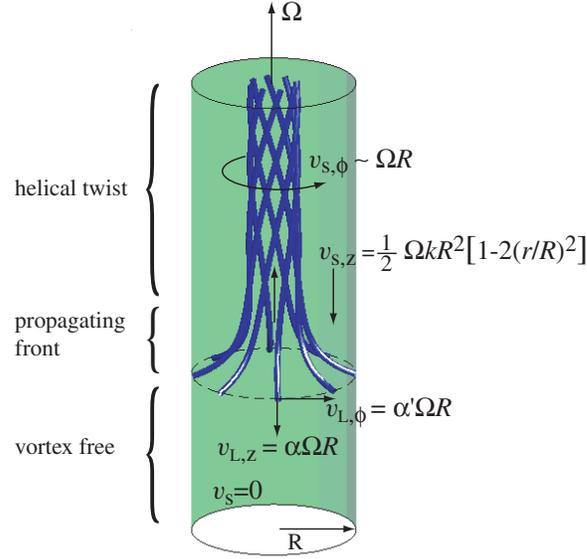}\\
\caption{The vortex front expands along the rotating column into
the region with vortex free counterflow with the longitudinal
velocity v$_\mathrm{L,z} \approx \alpha\Omega$R. In the NMR signal
the propagating vortex front is manifested by the rapid
disappearance of the counterflow signal and by the increase in
absorption in the Larmor region of the spectrum
(Fig.~\ref{sequence}). Behind the front the number of vortex lines
is close to that in equilibrium and also superfluid flow is much
closer to the ultimate state of solid body rotation. The vortices
in the front move also in the azimuthal direction with the
velocity $v_{{\rm L},{\phi}} \approx \alpha'\Omega R$ and thus
generate a helically twisted vortex bundle. The twist gives rise
to superflow in the z-direction with the velocity $v_\mathrm{s,z}$
such that at the center of the cylinder there is upward flow and
close to the wall downward flow (the case in this figure where the
front propagates down). The helical twist with its z-directional
flow is seen as an overshoot of the NMR absorption signal in the
Larmor region of the spectrum, while the relaxation of the twist
explains the exponential decay of the Larmor overshoot
(Fig.~\ref{sequence}). The velocities in this figure are marked as
viewed from the inertial laboratory frame and the flow velocities
according to the uniform twist model.}\label{front}
\end{center}
\end{figure}

Consider next the situation with a large number of loops (close to
$N_{\rm eq}$) in a configuration similar to that in
Fig.~\ref{onevort}: one end connecting to the top end plate, and
the other to the side wall. The local superfluid velocity at any
element on one of these loops is then modified by the contribution
induced by the other vortices. In particular, the vortex segments
at the top end plate would be expected to have ${\bf v}_{\rm s}
\approx {\bf v}_{\rm n}$, and be in solid-body rotation with the
container. Hence the azimuthal counterflow experienced by the
vortex elements changes from zero to $\sim \Omega r$ over the
volume occupied by vortices. On the other hand, both numerical
simulations and experimental data (see below for more details)
indicate that the flight time for a cluster of vortices remains
almost unchanged from the single-vortex value. The combined effect
of these two motions is to drive the vortex cluster into a
helically twisted configuration, as shown schematically in
Fig.~\ref{front}. Estimating $(1-\alpha')\Omega$ as the angular
velocity of azimuthal motion, and taking $\alpha \Omega R$ for the
axial velocity, we arrive at the approximate value $k \sim
(qR)^{-1}$ for the wave number of the helical twist.

A more extended discussion of the formation and properties of the
twisted vortex state can be found in Ref.~\cite{twist}. Most
importantly, however, the helical structure implies superflow in
the axial direction which, as we now proceed to discuss, explains
the peculiar transient overshoot in the NMR absorption near the
Larmor frequency, as observed in the measurement of
Fig.~\ref{sequence}.

\subsection{Superflow field of twisted state: model of uniform twist}

We will now construct a model for the superfluid-velocity profile
in the twisted vortex state. This is achieved by noting that the
equation of motion for superfluid hydrodynamics,
Eq.~(\ref{SuperfluidHydrodynamics}), allows stationary ($\partial
{\bf v}_{\rm s} /\partial t=0$) solutions which represent twisted
vorticity. The simplest one of these is translationally invariant
along the axial ($z$) direction and of the form
\begin{equation}
{\bf v}_{\rm s}=v_{\phi}(r)~\hat{\bphi}+v_z(r)~\hat{\bf z}.
\end{equation}
This situation corresponds to one where the wave vector $k$ of the
twist is independent of the radial coordinate $r$, or
\begin{equation}
\frac{\omega_\phi}{\omega_z}=kr. \label{twist-k}
\end{equation}
Here ${\bomega}_{\rm s}=\omega_{\phi}\hat{\bphi}+\omega_z\hat{\bf
z}$ is the local coarse-grained superfluid vorticity. The
condition (\ref{twist-k}) simply means that in the case of uniform
twist the azimuthal tilt of the vortices has to increase with
radial distance from the cylinder axis. In addition, we impose a
further condition on ${\bf v}_{\rm s}$ by requiring that the
mutual-friction-induced force on the vortices vanishes, {\it i.e.}
\begin{equation}
{\bomega}_{\rm s} \times ({\bf v}_{\rm s}-{\bf v}_{\rm n})=0,
\end{equation}
meaning that the vortices are aligned parallel to the counterflow
in a ``force-free" configuration.  An equivalent condition would
be to require that radial vortex motion is absent: $v_{{\rm
L},r}=0$ from Eq.~(\ref{vl}). Under these requirements, the
superfluid velocity field can be solved as
\begin{eqnarray}
\nonumber v_\phi(r)&=&\frac{\Omega r+(kr) v_0}{1+(kr)^2}, \\
v_z(r)&=&\frac{v_0-(kr)\Omega r}{1+(kr)^2}. \label{vz_twist}
\end{eqnarray}
The parameter $v_0$ is fixed to
$v_0=(\Omega/k)\{(kR)^2/\ln[1+(kR)^2]-1\}$ by requiring that the
net flow through any cross section of the cylinder vanishes.

The NMR response from a twisted vortex state with a flow profile
like that in Eq.~(\ref{vz_twist}) can be roughly understood in
terms of the following simple arguments. The local resonance
frequency depends on the orientation of the B phase order
parameter which, in turn, is affected by the magnitude -- and
orientation -- of the local counterflow velocity. This can be
described in terms of a free-energy contribution which has the
form $-[\hat{\bf l}_{\rm B} \cdot ({\bf v}_{\rm n}- {\bf v}_{\rm
s})]^2$, where $\hat{\bf l}_{\rm B}=[-\hat{\bf z}+5(\hat{\bf
z}\cdot \hat{\bf n})\hat{\bf n}+\sqrt{15}(\hat{\bf
z}\times\hat{\bf n})]/4$ is the orbital anisotropy axis (in the
presence of an axial magnetic field) corresponding to the
order-parameter vector $\hat{\bf n}$. Therefore, counterflow along
the cylinder axis should have a tendency to favour the orientation
$\hat{\bf l}_{\rm B}\parallel\hat{\bf n}\parallel \hat{\bf z}$.
This in turn corresponds to additional NMR absorption at the
Larmor frequency, as observed in Fig.~\ref{sequence}. As shown
below in Sec.~\ref{expdata}, these arguments are further supported
by more detailed comparison to order-parameter textures which are
associated with flow profiles of the form given by
Eq.~(\ref{vz_twist}).

\begin{figure}[t]
\begin{center}
\includegraphics[width=0.6\linewidth]{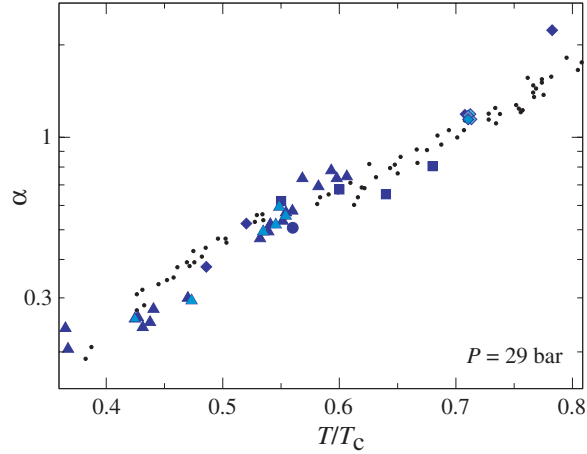}\\
\caption{The dissipative mutual friction coefficient $\alpha$,
plotted on a logarithmic scale as a function of temperature. It is
determined from the axial flight time, assuming
$v_\mathrm{L,z}=\alpha\Omega R$, regardless of the number of
vortex lines in the front. Flight time data have been used from
NMR measurements to extract $\alpha=d/(t_\mathrm{flight}\Omega R)$
where $d$ is the distance which the vortex lines have traveled.
These are marked with large size filled symbols, which correspond
to different methods of determining the flight time (for details
see \cite{flight_time_jltp}). The data marked ($\blacktriangle$)
is measured using the method explained in
Figs.~\ref{KH-Measurement} and \ref{sequence}. The small-size
symbols ($\bullet$) represent the data from Ref.~\cite{bevan} for
comparison. The agreement in $\alpha$ values between the different
sets of data is well within the accuracy of the individual
measurements, which can be argued to justify the model for the
flight time of the first vortices in the vortex front at $T
\gtrsim 0.45\,T_{\rm c}$.}\label{alpha}
\end{center}
\end{figure}


\subsection{Experimental results on twisted vortex state}
\label{expdata}

Next we note some further experimental details on the propagating
vortices and the twisted vortex state. Fig.~\ref{alpha} shows
measurements on the flight time of the vortex front as a function
of temperature. In these experiments, the Kelvin-Helmholtz
instability is triggered and the time of flight  $t_{\rm flight}$
for the vortex lines to reach the closer end of the pick-up coil
(a known distance $d$ away) is recorded. The data has been
presented in terms of the mutual-friction parameter $\alpha$, as
extracted from the relation $\alpha=d/(t_{\rm flight}\Omega R)$,
which is based on the assumption that the vortices propagate with
axial velocity $\alpha\Omega R$. The degree of validity of this
assumption can be assessed by comparing to the previously
published data of $\alpha(T)$ from Ref.~\cite{bevan}.

The division of vortex propagation into regular and turbulent
regimes as a function of temperature means that different
experimental procedures had to be followed, to achieve as similar
initial conditions for the measurements as possible. In the
low-temperature (low damping) regime, $T \lesssim 0.6 \, T_{\rm
c}$, a KH vortex injection always initiates turbulent vortex
formation, where nearly the equilibrium number of vortices ($N
\approx N_{\rm eq}$) is created and then propagates along the
column. However, in the regular regime at temperatures above $0.6
\, T_{\rm c}$ the vortex number is conserved and only the few
initially injected loops expand in the column. In this regime a
much larger number of vortices can be injected using the
Kelvin-Helmholtz method differently: the sample is rotated at high
$\Omega \gtrsim 2$ rad/s (well above the KH critical velocity)
with only B phase in the column. Next the barrier field is
increased by an incremental amount so that suddenly a narrow
sliver of A phase is created in the middle of the column. When the
A phase forms, the AB interfaces are unstable from the start,
until a large number of vortices have been transferred to the
surrounding B phase.

\begin{figure}[t]
\begin{center}
\includegraphics[width=0.45\linewidth,clip]{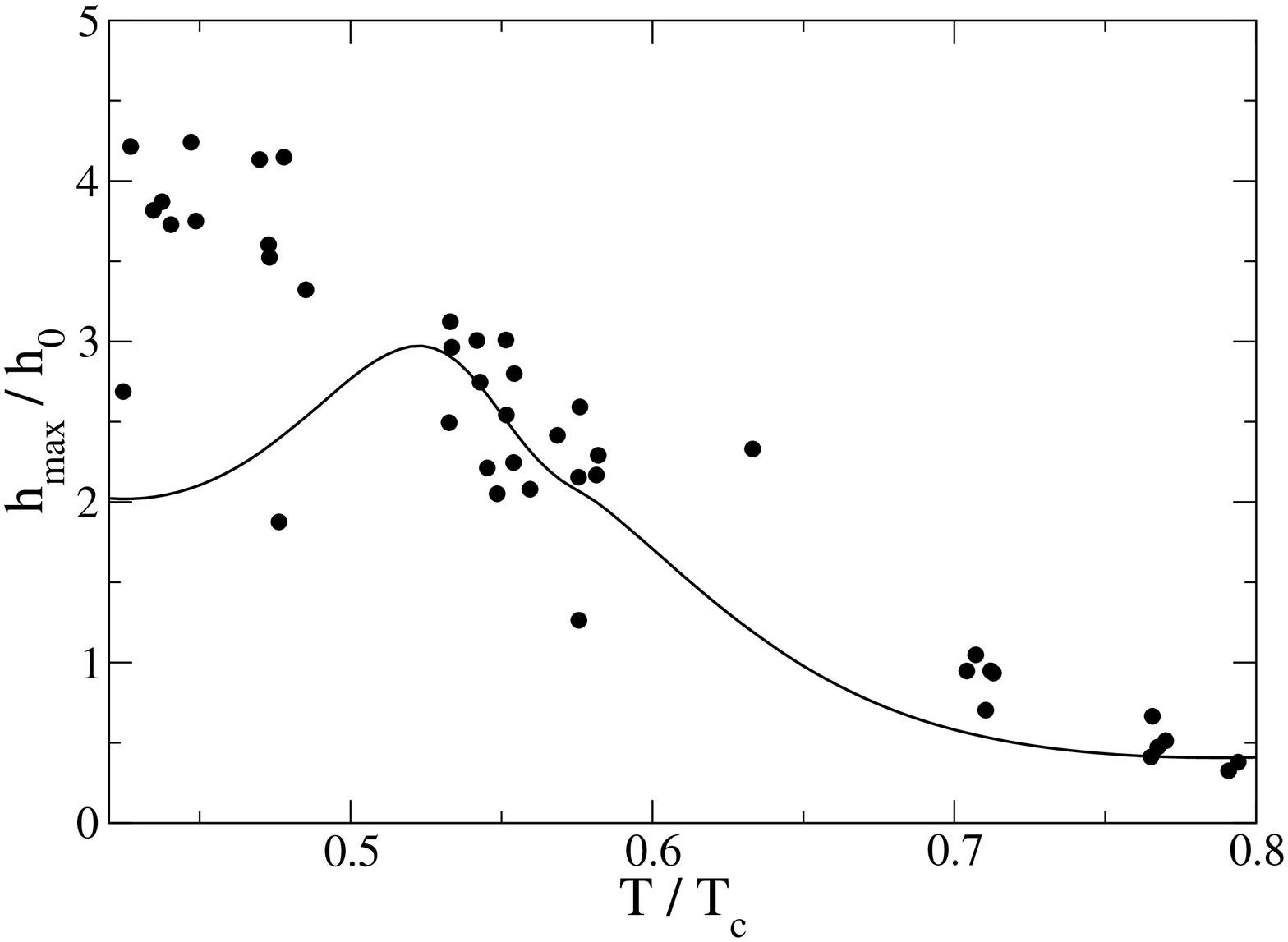}
\includegraphics[width=0.45\linewidth,clip]{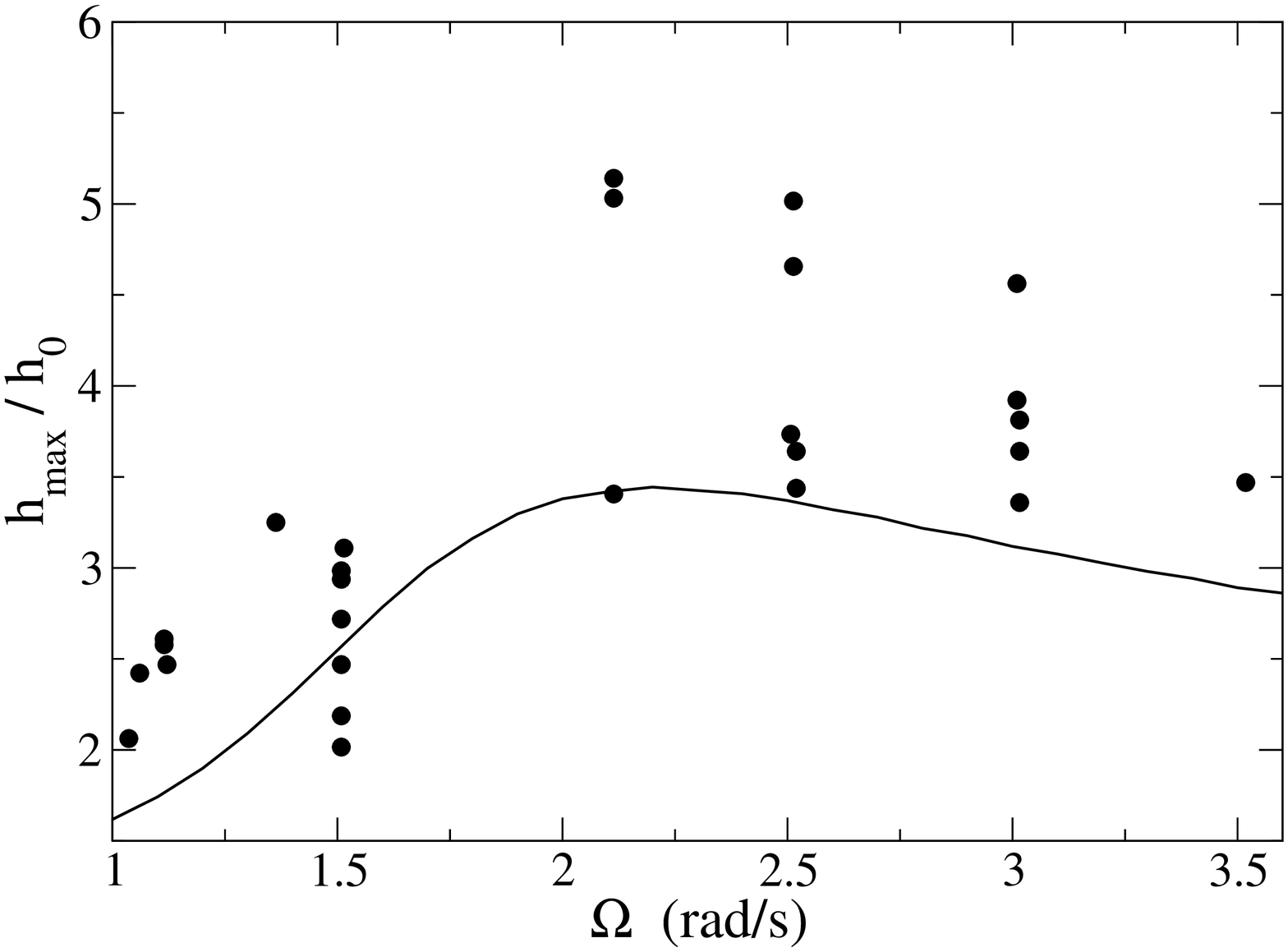}
\caption{{\em (Left)} The amplitude of the maximum Larmor
overshoot $h_{\rm max}$, divided by the maximum $h_0$ of the
corresponding absorption spectrum in the nonrotating state
$\Omega=0$, plotted as a function of temperature. The measurements
have been performed at a pressure of $P = 29.0\,$bar. Below
$0.60\,T_{\rm c}$ the rotation velocity $\Omega$ is as shown for
the KH instability in Fig.~\ref{KH-8AvsT}, while above
$0.7\,T_{\rm c}$ a higher rotation velocity of $\Omega =
2.5\,$rad/s is used. The solid line shows the corresponding
quantity from numerically calculated NMR absorption spectra, using
the uniform-twist model with $kR=1/q$. {\em (Right)} The amplitude
of the maximum overshoot divided by the maximum of the $\Omega=0$
spectrum as a function of $\Omega$. These measurements have been
performed in the temperature interval 0.50 -- $0.55\,T_{\rm c}$.
The solid line is the theoretical prediction of the uniform-twist
model with $kR=1/q$.} \label{overshoot}
\end{center}
\end{figure}

As mentioned above, the characteristic signature of the twisted
vortex state in the NMR spectrum is the absorption overshoot in
the vicinity of the Larmor frequency. A quantitative measure is
the maximum peak height of the overshoot during its time-dependent
evolution, which is denoted with the symbol $h_\mathrm{max}$ in
Fig.~\ref{sequence}. Fig.~\ref{overshoot} shows the dependence of
the maximum peak height on temperature. Here the peak height is
normalized to the corresponding peak height $h_0$ of the NMR
spectrum from the nonrotating state at $\Omega=0$.  The solid line
in the figure represents the corresponding quantity obtained from
numerically calculated NMR line shapes. In these calculations, the
equilibrium order-parameter distributions were determined by
minimizing the textural free energy in the presence of superflow
according to the uniform-twist model, Eq.~(\ref{vz_twist}). The
wave vector of the twist was chosen as $k=1/(qR)$ (see discussion
in Sec.~\ref{front_subsec}). Towards low temperatures the
temperature dependence of the Larmor overshoot is seen to reflect
the exponential dependence of $q^{-1}$ on $1/T$: the twist becomes
tighter with decreasing temperature and leads to an increased
superfluid velocity in the axial direction, and hence to more
excess absorption near the Larmor frequency. The dependence of the
overshoot on the angular velocity $\Omega$ is presented on the
right in Fig.~\ref{overshoot}.

The theoretical curves in Fig.~\ref{overshoot} have been
calculated without adjustable parameters. In view of this, the
agreement between experiment and theory can be considered as
remarkably good at higher temperatures where the model of uniform
twist could be expected to apply. Thus the calculations of the NMR
spectra corroborate the explanation of the NMR measurements in
terms of a traveling vortex front and a trailing twisted vortex
state. Since the $^3$He-B NMR absorption spectrum does not
uniquely define the order parameter texture (see Appendix), the
agreement in Fig.~\ref{overshoot} might not entirely exclude other
possibilities. Nevertheless, it appears safe to conclude that a
new dynamical vortex state has been identified.  In the following
section we describe numerical simulations of the propagation of
vortices in the rotating column. They provide additional
justification for both the vortex front and the twisted state.

\subsection{Propagating vortex state in simulations}\label{simu_front}

Our numerical simulations are performed using the vortex filament
model \cite{schwarz_85,schwarz_88} where vortices are included
individually as topologically stable line objects. The local
superfluid velocity at a particular vortex segment
$\hat{\mathbf{s}}$ is calculated from a Biot-Savart integral by
summing the flow contributions from all other vortex elements. A
discretized Laplace equation is solved to take into account
boundary conditions on vortices and their image fields in the
geometry of a finite cylinder (for details see
Ref.~\cite{Simulation_JLTP}).

Consider first the case of a single vortex expanding in rotation
(Fig.~\ref{onevort}). Ignoring the vortex curvature, the radially
oriented endpoint on the cylinder wall propagates along the
cylinder with the velocity $\alpha\Omega R$ and rotates with
angular velocity $\alpha'\Omega$ (in the laboratory frame). The
self-induced superfluid velocity of the curved vortex (together
with the boundary contribution due to the image field) is mainly
along $\mathbf{v}_\mathrm{n}$ and reduces the counterflow
velocity. This does not affect the longitudinal propagation
velocity substantially if the rotation velocity is large, but at
low $\Omega$ the difference is easily observable, and a somewhat
slower propagation speed than $\alpha\Omega R$ is realized.
Similarly, the self-induced velocity tends to speed up the
azimuthal rotation of the vortex. This is especially visible at
low temperatures where $\alpha'$ is small and the self-induced
velocity gives the main contribution to the rotation. A smaller
cylinder radius enhances these effects further.

\begin{figure}[t]
\centerline{\includegraphics[width=0.9\linewidth]{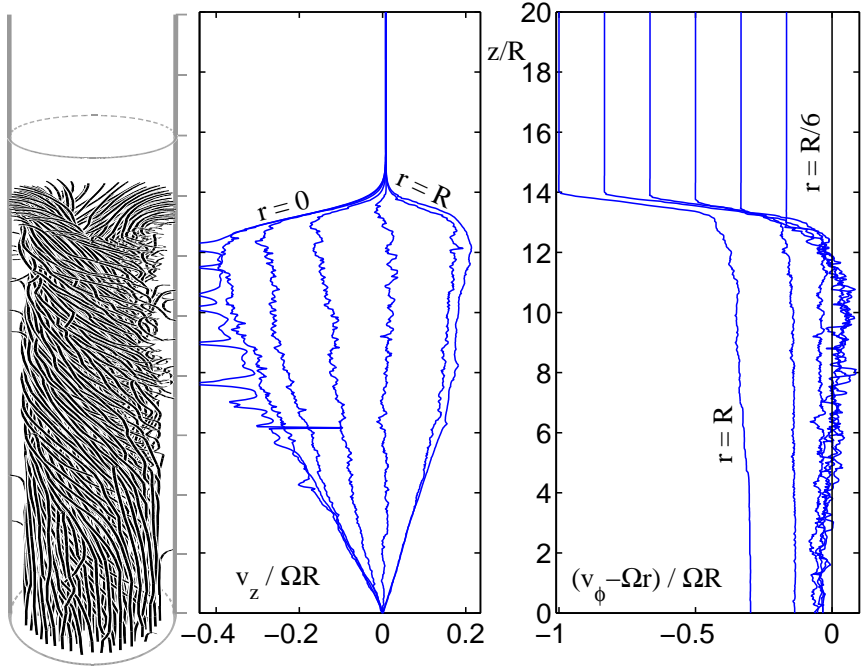}}
\caption{Vortex configuration \emph{(left)}, longitudinal
counterflow velocity $v_{\rm z}(r)$ {\em (center)}, and azimuthal
counterflow velocity $v_{\phi}(r)$ {\em (right)} at $t$ = 60\,s
after the start of the propagation along a cylinder of radius $R$
= 1.5\,mm and length $L$ = 40\,mm at $0.40\,T_{\rm c}$ and
$\Omega$ = 1.0\,rad/s. Only a  30\,mm  long section of the lower
end of the cylinder is shown. For clarity the axial and radial
dimensions have different scales in the left panel. The velocities
in the center and right panels are shown at radii $r = nR/6$, with
integer $n$. The spikes on the $r=0$ traces originate from the
noise in the position of the center-most vortex with respect to
the axis of the cylinder. The boundary condition on the vortices
at the bottom end plate of the cylinder at $z=0$ causes the twist
and $v_{\rm z}(r)$ to vanish there. A stable time-invariant vortex
front travels along the column and has here reached a height $z
\approx 12$ -- $14 \, R$. This is evident from the fact that the
radial velocity profiles $v_{\rm z}(r)$ and $v_{\phi}(r)$ do not
change as a function of time $t$.} \label{simufrontT40}
\end{figure}

\begin{figure}[t]
\centerline{\includegraphics[width=0.9\linewidth]{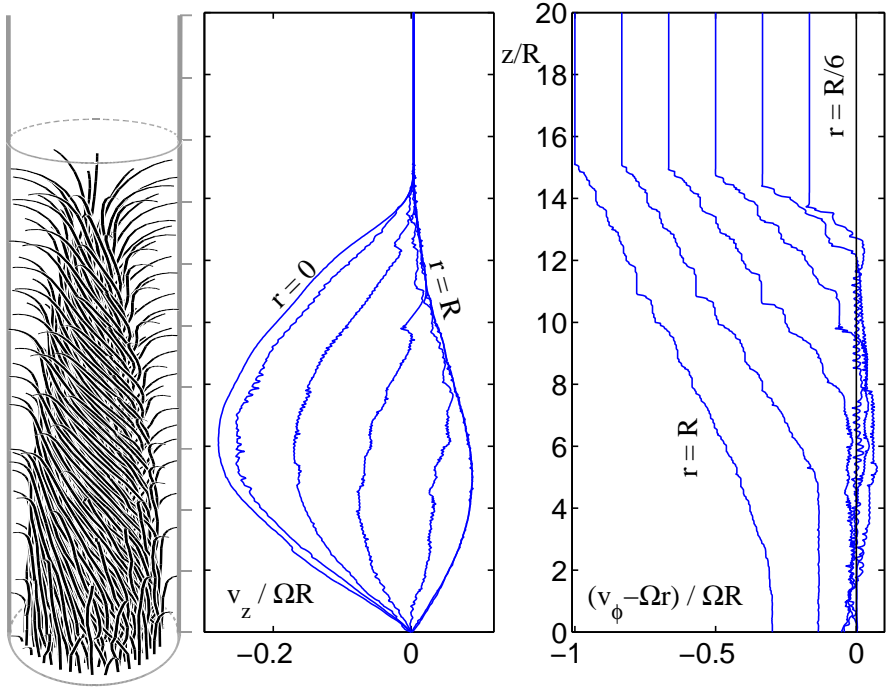}}
\caption{Same plots as in Fig.~\protect\ref{simufrontT40}, but
calculated at $0.60\,T_{\rm c}$ and $t=25\,$s. Here the upper ends
of the vortices are distributed approximately evenly along the
length of the cylinder and it is not possible to identify a time
invariant structure for the front.} \label{simufrontT60}
\end{figure}

The same principles are at work when many vortices expand along
the column. Such calculations are started by placing
$N_\mathrm{eq}$ curved seed vortices in the cylinder. The usual
starting configuration is that with one end of the vortex on the
bottom end plate of the cylinder and the other end bent to the
cylindrical side wall. The locations of the vortex ends on the
bottom plate are the equilibrium positions of rectilinear vortex
lines in solid-body rotation. The choice of these locations and
the way how the vortices are bent to the cylindrical wall prove to
have a minor effect on the outcome from the simulations. This is
because during the first few seconds the vortices move rapidly
around owing to their strong curvature, which upsets, for
instance,  the ordered initial configuration on the bottom end
plate.

In the measurements $ N_\mathrm{eq}$ is  of order $10^3$. For
numerical calculations such a large number of vortices is too
time-consuming, since the spatial resolution has to be kept below
the average inter-vortex spacing ($\sim (2\Omega/\kappa)^{-1/2}
\sim 0.2\,$mm). In practice this limits the calculations to low
rotation velocities or, as was done here, to smaller cylinder
radii than in the experiments. With a large number of vortices
there are more reconnections between vortices, which partly
disrupts and straightens the twisted state.
Figs.~\ref{simufrontT40} and \ref{simufrontT60} illustrate the
characteristics of the propagating front and the twist which the
former generates at $0.40\,T_{\rm c}$ and $0.60\,T_{\rm c}$
temperatures. At the lower temperature of $0.40\,T_{\rm c}$ a
sharp and stable vortex front is formed, but at the higher
temperature of $0.60\,T_{\rm c}$ this is not the case. Thus the
stability of the front is strongly temperature dependent, although
in both cases the trailing vortex bundle behind the front is
clearly twisted.

The stability of the front is governed by the axial counterflow
velocity and thus by the pitch of the twist immediately behind the
front. At high temperatures the axial counterflow is weak. Here
the vortices which happen to fall behind the front feel a reduced
azimuthal counterflow $v_{\rm n}-v_{{\rm s},\phi}$
(Fig.~\ref{simufrontT60} {\em (right)}) and, as a result, their
longitudinal propagation slows down. This is seen in simulations
at high temperatures where the longitudinal spread of the vortex
ends on the cylindrical wall increases continuously with time. At
lower temperatures the axial flow is larger, reaches a clear
maximum just behind the front (Figs.~\ref{simufrontT40} {\em
(center)}), and provides a compensating longitudinal force on the
vortices which fall behind. This is seen by considering the motion
of a radially oriented vortex segment ending on the cylindrical
wall. The longitudinal velocity of this vortex element is given by
$v_{L,z} = \alpha(v_{\rm n}-v_{{\rm s},\phi})+(1-\alpha')v_{{\rm
s},z}$. On moving towards low temperatures, where both $\alpha$
and $\alpha'$ approach zero, but are similar in magnitude, the
increasing axial velocity $v_{{\rm s},z}$ behind the front is
sufficient to compensate for a reduced azimuthal counterflow, if a
vortex falls behind the front.

\begin{figure}[t]
\begin{center}
\includegraphics[width=0.46\linewidth]{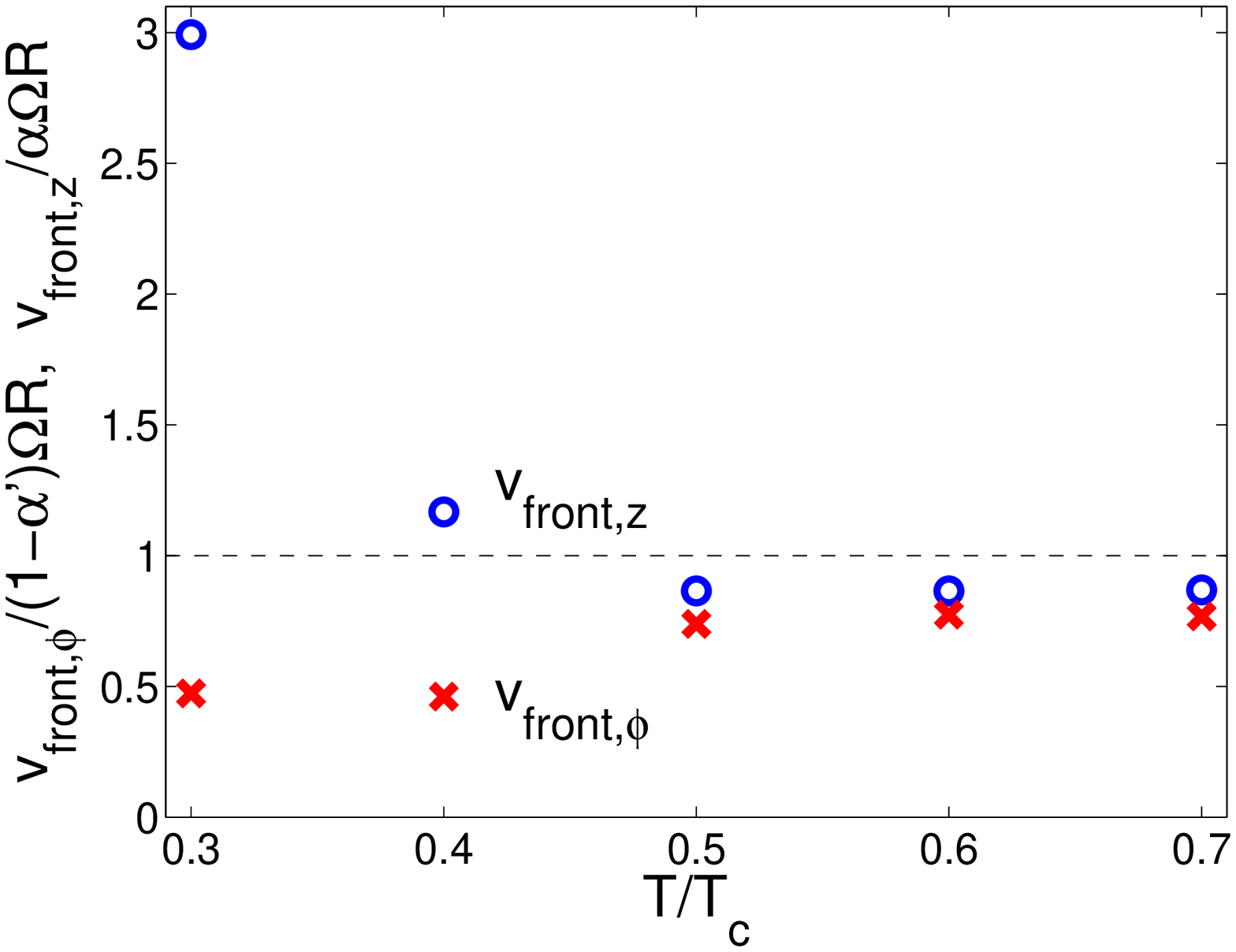}
\includegraphics[width=0.45\linewidth]{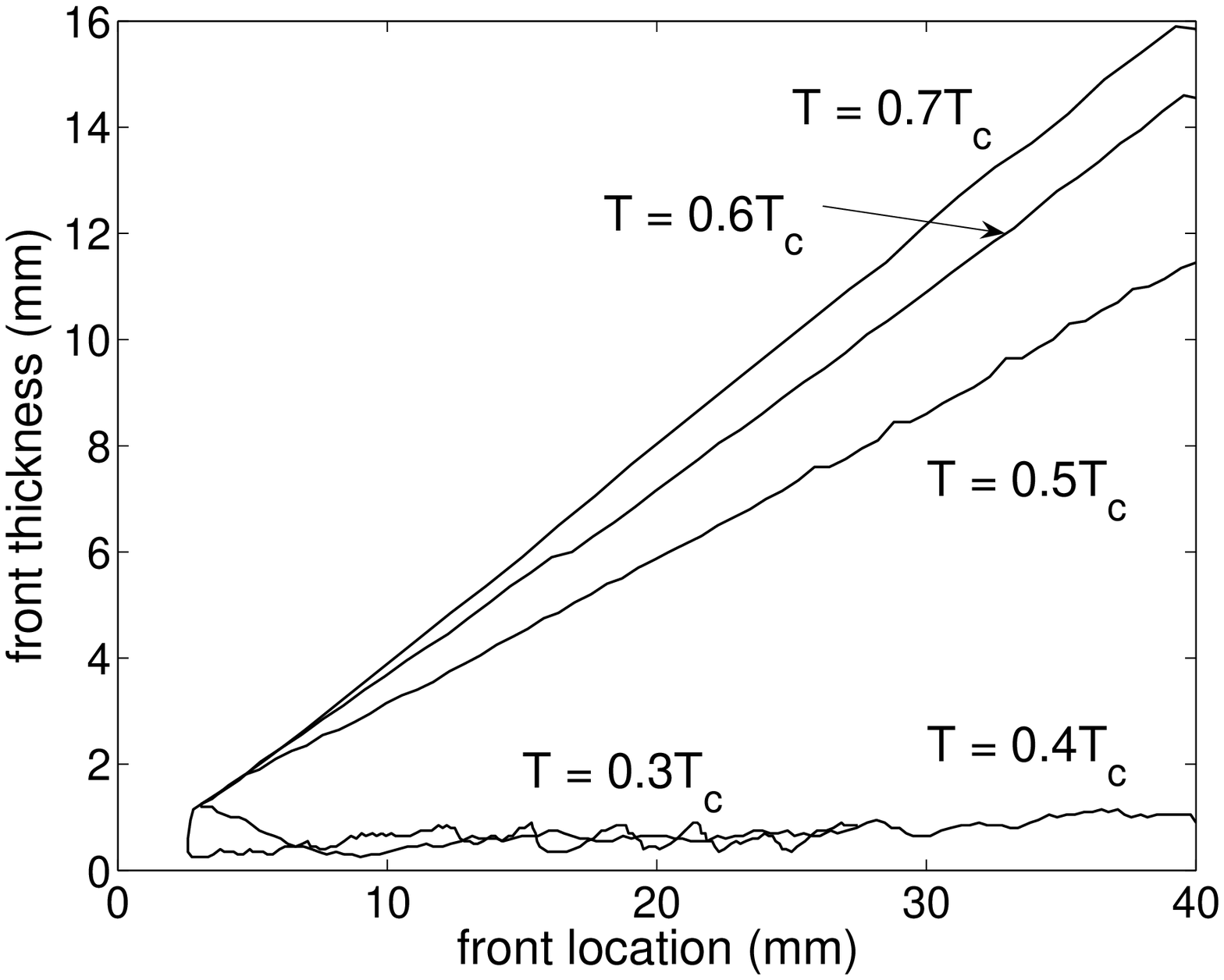}
\caption{{\it (Left)} Vortex-front velocity in simulations with a
cylinder of radius $R$ = 1.5\,mm and length $L$ = 40\,mm rotating
at $\Omega$ = 1.0\,rad/s (expressed in the rotating frame). The
axial velocity $v_{\rm front,z}$ (denoted with $\circ$) was
obtained from the location $z$ of the front at different points in
time. The front was defined to be at the location $z$ where the
azimuthal counterflow at $r = R$ had decreased 2\% from its
maximum value. The azimuthal velocity $v_{\rm front,\phi}$
(denoted with $\times$) is the average velocity of the vortex
endpoint at $r = R$ when the front is at $z \approx L/2$. In the
averaging only the vortices within 5\,mm from the front were taken
into account. Note that the enhancement of the axial velocity
above the value $v_{\rm front,z} \approx \alpha\Omega R$ at
temperatures below $0.45\,T_{\rm c}$ is caused by the longitudinal
velocity contribution from the twist. {\it (Right)} Longitudinal
width or thickness of the front in the same simulation
calculations. At low temperatures below $0.45\,T_{\rm c}$ the
tight twist generates a large longitudinal velocity contribution
which stabilizes a narrow front. At high temperatures vortices lag
more and more behind the front and its width slowly increases
during the propagation along the cylinder. In this regime the
propagating structure does not qualify as a time invariant
traveling solution.} \label{simufront}
\end{center}
\end{figure}

Above $0.45\,T_{\rm c}$ the width of the front increases with
time. This is seen in the right panel of Fig.~\ref{simufront},
which illustrates the front thickness as a function of its axial
location. A time-invariant propagating front solution is only
obtained at temperatures below about $0.45\,T_{\rm c}$. Thus the
definition of the front at temperatures above $0.45\,T_{\rm c}$
becomes arbitrary. However, to characterize the propagation we
continue assigning widths to fronts up to $0.7\,T_{\rm c}$ in
Fig~\ref{simufront}. We locate the position of the front to where
the azimuthal counterflow at $r$ = $R$ has decreased 2\% from its
maximum value $\Omega R$. The rear of the front is chosen to be
located at the position where the azimuthal counterflow at $r$ =
$R$ falls by one half of the difference between its maximum and
minimum values (the latter is nonzero because of the vortex-free
region outside the equilibrium vortex bundle). We then define the
front width as the difference between these two positions.

In the left panel of Fig~\ref{simufront} the longitudinal and
azimuthal propagation velocities of the vortex front are shown at
different temperatures above $0.30\, T_{\rm c}$. Above
$0.45\,T_{\rm c}$ the longitudinal propagation velocity is
approximately given by $\alpha\Omega R$ but on cooling below
$0.45\,T_{\rm c}$ the speed starts to deviate from this value. The
rapid rise of the normalized propagation velocity $v_{\rm front,z}
/ (\alpha \Omega R)$ in this regime is a clear indication from the
strong pushing action which arises from the axial counterflow,
generated by the tightening twist. This enhancement in the
longitudinal velocity agrees with recent measurements at
temperatures below $0.40\,T_{\rm c}$.

At high temperatures above $0.7\,T_{\rm c}$ the twist is barely
visible, but it grows rapidly tighter with decreasing temperature.
One of the evident difficulties in the present interpretations is
the stability of the twisted state on cooling to temperatures
below $0.30\,T_{\rm c}$. The wave vector of the twist is
approximately proportional to $1/q = (1-\alpha')/\alpha$.
Ultimately at the lowest temperatures the helical vortices may
become unstable with respect to Kelvin wave formation, which will
lead to vortex reconnections and vortex multiplication. The
details of such breakup of the front depend on temperature, vortex
number, and rotation velocity. With $R$ = 3\,mm, $\Omega$ =
1\,rad/s and with 23 vortices the twist is stable at $0.40\,T_{\rm
c}$, while at $0.30\, T_{\rm c}$ new vortices are generated owing
to the break up in the helical structure by vortex reconnections,
if the original number of vortices is much less than in
equilibrium. From these examples it becomes evident that the
spiralling propagation of vortices in a rotating column cannot be
extrapolated far below $0.30\,T_{\rm c}$ before other phenomena
can be expected to intervene. Therefore it becomes more and more
pressing to understand the correct behavior for vortex propagation
on approaching the zero temperature limit. A second related
problem is the fact that in numerical calculations vortices
connecting to the cylindrical side wall of the rotating column
tend to be more stable than in experiments. The origin for this
difference has not yet been explained.

\section{Concluding remarks}

We have discussed new developments in superfluid hydrodynamics,
which have been identified from measurements on $^3$He
superfluids. These measurements pertain to situations where the
velocity of the externally applied counterflow dominates over
other contributions, which might arise from other existing
vortices. The applied flow is generated with uniform rotation.
This is chosen for experimental convenience, it does not dictate
the existence of the observed phenomena, except for the twisted
vortex state which is a true peculiarity of rotation. Partly the
new phenomena appear owing to the multicomponent order parameter
structure of the \he superfluids. Such effects cannot be
reproduced with the traditional $^4$He-II superfluid, where only
the $U(1)$ symmetry is spontaneously broken. Other phenomena were
revealed because of the different measuring techniques which are
employed in superfluid \he research, in this case uniform rotation
of a long cylindrical sample with multiple detectors. The common
link, which connects the different observations, is the
possibility to stabilize in the long sample one or two AB
interfaces in a two-phase arrangement with vortex-free counterflow
in $^3$He-B.

The starting point is the superfluid Kelvin-Helmholtz instability
of the AB interface, which becomes possible owing to the
difference of three orders in magnitude in the vortex core radii
of $^3$He-B and $^3$He-A. In the former case quantized vortices
have more traditional structure while in the latter case
continuous vorticity is generally formed which is composed of
doubly quantized vortex-skyrmions or of a meandering vortex sheet
where the vortex-skyrmions have been confined as a linear periodic
chain within a domain-wall-like soliton sheet. The critical
velocity for the formation of continuous vorticity is more than an
order of magnitude smaller than for singular $^3$He-B vortices.
This large difference makes it possible to have a shear flow state
in the rotating cylinder, where an equilibrium vortex state of
$^3$He-A coexists with the vortex-free Landau state of $^3$He-B
across the AB interface.

Experimentally the Kelvin-Helmholtz instability of the AB
inter\-face is an un\-usual\-ly robust phenomenon, with a
well-behaved critical velocity for the nondissipative shear flow
state of the two superfluids. Theoretically the instability can be
associated with two critical thresholds of different nature. The
higher threshold corresponds to the traditional critical velocity
in the relative motion of two ideal inviscid liquids, {\it i.e.}
it follows the expression for the classical Kelvin-Helmholtz
instability in the absence of viscosity. This threshold might be
the appropriate one in the limit of zero temperature far from the
wall of the container with no contact to a reference frame. Under
the conditions of the present measurements at temperatures above
$0.3\,T_{\rm c}$, the contact is established by the equilibrium
distribution of quasiparticle excitations and the AB interface
instability occurs at lower critical velocity. This second lower
threshold is also fundamental in the sense that it does not depend
on the magnitude of the interaction with the environment, the only
requirement is the presence of a reference frame imposed by the
environment. In equilibrium this is the frame where the normal
component is at rest. As a result, the lower threshold is not
determined by the relative motion of the superfluid components,
but by the relative motion of each component with respect to the
normal component. Interestingly, both thresholds can be compared
to the instability of quantum vacuum around the black hole. While
the lower threshold corresponds to the instability of quantum
vacuum in the ergoregion close outside the event horizon, the
upper threshold is related to the instability within the event
horizon owing to the black-hole singularity.

The Kelvin-Helmholtz instability of the AB interface has become
important as a tool to inject reproducibly many small vortex loops
in tight proximity with each other in B-phase flow. This process
was the key to the identification of the transition to turbulence
as a function of mutual friction, controlled by the intrinsic
damping parameter $1/q=(1-\alpha')/\alpha$, which here is the
equivalent of the Reynolds number. Kelvin-Helmholtz injection
starts the transition at the lowest value of $1/q$ and is now
believed to be the closest example to the limiting case of a
velocity independent direct transition into turbulence where the
initially injected seed vortices interact and immediately start a
turbulent increase in vortex number. At higher temperatures with a
lower value of $1/q$ no examples exist of vortex multiplication,
even if large numbers of vortices are injected in the applied
flow. In contrast, at lower temperatures and higher values of
$1/q$ the transition becomes increasingly more probable when the
applied flow velocity is increased and the configuration of the
injected vortices is more favourable for interactions between the
expanding seeds. Eventually at the lowest temperatures $\sim
0.30\, T_{\rm c}$ a vortex injection of any kind will always lead
to turbulence. Overall the most unexpected new result is the fact
that in the limiting case of a direct transition to turbulence,
like Kelvin-Helmholtz injection in Fig.~\ref{turb_29bar}, the
threshold becomes velocity independent, as can also be concluded
from numerical simulations and theoretical models. These models
now show that in the theory of homogeneous developed superfluid
turbulence the Kolmogorov-Richardson cascade is modified by the
influence of mutual friction.

In \heb the transition to turbulence at $1/q \sim 1$ is in the
middle of the experimental temperature range in stark contrast to
$^4$He-II. This is a property expected for Fermi liquids in
general. In most superfluid and superconducting systems with
Cooper pairing it is to be expected that $q(T)^{-1}$ crosses unity
at $T \sim 0.5\, T_{\rm c}$, if the spectrum of fermionic
quasiparticle excitations is fully gapped and the system is in the
clean limit where impurity effects can be neglected. The
underlying mechanism for mutual friction is in this case analogous
to the chiral anomaly in relativistic quantum field theories
\cite{volovik_droplet}.

The final topic in this review concerns the evolution of the
vorticity after injection in the long rotating column. Turbulence
in the rotating column is a short burst where the number of vortex
lines rapidly increases. These are immediately polarized by the
rotating flow. After the burst the vortices expand towards the
unstable vortex-free flow in the form of a sharp front which is
followed by a twisted vortex bundle. The expansion is governed by
the mutual friction controlled motion of the vortex lines in the
front which travel in cork-screw-like manner along the cylinder.
These observations raise a host of new questions which await
further work, especially at lower temperatures, to elucidate the
characteristics of the $T \rightarrow 0$ limit. Can the spiralling
propagation of vortices along the rotating column be expected to
remain stable on approaching the zero temperature limit? Is there
some new form of dissipation present at the lowest temperatures,
as suggested by recent measurements \cite{LatestLancaster} of the
decay of vibrating wire or grid generated turbulence at
temperatures below $0.2\,T_{\rm c}$ in $^3$He-B? It is also to be
expected that studies of vortex dynamics in rotating Bose-Einstein
condensates will shed more light on these questions
\cite{BEC-Turbulence,BEC-MutualFriction}.

Finally we wish to emphasize that, as distinct from the
traditional superfluid example case of $^4$He-II, $^3$He
superfluids incorporate interacting Fermi and Bose quantum fields.
This is similar to the situation in relativistic quantum field
theory. In a multi-component order parameter field the analogy
becomes stronger such that quasiparticle excitations become the
counterparts of elementary particles (electrons, neutrinos,
quarks), while the order parameter collective modes resemble the
gauge fields and gravity. Such analogies provide contacts to other
fields of physics and a strong incentive to learn more about the
coherent quantum systems in condensed matter. They also present
astonishing evidence for the underlying universality of physical
principles in seemingly different theories.

\section*{Acknowledgements}

This work was supported in part by the Academy of Finland, Magnus
Ehrnrooth Foundation, and by the ESF research program COSLAB. AF
acknowledges study grants from the Jenny and Antti Wihuri
Foundation, RH from the Foundation for the Advancement of
Technology in Finland, and NBK and GEV from the Russian Foundation
for Basic Research (grant 06-02-16002-a). MT is supported by a
Grant-in-Aid for Scientific Research from JSPS (Grant No.
18340109) and a Grant-in-Aid for Scientific Research on Priority
Areas from MEXT (Grant No. 17071008). We thank W.F. Vinen for
critical examination of an early version of this work.

\appendix

\section*{Appendix: Nuclear magnetic resonance in \heb}\label{NMR}
\addcontentsline{toc}{section}{Appendix: Nuclear magnetic
resonance in \heb}

\setcounter{section}{1}

Continuous wave nuclear magnetic resonance measurement (cw-NMR)
can be used both in $^3$He-A and $^3$He-B to monitor the state of
the rotating sample and to retrieve information about the
structure and configuration of quantized vorticity. In both phases
the measurement entails the task of working out the order
parameter texture from the recorded NMR absorption spectrum,
albeit in different ways. The B-phase spectrum provides a mapping
of the spatial variation of the global order parameter texture.
This texture is affected by the presence or absence of vortices.
Thus indirectly from the changes in the global texture one can
infer information about vortices, up to the point that in certain
situations the vortices can be detected individually, as seen in
Figs.~\ref{VortexSteps} and \ref{stairs}. This is different from
the A phase where NMR absorption is sensitive to the local order
parameter distribution, in particular in locations where its
alignments differ from the spin-orbit equilibrium configuration.
Here each type of topological defect leaves a characteristic
satellite peak in the absorption spectrum
\cite{AphaseSpectrometry}.  The frequency shift of the satellite
identifies the type of defect and the height of the peak the
number of these defects. Thus ``spectrometry'' of different types
of defects in the A-phase order parameter texture is even more
straightforward and powerful than in the B phase. Since B-phase
measurements are in the forefront in this context, in the
following we concentrate on the main characteristics of B-phase
order parameter textures: the influence of vortices on the texture
and how this is reproduced in the NMR absorption spectrum
\cite{Schanen_JLTP}. Technical requirements for single-vortex
resolution and cw-NMR spectrometer design can be found, for
instance, in Ref.~\cite{elektroniikkaa}.

The bulk tensor order parameter of undisturbed $^3$He-B can be
written in the form ${\bf A}= \Delta~e^{i\phi} {\bf R}(\hat{\bf
n},\theta)$, where the (real-valued) quantities $\Delta (T,P)$ and
$\phi$ are the magnitude and the phase of the order parameter, and
${\bf R}$ is a rotation matrix which can be parametrized in terms
of a rotation axis $\hat{\bf n}$ and an angle $\theta$ as $R_{\mu
j} =\cos\theta~\delta_{\mu j}+(1-\cos\theta)~\hat{n}_\mu \hat{n}_j
-\sin\theta~\epsilon_{\mu jk}\hat{n}_k$. While $\Delta$ and $\phi$
are usually fixed by external constraints and $\theta$ by the
dipolar spin-orbit interaction, the orientation of the unit vector
$\hat{\bf n}$ is influenced by various orientational interactions,
giving rise to $\hat{\bf n}({\bf r})$ textures. The appropriate
equilibrium texture  is the  minimum of a free-energy functional,
the main contributions to which in the presence of an external
magnetic field ${\bf H}$~are the field anisotropy energy

\begin{equation}
F_{\rm DH}=-a\int d^3{\bf r}~(\hat{\bf n}\cdot{\bf H})^2,
\label{fdh}
\end{equation}
the surface energy (the unit vector $\hat{\bf s}$ denotes the
surface normal)

\begin{equation}
F_{\rm SH}=-d\int d^2{\bf r}~({\bf H}\cdot {\bf R} \cdot \hat{\bf
s})^2, \label{fsh}
\end{equation}
the energy due to the counterflow velocity field ${\bf v}({\bf
r})= {\bf v}_{\rm n}({\bf r})-{\bf v}_{\rm s}({\bf r})$

\begin{equation}
F_{\rm HV}=-\lambda_{\rm HV}\int d^3{\bf r}~({\bf H}\cdot {\bf R}
\cdot {\bf v})^2, \label{fhv}
\end{equation}
and the vortex contribution

\begin{equation}
F_{\rm LH}=\int_L d^3{\bf r}~\lambda_{\rm LH} ({\bf H}\cdot {\bf
R} \cdot \hat{\bf l})^2, \label{flh}
\end{equation}
where $\hat{\bf l}$ is a unit vector directed along the vortex
line and the integration extends over the volume occupied by
vortices (the information on the vortex density is contained in
$\lambda_{\rm LH}$). Our notation follows that of
Ref.~\cite{erkki_jltp}. In addition to the above energy terms,
spatially varying order-parameter distributions are associated
with a gradient (or bending) energy \noindent
\begin{eqnarray}
F_{\rm G}&=&\int d^3{\bf r}~\left[\lambda_{\rm G1}\frac{\partial
R_{\mu i}}{\partial r_i} \frac{\partial R_{\mu j}}{\partial r_j}+
\lambda_{\rm G2}\frac{\partial R_{\mu j}}{\partial r_i}
\frac{\partial
R_{\mu j}}{\partial r_i}\right]  \nonumber  \\
&+&\lambda_{\rm SG} \int d^2{\bf r}~\hat{s}_j R_{\mu j}
\frac{\partial R_{\mu i}}{\partial r_i} \,, \label{fg}
\end{eqnarray}
where the first integral is taken over the volume and the second
over the surface of the sample. The characteristic length scale of
the B-phase textures can be obtained by balancing the gradient
energy with the bulk magnetic energy $F_{\rm DH}$: the resulting
magnetic coherence length is defined as $\xi_{\rm
H}=\sqrt{65\lambda_{\rm G2}/(8aH^2)}$, and is inversely
proportional to the magnitude of the external magnetic field.
Since the typical values of $\xi_{\rm H}$ are of the order of a
millimeter, the textures are extended and in a usual experimental
setup the finite container size and the associated boundary
effects become important and need to be taken into account.

\begin{figure}[t]
\begin{center}
\includegraphics[width=0.7\linewidth]{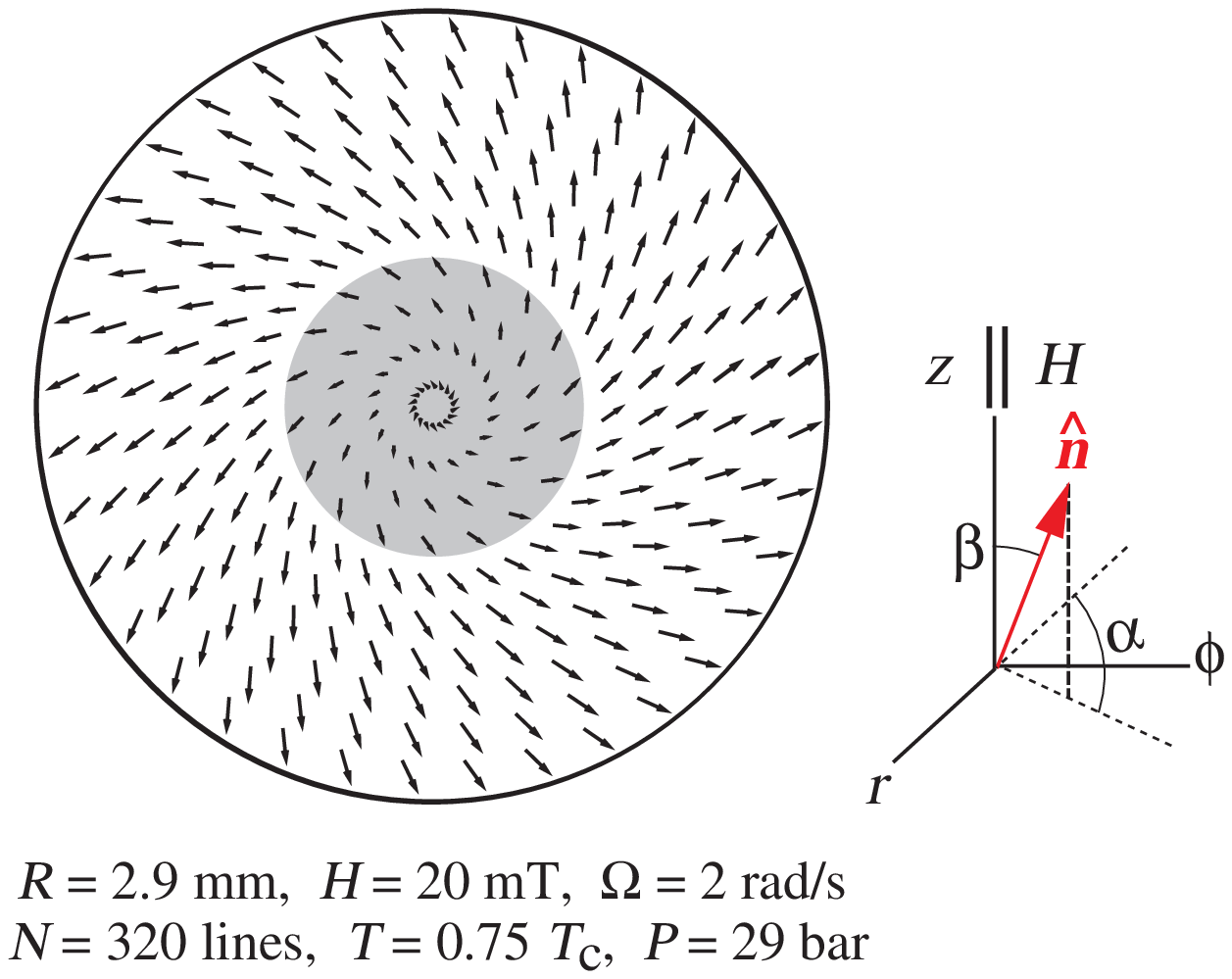}\\
\caption{Axially symmetric flare-out ${\hat \mathbf{n}}$ texture
in an infinitely long cylinder, when the applied field is oriented
along the symmetry axis of the cylinder. The arrows denote the
projection of ${\hat \mathbf{n}}$ in the  plane perpendicular to
the cylinder axis.  The shaded region in the center is the vortex
cluster with 320 rectilinear vortex lines. The texture has been
calculated with the parameters given under the cross section
through the cylinder.} \label{FlareOut}
\end{center}
\end{figure}

Also, it should be noted that the presence of vortices in the
sample volume modifies the texture in several different ways. The
free-energy term $F_{\rm LH}$ contains both the effect of
order-parameter suppression at the vortex cores, and the
orientational effect of the quantized superflow fields circulating
the cores. Additionally, the vortices modify the global
counterflow velocity field ${\bf v}({\bf r})$ entering the
expression of $F_{\rm HV}$; this is typically the dominating
effect.

\begin{figure}[t]
\begin{center}
\includegraphics[width=0.8\linewidth]{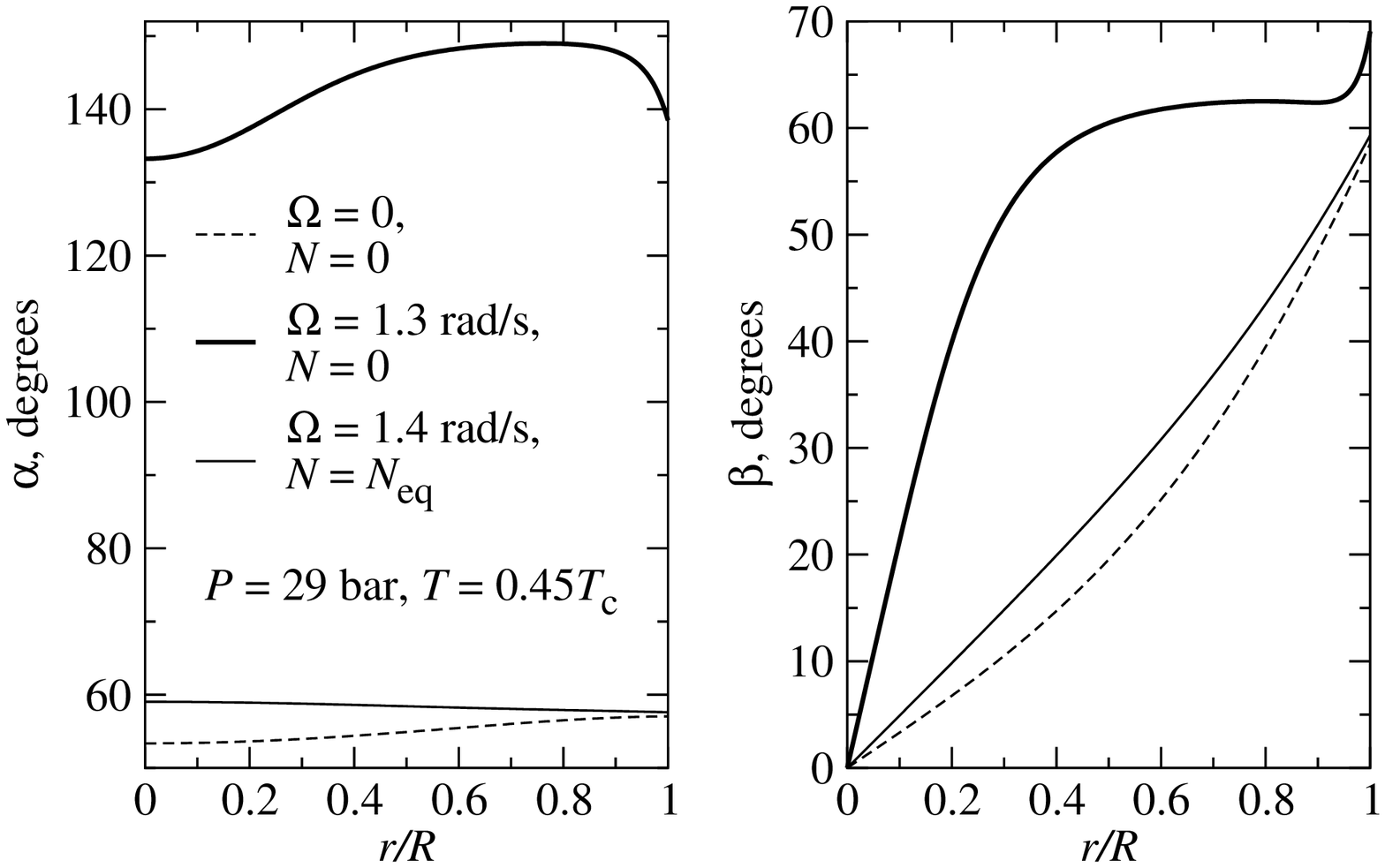}\\
\caption{Calculated radial distributions of the azimuthal angle
$\alpha (r)$ {\it (left)} and polar angle $\beta (r)$ {\it
(right)} of the B-phase order parameter orientation, the unit
vector ${\hat \mathbf{n}}$, in the flare-out textures of the three
NMR spectra in Fig.~\protect\ref{Bspectrum}. }
\label{n_TextureOrientations}
\end{center}
\end{figure}

In a long cylindrical container with the external magnetic field
directed along the axis, the minimum-energy $\hat{\bf n}$ texture
is axially symmetric of ``flare-out'' form \cite{flareout_1,
flareout_2}. In cylindrical coordinates, this can be represented
in the form

\begin{equation}
\hat{\bf n}({\bf r}) =-\sin\beta(r)\cos\alpha(r)~\hat{\bf r}
+\sin\beta(r)\sin\alpha(r)~\boldsymbol{\hat{\phi}}
+\cos\beta(r)~\hat{\bf z}.
\end{equation}
The NMR resonance frequency $\nu$ depends on the polar angle
$\beta$ as

\begin{equation}
\nu(\beta) \approx \nu_0 + \frac{\nu_{\rm B}^2}{2\nu_0}
\sin^2\beta~, \label{resonance}
\end{equation}
where $\nu_0=\gamma H/(2\pi)$ and $\nu_{\rm B} (T,P)$ are the
Larmor frequency and the B-phase longitudinal resonance frequency
\cite{Ahonen,hakonen_spectra}, respectively. This approximate
expression is valid at high external fields, when $\nu_{\rm B} \ll
\nu_0$, which is the usual experimental regime. In the local
oscillator picture (excluding line broadening effects), a
spatially varying $\beta(r)$ leads to a distribution of resonance
frequencies, which then define the absorption spectrum
\cite{hakonen_spectra}
\begin{equation}
P(\nu) \propto \int dr~r~\delta[\nu-\nu(r)], \label{spec}
\end{equation}
where $\nu(r)\equiv\nu[\beta(r)]$. This defines the connection
between the order-parameter texture $\beta(r)$, found by
minimizing the sum of Eqs.~(\ref{fdh})--(\ref{fg}), and the
measured line shape $P(\nu)$.

\begin{figure}[t]
\begin{center}
\includegraphics[width=0.8\linewidth]{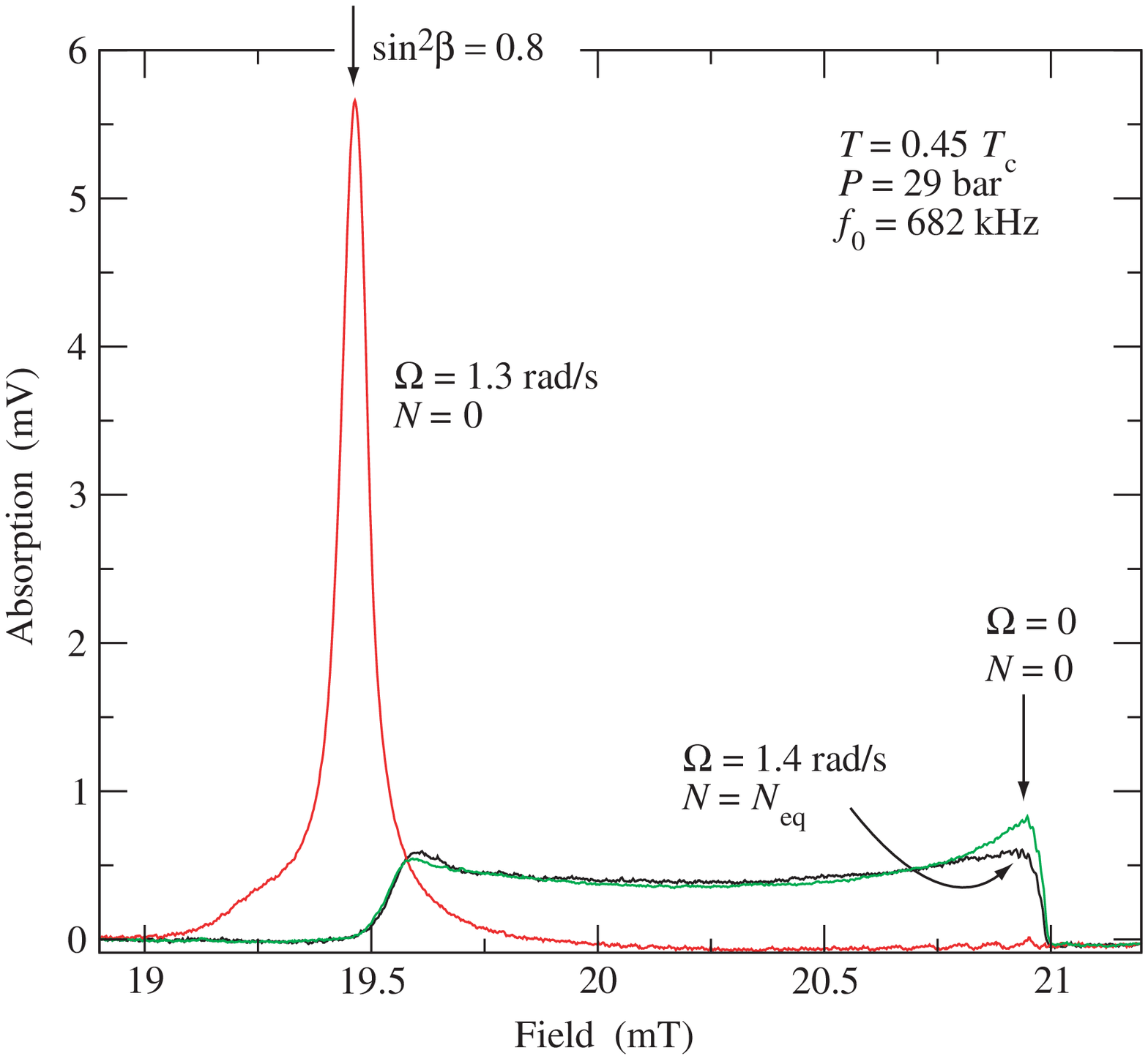}\\
\caption{Examples of NMR absorption spectra measured with the
bottom spectrometer in the setup of Fig.~\ref{setup}. The
measurements are performed with a highly tuned resonance tank
circuit, so that the excitation frequency is kept constant while
the external polarizing field is swept to record the line shape.
The Larmor field is at 21\,mT. This is the value around which the
NMR absorption is centered in the normal phase at temperatures
above $T_{\rm c}$. In the B-phase the Larmor value becomes a sharp
edge which borders the absorption towards high fields (or
equivalently, low frequencies). The oscillations close to the
Larmor edge are spin wave resonances which grow larger in
amplitude on cooling to lower temperatures. The three line shapes
have been measured for (1) vortex-free rotation ($N = 0$, $\Omega
\neq 0$), (2) equilibrium vortex state ($N = N_{\rm eq}$), and (3)
stationary state (nonrotating, $\Omega = 0$). } \label{Bspectrum}
\end{center}
\end{figure}

The form of the relevant free-energy functional in rotating \heb
is complicated, giving rise to several different types of
equilibrium textures depending on the magnitude and direction of
the counterflow in the sample \cite{texture_prl,Kopu_JLTP}.
However, a rough qualitative view of the textures can be obtained
relatively easily. First, consider the situation in axial magnetic
field and zero counterflow: through Eq.~(\ref{fdh}) the field
tends to align $\mathbf{\hat{n}}\parallel{\bf H}$, or
$\sin^2\beta=0$, in bulk. Through Eq.~(\ref{resonance}), this
corresponds to NMR absorption at the Larmor frequency $\nu_0$.
However, the surface energy at the cylindrical sidewall,
Eq.~(\ref{fsh}) with $\hat{\bf s}=-\hat{\bf r}$, favors the
orientation $\sin^2\beta=4/5$ and shifts the NMR absorption to
higher frequencies. A combination of these two orienting effects
and of the gradient energy creates the characteristic even
distribution of NMR  absorption in the nonrotating state of
Fig.~\ref{Bspectrum}.

If the sample is rotated with angular velocity $\Omega$ in the
vortex-free state, then azimuthal counterflow ${\bf v}(r)=\Omega
r~\hat{\bphi}$ is created. Since the flow energy $F_{\rm HV}$ in
this case is also minimized by having $\sin^2\beta=4/5$, this
leads to the formation of the so-called counterflow peak in the
NMR spectrum (Fig.~\ref{Bspectrum}). In contrast, in the
equilibrium vortex state the sample is filled with rectilinear
vortices oriented parallel to the axis of rotation, evenly
distributed with an areal density $n_{\rm v} = 2\Omega/\kappa$,
such that the counterflow vanishes on an average. In this case,
the NMR line shape is similar to that of the nonrotating sample;
the small shift of absorption away from the Larmor region seen in
Fig.~\ref{Bspectrum} arises from the local contribution $F_{\rm
LH}$.

In general, metastable states containing a cluster of any number
of vortices $N$ between zero and the equilibrium number $N_{\rm
eq}$ can be observed in rotating \heb (Fig.~\ref{FlareOut}). In
such a case, the magnitude of the azimuthal counterflow is

\begin{equation}
v(r)=\left\{ \begin{array}{ll}
0, & 0<r<R_{\rm c}, \\
\Omega r - \Omega R_{\rm c}^2/r, & R_{\rm c}<r<R, \end{array}
\right.
\end{equation}
where the cluster radius $R_{\rm c}=R\sqrt{N/N_{\rm eq}}$\,. The
number of vortex lines in a given cluster can be determined by
analyzing the counterflow peak height, either by comparing to
numerically calculated equilibrium textures and their
corresponding spectra \cite{Schanen_JLTP}, or with purely
experimental calibration techniques, which have been described in
Refs.~\cite{VortexFormation,neutron_jltp}. However, in transient
conditions, for instance after the turbulent burst, also more
complicated flow profiles can arise, which are no longer
necessarily purely azimuthal. This is the case with a twisted
vortex cluster which is left behind by an advancing vortex front
(Fig.~\ref{sequence}), as discussed in Sec.~\ref{HelicalBundle}
and Ref.~\cite{Kopu_JLTP}: the nonzero axial component of
counterflow gives rise to excess absorption close to the Larmor
frequency. Thus NMR in \heb can be used to distinguish between
different flow states and vortex configurations, when good models
for the corresponding counterflow profiles are available. Such
models can be obtained from either analytical arguments, or from
detailed numerical simulations of vortices. The latter ones are
especially valuable in complicated dynamic situations. Thus, if
the order parameter texture is known, the NMR absorption spectrum
is obtained from Eq.~(\ref{spec}). However, the reverse is not
generally true: a measured spectrum does not uniquely fix the
order parameter texture, only once many more details are known
about the measuring situation.

\section*{References}
\bibliographystyle{iopart-num-titles}

\bibliography{rpp_finne-b}

\end{document}